\documentclass{emulateapj}

\usepackage{amsmath}
\usepackage{graphicx}
\usepackage{psfig}
\usepackage{apjfonts}
\usepackage{color}

\def\cm{\textrm{cm}}
\def\km{\textrm{km}}

\def\erg{\textrm{ergs}}
\def\kpc{\textrm{kpc}}
\def\pc{\textrm{pc}}
\def\Mpc{\textrm{Mpc}}

\def\Kelv{\textrm{K}}
\def\sr{\textrm{sr}}
\def\Jy{\textrm{Jy}}
\def\sec{\textrm{s}}

\def\ergps{\textrm{ergs}~\textrm{s}^{-1}}

\def\kms{\textrm{km}~\textrm{s}^{-1}}
\def\gcm2{\textrm{g}~\textrm{cm}^{-2}}
\def\ergscm3{\textrm{erg}~\textrm{s}^{-1}~\textrm{cm}^{-3}}
\def\ergcm3{\textrm{erg}~\textrm{cm}^{-3}}
\def\gscm2{\textrm{g}~\textrm{s}^{-1}~\textrm{cm}^{-2}}
\def\ergcmK34{\textrm{erg}~\textrm{cm}^{-3}~\textrm{K}^{-4}}

\def\cms31{\textrm{cm}^{-3}~\textrm{s}^{-1}}
\def\cmg21{\textrm{cm}^{2}~\textrm{g}^{-1}}

\def\phFluxUnits{\textrm{ph}~\textrm{cm}^{-2}~\textrm{s}^{-1}}

\def\keV{\textrm{keV}}
\def\MeV{\textrm{MeV}}
\def\GeV{\textrm{GeV}}
\def\TeV{\textrm{TeV}}
\def\PeV{\textrm{PeV}}

\def\GHz{\textrm{GHz}}

\def\yr{\textrm{yr}}
\def\Myr{\textrm{Myr}}

\def\mbarn{\textrm{mb}}
\def\muGauss{\mu\textrm{G}}

\def\GeVs1cm3{\textrm{GeV}~\textrm{s}^{-1}~\textrm{cm}^{3}}
\def\log{\textrm{log}}
\def\2phn{\phn\phn}
\def\Msun{\textrm{M}_{\sun}}
\def\Lsun{\textrm{L}_{\sun}}

\newcommand{\SciNot}[2]{\ensuremath{#1 \times 10^{#2}}}
\def\bfnop{}

\begin{document}

\title{The Star-Forming Galaxy Contribution to the Cosmic M\lowercase{e}V and G\lowercase{e}V Gamma-Ray Background}
\author{Brian C. Lacki\altaffilmark{1,2}, Shunsaku Horiuchi\altaffilmark{3}, and John F. Beacom\altaffilmark{4,5,6}}
\shorttitle{MeV BACKGROUND FROM STAR-FORMING GALAXIES}
\shortauthors{LACKI, HORIUCHI, \& BEACOM}
\altaffiltext{1}{Jansky Fellow of the National Radio Astronomy Observatory}
\altaffiltext{2}{Institute for Advanced Study, Einstein Drive, Princeton, NJ 08540, USA; brianlacki@ias.edu}
\altaffiltext{3}{Center for Cosmology, Department of Physics and Astronomy, 4129 Frederick Reines Hall, University of California, Irvine, CA 92697-4575 USA}
\altaffiltext{4}{Department of Astronomy, The Ohio State University, 140 West 18th Avenue, Columbus, OH 43210, USA}
\altaffiltext{5}{Center for Cosmology \& Astro-Particle Physics, The Ohio State University, Columbus, Ohio 43210, USA}
\altaffiltext{6}{Department of Physics, The Ohio State University, 191 W. Woodruff Avenue, Columbus, OH 43210, USA}

\begin{abstract}
While star-forming galaxies could be major contributors to the cosmic GeV $\gamma$-ray background, they are expected to be MeV-dim because of the ``pion bump'' falling off below $\sim 100$ MeV.  However, there are very few observations of galaxies in the MeV range, and other emission processes could be present. We investigate the MeV background from star-forming galaxies by running one-zone models of cosmic ray populations, including Inverse Compton and bremsstrahlung, as well as nuclear lines (including $^{26}$Al), emission from core{\bfnop-}collapse supernovae, and positron annihilation emission, in addition to the pionic emission.  We use the Milky Way and M82 as templates of normal and starburst galaxies, and compare our models to radio and GeV--TeV $\gamma$-ray data. We find that (1) higher gas densities in high-$z$ normal galaxies lead to a strong pion bump, (2) starbursts {\bfnop may} have significant MeV emission if their magnetic field strengths are low, and (3) cascades can contribute to the MeV emission of starbursts if they emit mainly hadronic $\gamma$-rays.  Our fiducial model predicts that most of the unresolved GeV background is from star-forming galaxies{\bfnop, but this prediction is uncertain} by an order of magnitude.    About $\sim 2$\% of the claimed $1$ MeV background is diffuse emission from star-forming galaxies; we place {\bfnop a} firm upper limit {\bfnop of $\la 10\%$ based on the spectral shape of the background}. {\bfnop The star-formation contribution is constrained to be small, because its spectrum is peaked, while the observed background is steeply falling with energy through the MeV-GeV range.}
\end{abstract}


\section{Introduction}
\label{sec:Introduction}

\begin{figure*}
\centerline{\includegraphics[width=14cm]{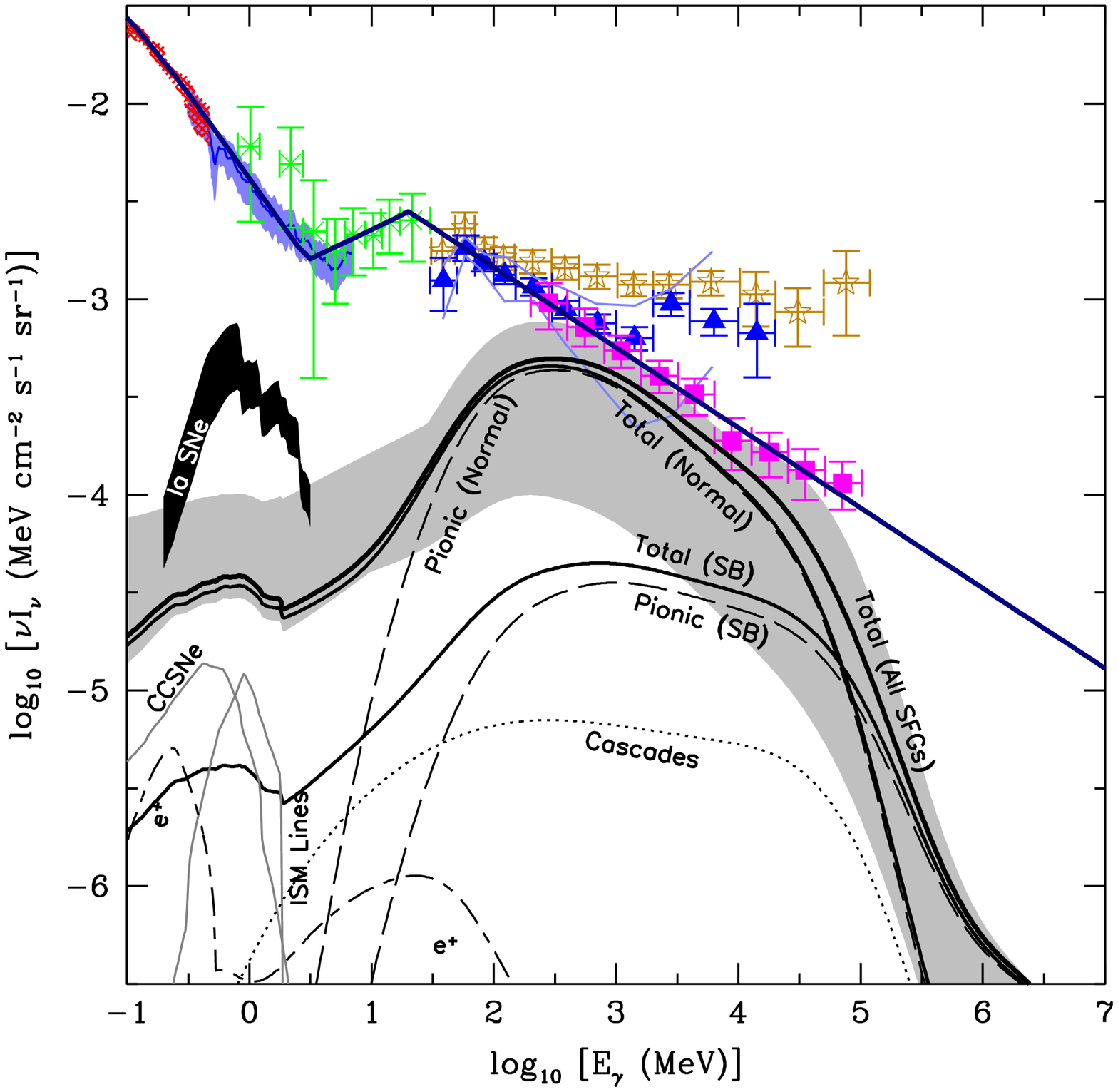}}
\figcaption[figure]{The observed unresolved extragalactic $\gamma$-ray background and the predicted $\gamma$-ray background from star-forming galaxies with our standard assumptions.  Plotted data include the compilation in \citet{Gilli07} (red cross-hatching), SMM (blue shading),  COMPTEL (green X's), EGRET (gold stars for \citet{Sreekumar98}; blue triangles with lines showing the estimated systematic errors from \citealt{Strong04a}), and \emph{Fermi}-LAT (magenta squares).   We show the total predicted background from star-forming galaxies (black, thick solid), and the contributions from normal galaxies and starbursts.  The grey shading encompasses the estimated uncertainty in the background.  We also show various components of the $\gamma$-ray background, including the pionic emission (long-dashed) from starbursts and normal galaxies, positron annihilation radiation (long-dashed/short-dashed), nucleosynthetic $\gamma$-ray lines and radiation from CCSNe (grey solid), and cascade emission (from starbursts only; dotted).  The separate contributions from IC and bremsstrahlung contribution are not shown, but they make up most of the difference between the total and pionic emission.  For comparison, the MeV background from type Ia supernovae from \citet{Horiuchi11} is shown as black shading.  Our fiducial model for the Milky Way is the $h = 1000\ \pc$ model with redshift evolution; for starbursts, we use the fiducial ``high-$B$'' M82 template; and we assume that 15\% of the cosmic star-formation rate at all redshifts is in starbursts.\label{fig:GammaBackground}}
\end{figure*}

The origin of the cosmic $\gamma$-ray background (Figure~\ref{fig:GammaBackground}) remains a mystery.  The unresolved extragalactic $\gamma$-ray background appears to have a single power law spectrum extending from $\sim 50\ \MeV$ to 100 GeV, as determined from {\bfnop the Energetic Gamma Ray Experiment Telescope (EGRET)} and \emph{Fermi} {\bfnop Large Area Telescope} ({\bfnop \emph{Fermi}-LAT;} \citealt{Sreekumar98,Strong04a,Abdo10a}).  There has been much recent debate about the origin of this emission, particularly whether it comes from blazars or star-forming galaxies.  The calculated blazar contribution (dominated by the soft Flat Spectrum Radio Quasars at low energies and the hard BL Lacs at high energies; \citealt{Pavlidou08,Venters11}) has a similar spectrum to the $\gamma$-ray background and, if they make up most of the unresolved background, the remaining blazars should be resolved fairly easily \citep{Abazajian10}.  Star-forming galaxies are expected to have a ``bump'' in their emission at around a GeV, because most of their $\gamma$-ray emission comes from pion production by cosmic ray protons with a threshold of a few hundred MeV \citep{Prodanovic04}.  In contrast to blazars, very few star-forming galaxies will be resolved with \emph{Fermi} \citep{Pavlidou01}.  The contribution of both source classes at GeV energies is still unclear.  The MeV background potentially is a powerful way of discriminating the two sources: the SEDs of blazars {\bfnop are not expected to have a pion bump so that they are brighter at MeV energies than at $\sim 1\ \GeV$, while star-forming galaxies with a pion bump should be fainter at MeV energies than at $\sim 1\ \GeV$} \citep[e.g.,][]{Stecker10}.

At GeV energies, \emph{Fermi} has found that the unresolved extragalactic $\gamma$-ray background is a power law $dN/dE \propto E^{-2.41}$ \citep{Abdo10a}.  Before \emph{Fermi}, the GeV $\gamma$-ray background was usually attributed to blazars, which dominate the population of detected $\gamma$-ray sources \citep{Padovani93,Stecker93,Salamon94,Stecker96,Muecke00}.  {\bfnop {\emph Fermi} resolved many more blazars than EGRET, and indeed most of the \emph{Fermi}-LAT sources are blazars \citep{Nolan12}.  But there remains a $\gamma$-ray background unresolved with \emph{Fermi} that is steeper than the EGRET-resolved background; sources resolved by EGRET increase the background by $\la 1/3$ \citep{Abdo10a}.}  \citet{Abdo10b} claim that the \emph{Fermi} source counts are inconsistent with a blazar origin of the unresolved GeV background, although this conclusion has been disputed (\citealt{Stecker10}; compare with \citealt{Singal11} and \citealt{Malyshev11}), and some estimates find that blazars are a minority of the unresolved GeV background \citep{Dermer07,Inoue09}.  Studies of blazars detected in the flux-limited Australia Telescope 20 GHz survey also find that these blazars contribute a minority of the GeV background, based on an observed $\gamma$-ray--radio correlation \citep{Ghirlanda11} and \emph{Fermi} image stacking of GeV undetected blazars \citep{Zhou11}.  Other proposed alternatives to a blazar origin for the GeV $\gamma$-ray background include dark matter annihilation and decay \citep[e.g.,][]{Bergstroem01,Ullio02,Bertone07}, intergalactic shocks \citep{Loeb00,Miniati02,Keshet03}, non-blazar AGNs \citep[Active Galactic Nuclei; e.g.,][]{Inoue11,Teng11}, galaxy clusters \citep{Dar95,Colafrancesco98}, cascades from ultrahigh energy cosmic rays \citep[e.g.,][]{Wdowczyk72,Coppi97,Murase12}, cosmic rays accelerated by type Ia supernovae \citep{Lien12}, and Galactic foreground emission from millisecond pulsars \citep{FaucherGiguere10}, cosmic rays in the Galactic corona \citep{Feldmann12}, and unsubtracted diffuse emission \citep{Keshet04,Malyshev11}.

Star-forming galaxies are also expected to contribute to the GeV $\gamma$-ray background.  Supernovae and other processes accelerate cosmic rays (CRs), particularly protons but also other nuclei and electrons; these CRs can then interact with the ambient gas, radiation, and magnetic fields to emit broadband radiation.  EGRET only detected the Milky Way and the LMC \citep{Sreekumar92,Abdo10c}, but \emph{Fermi} has also detected the SMC \citep{Abdo10d}, M31 \citep{Abdo10e}, and the starburst galaxies M82 and NGC 253 \citep{Abdo10f}, as well as the starburst-Seyferts NGC 4945, NGC 1068, and the Circinus Galaxy \citep{Abdo10g,Lenain10,Ackermann12,Hayashida13}.  The TeV telescopes VERITAS (Very Energetic Radiation Imaging Telescope Array System) and HESS (High Energy Stereoscopic System) have also detected M82 and NGC 253 \citep{Acciari09,Acero09,Abramowski12}.  The GeV emission is thought to come from the inelastic collisions of cosmic ray protons with protons in the interstellar medium (ISM), creating pions which in turn decay into $\gamma$-rays, neutrinos, and secondary electrons and positrons.  Leptonic processes, particularly Inverse Compton (IC) and bremsstrahlung, may also contribute at lower energies.  The sheer number of normal star-forming galaxies indicates that they contribute to the $\gamma$-ray background, even though they are individually faint, with pre-\emph{Fermi} estimates placing the contribution anywhere from a few percent to most of the {\bfnop total (resolved plus unresolved)} GeV background \citep{Strong76,Lichti78,Pavlidou02}.  

A special subset of star-forming galaxies are the less numerous but individually luminous starburst galaxies, which have high star-formation rates and thus high CR energy densities.  The high gas densities can enhance the pionic $\gamma$-ray emission from starbursts \citep{Torres04a} so that they are a major component of the GeV $\gamma$-ray background (\citealt{Thompson07}; see also \citealt{Loeb06}).  It is even possible that starbursts are  ``proton calorimeters'', converting the power injected as CR protons into pionic losses \citep{Pohl94,Loeb06,Thompson07,Lacki10c}, but the extent to which this is true is disputed \citep{Acero09,Lacki10a,Ohm12}, as is the fraction of star-formation in calorimetric galaxies \citep{Stecker07,Thompson06a,Stecker10}.  Recent estimates of the star-formation contribution to the GeV $\gamma$-ray background find values ranging from 10\% to 100\% \citep[e.g.,][]{Bhattacharya09,Fields10,Makiya11,Stecker10,Ackermann12,Chakraborty13}, with starbursts alone making up 1\% to 50\% of the GeV $\gamma$-ray background \citep{Makiya11,Lacki10a,Stecker10}.  

Although \emph{Fermi} continues to advance our understanding of the GeV $\gamma$-ray sky, very little is known about the MeV $\gamma$-ray background.  The available data on the MeV background comes from Solar Maximum Mission (SMM; 0.3 -- 7 MeV; \citealt{Watanabe99}), Imaging Compton Telescope (COMPTEL; 1 -- 20 MeV; \citealt{Weidenspointner00}), and EGRET (30 -- 20000 MeV; \citealt{Sreekumar98,Strong04a}).  Blazars may constitute most of the MeV background \citep{Bloemen95,Zdziarski96,Giommi06,Ajello09}.  Additionally, $\gamma$-ray line emission and the accompanying Compton-downscattered continuum emission from type Ia supernovae is expected to contribute \citep{Clayton69,Clayton75,The93,RuizLapuente01}, but more recent estimates show that supernovae account for only $\sim 10\%$ of the MeV background \citep{Iwabuchi01,Strigari05,Ahn05a,Horiuchi10}.  Other proposed sources include nonthermal electrons in AGN coronae \citep{Inoue08}, the cores of misaligned radio loud AGNs \citep{Inoue11}, the radio lobes of AGNs \citep{Massaro11}, dark matter decay \citep{Olive85}, and dark matter annihilation \citep{Ahn05b}.  It is not even clear that the observed background is real; it could be caused by detector backgrounds (e.g., caused by incident CRs or radioactive decays in the satellite).

Not only the normalization but the spectral shape of the MeV background is mysterious.  We plot energy flux per natural log energy bin in Figure~\ref{fig:GammaBackground} and similar figures; the area under a curve in an energy range in these figures is directly proportional to the power coming out within that energy range\footnote{Since $\nu I_{\nu} = E^2 dN/dE = E dN/d\ln E$, the area under the curve as plotted is $\propto \int \nu I_{\nu} d\ln E = \int E^2 dN/dE d\ln E = \int E dN/dE dE$.}.  It is clear that the MeV background represents more power than the GeV background{\bfnop: the Universe is much more luminous in MeV $\gamma$-rays than in GeV $\gamma$-rays, whereas star-forming galaxies are thought to be brighter at GeV than MeV.}  Furthermore, {\bfnop there appears to be a break between the power law tail of the X-ray background and the MeV-GeV $\gamma$-ray background in the region between $\sim$3 and $\sim$20 MeV; the cause of this feature is unknown.}

We present here constraints and estimates of the star-forming galaxy contribution to the $\gamma$-ray background, with an emphasis on the MeV background.  \citet{Soltan99} first suggested that starburst galaxies contribute to the MeV background through their IC emission, by scaling the IC emission from the observed synchrotron radio emission, assuming that starburst magnetic fields were the same as in the Milky Way (see also \citealt{Soltan01}).  With the notable exception of \citet{Chakraborty13}, most of the other estimates of the $\gamma$-ray background focus on pionic emission above 100 MeV.  In the meantime, there have been a number of models of the $\gamma$-ray emission from NGC 253, M82, and Arp 220 that take into account radio observations \citep{Paglione96,Torres04b,Domingo05,Persic08,deCeaDelPozo09,Rephaeli09,Lacki10c,Paglione12,YoastHull13-M82,YoastHull13}.  On the observational side, studies of the MeV emission from star-forming galaxies and starbursts in particular are hampered by the very low sensitivity of current MeV instruments.  Only weak constraints have been set on the MeV emission from starburst galaxies with OSSE and INTEGRAL \citep{Bhattacharya94,Dermer97}.  We model the MeV emission for starburst and normal galaxies, and use the spectral shape to constrain the star-forming galaxy contribution to the MeV background.  

In Section~\ref{sec:Understanding}, we review the MeV emission processes of star-forming galaxies.  In Section~\ref{sec:Modelling}, we present one-zone models of the Milky Way and M82 that serve as templates for our $\gamma$-ray background calculation.  In Section~\ref{sec:Cascades}, we consider the effects of $\gamma\gamma$ cascades in the intergalactic medium on the shape of star-forming galaxy spectra.  In Section~\ref{sec:MeVConstraints} we present our results for the star-formation contribution to the MeV background, both the calculation of the MeV background and constraints from the spectral shape.  We assume $\Omega_M = 0.25$, $\Omega_\Lambda = 0.75$, and $H_0 = 70\ \km\ \sec^{-1}\ \Mpc^{-1}$. 

\section{Review of M\lowercase{e}V Emission Processes}
\label{sec:Understanding}

At MeV energies, corresponding to temperatures $\ga 10^{10}\ \Kelv$, there are essentially no diffuse, thermal emission processes that can contribute to the $\gamma$-ray background.  MeV emission from star-forming galaxies therefore either (1) {\bfnop involves} nonthermal cosmic rays, or (2) directly comes from nuclear reactions, and sometimes both.  We review the known MeV emission processes that correlate with star-formation here.

\subsection{Leptonic CR Emission}
Most of the observed $\gamma$-ray emission from galaxies comes from CRs.  The GeV-TeV emission is probably pionic, because the injection rate of CR protons exceeds that of the primary CR electrons by a factor $\sim 50 - 100$ at GeV energies in the Milky Way.  Even weak pionic losses as in the Milky Way are sufficient to overwhelm the leptonic emission at these energies \citep[e.g.,][]{Strong10}.  However, the kinematics of the pion production process leads to a rapid decline in the pionic luminosity below $\sim 100\ \MeV$ (the ``pion bump''), below which leptonic CR emission dominates.  At a minimum, secondary $e^{\pm}$ accompany pionic $\gamma$-rays; their leptonic emission can flatten out the pion bump \citep{Schlickeiser82}.

\emph{Inverse Compton (IC) emission} -- Inverse Compton arises when a CR $e^{\pm}$ Compton scatters an ambient photon, boosting its energy in the observer frame.  The IC emission is expected to form a broad continuum, because the upscattered photon energy scales as the square of the electron energy (in the Thomson limit).  It likely dominates the Milky Way nonthermal continuum at 100 keV \citep{Porter08}.  Below the Klein-Nishina cutoff, electrons of energy $E_e$ will typically upscatter photons of energy $\epsilon$ to $4 \epsilon E_e^2 / (3 m_e c^2)^2$. Ambient photons can come from the CMB, far-infrared (FIR) emission from dust, or intragalactic stellar UV/optical/infrared backgrounds; in the Milky Way, these all have similar energy densities, while in starburst galaxies, the FIR emission dominates.  Each of the radiation fields are roughly greybody fields with a temperature $T$ and average photon energy $\sim 2.7 k T$.  Therefore{\bfnop,} typical upscattered photon energies are $E_{\rm IC}^{\rm CMB} \approx 0.32 (1 + z) E_{10}^2\ \MeV$, $E_{\rm IC}^{\rm FIR} \approx 2.4 (T_{\rm FIR} / 20\ \Kelv) E_{10}^2\ \MeV$, and $E_{\rm IC}^{\rm opt} \approx 480\ \MeV (T_{\rm opt} / 4000\ \Kelv) E_{10}^2\ \MeV$ for electrons of energy $10 E_{10}\ \GeV$ upscattering the CMB, FIR, and optical radiation fields.  IC from each of the components is expected to have a $\Gamma = {\cal P}/2 + 1/2 \approx 2.0 - 2.2$ photon index, if the CR $e^{\pm}$ have a steady-state power law spectrum $E^{-\cal P}$.  

The CMB energy density grows rapidly with redshift.  IC upscattering off a strong high-$z$ CMB can take energy away from other cooling processes, essentially quenching the radio luminosity of a galaxy and transferring the power into MeV emission \citep{Carilli99,Murphy09}.  At high energies, the CMB must overcome the Inverse Compton losses off the starlight and the synchrotron losses in the magnetic field.  Thus, this effect is only important when $U_{\rm CMB} \ga U_B$, or 
\begin{equation}
B \la 3.2\ \muGauss\ (1 + z)^2.
\end{equation}
In Milky Way-like galaxies with $B \approx 6 - 10\ \muGauss$ and starlight energy density $U_{\star} \approx U_B$, the CMB overwhelms the other losses before $z \approx 1$.  However, in starburst galaxies with $B \ga 100\ \muGauss$, the CMB remains unimportant out to $z \ga 5$.  At low electron energies, escape, bremsstrahlung, and ionization losses ``buffer'' the radio spectrum against the CMB effect to even higher redshift \citep{Lacki10zHigh}.

Finally, CR $e^{\pm}$ can escape into the halos of galaxies, where they cool by Inverse Compton emission.  Indeed, \citet{Strong10} find that about $\sim 30 - 60\%$ of the power injected into CR electrons in the Milky Way is lost to Inverse Compton cooling, with greater IC luminosities for larger halos.  The majority of this emission is below 100 MeV in most of their models, although the total $\le 100\ \MeV$ emission is still about a third or less of the total $\gamma$-ray emission \citep{Strong10}.  Starburst galaxies are expected to be ``electron calorimeters'', with most of the $e^{\pm}$ cooling in the starburst disk proper \citep[e.g.,][]{Volk89-Calor}, but radio haloes are observed in starbursts, indicating that some CR $e^{\pm}$ reach large heights off the plane \citep{Seaquist91}.  We do not consider these haloes further in this work.

\emph{Bremsstrahlung} -- Bremsstrahlung radiation is emitted by CR $e^{\pm}$ when they are deflected by the electric field of a nucleus.  A typical bremsstrahlung photon has half the energy of the CR $e^{\pm}$ \citep{Schlickeiser02}; thus the bremsstrahlung spectrum has the same shape as the CR $e^{\pm}$ spectrum, which is set by the dominant cooling or escape mechanism for CR $e^{\pm}$.  The bremsstrahlung loss time of an $e^{\pm}$ with energy $E_{\rm GeV}\ \GeV$ traversing pure hydrogen of number density $n$ is $t_{\rm brems} \approx 40\ \Myr\ (n / \cm^{-3})^{-1}$, but the ionization loss time is $t_{\rm ion} \approx 1380\ \Myr\ E_{\rm GeV} (n / \cm^{-3})^{-1} [\ln E_{\rm GeV} + 14.4]^{-1}$ where $E_{\rm GeV} = E / \GeV$ \citep{Strong98}.  Thus ionization losses therefore quickly softens both the CR $e^{\pm}$ and bremsstrahlung emission spectrum below $\sim 400\ \MeV$, independent of density.  Likewise, IC and synchrotron losses harden the CR $e^{\pm}$ spectrum at high energies, since they go as $t \propto E^{-1}$.  The result is a ``bremsstrahlung bump'' peaking at a few hundred MeV, where the energy-independent bremsstrahlung losses are most important.  Thus, leptonic emission may itself be peaked near 100 MeV, even without hadronic emission.  A low energy bremsstrahlung tail can be generated only if the injected electron spectrum is very steep \citep[e.g.,][]{Sacher83}.

\emph{Synchrotron and other emission from TeV-PeV $e^{\pm}$} -- Synchrotron emission from CR $e^{\pm}$ deflected in a galaxy's magnetic field is responsible for the low frequency radio emission of galaxies, but it can also extend to X-rays and $\gamma$-rays.  At MeV-emitting energies, there are few $e^{\pm}$ injected, but there are no competing cooling processes (IC is suppressed by Klein-Nishina effects).  The typical energy of synchrotron photons is $1.2\ \MeV E_{\rm PeV}^2 B_{10}$, where $E_{\rm PeV}$ is the $e^{\pm}$ energy in PeV and $B_{10} = B / (10\ \mu G)$ is the magnetic field strength.  In the Milky Way, the low magnetic field strength combined with the lack of IC emission observed at $\sim \PeV$ energies implies that there is little large scale high-energy synchrotron emission \citep{Aharonian00}.  The magnetic fields of starbursts are constrained to be much higher in starbursts like those in M82 and NGC 253 ($\sim 100 - 300\ \muGauss$; e.g., \citealt{Thompson06,deCeaDelPozo09-Obs}), so that synchrotron may be important at X-ray \citep{LackiXRay} and perhaps $\gamma$-ray energies.  Synchrotron has a photon index of $\Gamma \ga 2.0$, steepening greatly at the tail of the CR $e^{\pm}$ spectrum.

TeV CR $e^{\pm}$ deflected in electrostatic waves will also emit MeV radiation through electrostatic bremsstrahlung, which occurs at higher frequencies for electrons of a given energy \citep{Schroeder05}.  The typical electrostatic bremsstrahlung photon energy for an electron of Lorentz factor $\gamma$ is $\gamma^2 \nu_P$ compared to the synchrotron photon energy of $\gamma^2 \nu_B$, where $\nu_P$ is the plasma frequency, $\nu_B$ is the cyclotron frequency, and typically $\nu_P \gg \nu_B$.  However, electrostatic bremsstrahlung is only important if the energy density in electrostatic waves reaches the magnetic energy density; since this is not known, we ignore the process for the rest of the work.  
  
\subsection{Positron Annihilation}
Nearly thermalized positrons can annihilate with ISM electrons into 511 keV photons, or they can form positronium atoms which can emit 3-photon annihilation continuum at energies below 511 keV as well as 511 keV photons.  A bright source of annihilation radiation is detected from the Galactic Center bulge, with a flux and spatial distribution not consistent with star formation \citep{Weidenspointner08}, but additional positron emission is detected from the Galactic disk \citep{Weidenspointner08b}.  The positron generation rate in the Galactic disk is consistent with a nucleosynthetic origin in the fission of isotopes like $^{26}$Al produced by massive stars \citep[e.g.,][]{Prantzos10}.  We consider this star-formation component of positrons, but not the (dominant in the Milky Way) bulge source of positrons in this paper.  The bulge positrons may be due to activity by Sgr A$^{\star}$ and there {\bfnop may} not be an analogous population in all star-forming galaxies.

In addition, CR positrons can annihilate while they are still relativistic (in-flight annihilation or IA), although most CR positrons survive to rest.  Pionic positrons, produced alongside pionic $\gamma$-rays, are injected at energies $\ga 100\ \MeV$ \citep{Beacom06}.  If we approximate the hadronic positron injection spectrum as $Q(E) = Q_0 \delta(E - E_0)$ where $E_0 \approx 100\ \MeV$, and since $e^+$ are cooled primarily by ionization (which has a cooling rate roughly independent of energies) at $E \la E_0$, the steady-state low-energy hadronic positron spectrum should have a nearly constant $dN/dE$.  It can then be shown that the annihilation radiation from hadronic positrons is hard, with $dN_{\gamma}/dE \propto E^{-1} (\ln (2\gamma) - 1)$ \citep{Aharonian04-Book}, so that $\nu L_{\nu} \sim E^2 dN_{\gamma}/dE \propto E (\ln(2\gamma) - 1)$ peaks near 100 MeV.

\citet{Watanabe99} calculated the 511 keV line contribution to the $\gamma$-ray background, but did not include positronium continuum or inflight annihilation.

\subsection{Nuclear $\gamma$-Ray Lines}

\emph{Nucleosynthetic ISM $\gamma$-ray lines} -- Some radioactive isotopes synthesized in young, massive stars emit nuclear $\gamma$-rays when they decay.  The $^{26}$Al decay line at 1.809 MeV is the brightest $\gamma$-ray line above 1 MeV detected from the Galactic plane \citep{Mahoney84,Diehl95,Diehl06a}.  The spatial distribution of the line is correlated with massive star-formation in the Galaxy, including individual star-forming regions such as Cygnus \citep{delRio96,Knoedlseder99-Maps,Diehl06a}, although older sources may contribute $^{26}$Al to the ISM.  The only other detection of $\gamma$-ray lines above 1 MeV from the diffuse Galaxy are the $^{60}$Fe decay lines at 1.173 and 1.333 MeV \citep{Harris05,Wang07}.  However, the 1.157 MeV $^{44}$Ti $\gamma$-ray line has been detected from the Cas A supernova remnant.  The injection sites of $^{26}$Al and $^{60}$Fe are thought to be supernovae with some contribution from Wolf-Rayet star winds \citep[e.g.,][]{Knoedlseder99-Implications,Palacios05,Limongi06}, so their yields should scale with massive star-formation rate.  However, some processes in old stellar populations, particularly novae, may also synthesize isotopes that emit $\gamma$-ray lines \citep[e.g.,][]{Jose97}.  

Gamma-ray lines from nuclear decay in the ISM are narrow; Galactic observations of the $^{26}$Al line constrain line widths to be less than $\sim 3\ \keV$ \citep{Diehl06b}, but the line emission from a redshift distribution of galaxies will be smeared out into a continuum.

The MeV background from these lines was also calculated in \citet{Watanabe99}.

\emph{Core collapse supernovae} -- In addition to the long-lasting radioisotopes visible in the ISM, core collapse supernovae (CCSNe) themselves glow in $\gamma$-rays from short-lived isotopes.  The CCSNe are powered by the decay chain of $^{56}$Ni into $^{56}$Co (half-life of 6 days), and then $^{56}$Co into $^{56}$Fe (half-life of 78 days).  Each decay can produce $\gamma$-ray lines.  However, CCSNe are highly Compton-thick, especially at early times, which downgrades much of the line emission into weaker continuum emission \citep[e.g.,][]{Gehrels87,Arnett89,RuizLapuente01}.  

The $\gamma$-ray yield and MeV background from CCSNe was calculated by \citet{Watanabe99}, \citet{RuizLapuente01}, and \citet{Iwabuchi01}.

Besides core collapse supernovae, there will be $\gamma$-ray line emission from type Ia supernovae, some of which are ``prompt'' after the formation of the progenitors.  In this sense, some of the type Ia supernova emission can be associated with star-formation.  However, including this component would require a distribution of delay times of star-formation, which is beyond the scope of this paper.  In any case, the most recent predictions for this background show that it is only $\sim 10\%$ of the observed MeV background \citep{Iwabuchi01,Strigari05,Ahn05a,Horiuchi10}.  We refer the reader to \citet{Horiuchi10} for discussion of the MeV background from type Ia supernovae.

\emph{CR de-excitation lines} -- CRs can also emit $\gamma$-ray lines through nuclear interactions with nuclei in the ISM \citep[e.g.,][]{Meneguzzi75,Ramaty79}.  In addition to narrow line components from the excitation of nuclei in the ISM, broader components are expected when CR nuclei themselves are excited.   The strongest lines expected are the 4.438 MeV $^{12}$C and 6.129 MeV $^{16}$O de-excitation lines \citep[e.g.,][]{Ramaty79}.  For these lines, the broad component should actually be more luminous than the narrow lines, because CRs have much higher abundances of C and O relative to H and He than the ISM \citep[e.g.,][]{Meyer98}. 

CR-excited lines will be much fainter than the pionic emission unless there are many low energy CR nuclei, with only $0.01 - 0.1$ line $\sim\MeV$ photon but $\sim 1$ pionic photon per $\sim 100\ \MeV/n$ nucleus if losses are unimportant \citep{Parizot02}.  A low energy component of CR nuclei, peaking at only a few MeV, is suggested by some CR ionization studies \citep[e.g.,][]{Indriolo09}, but CR-excited nuclear lines have not been detected in the Galaxy \citep{Teegarden06}.  A claimed detection of the 4.438 MeV and 6.129 MeV lines in the Orion star-forming region \citep{Bloemen94} later proved spurious \citep{Bloemen99}. 

\subsection{Emission from Discrete Sources}
Star-forming galaxies are unresolved spatially in $\gamma$-rays, except for the Milky Way and the Magellanic Clouds.  However, in addition to the diffuse $\gamma$-ray emission, compact $\gamma$-ray sources as observed in the Milky Way may contribute to the total galactic $\gamma$-ray luminosity.  Many of these sources -- including supernova remnants, pulsar wind nebulae, and molecular clouds located near CR accelerators -- are directly related to star-formation, and could increase the contribution of star-forming galaxies to the $\gamma$-ray background.  A study by INTEGRAL indicates that at $\ga 1\ \MeV$, most of the Galactic emission is an unresolved diffuse component, unlike at lower energies where point sources dominate \citep{Bouchet08}.  However, given how poorly the MeV regime is understood even in the Milky Way, with even the advanced GALPROP models failing to account for all emission in the $\sim 30\ \MeV$ range \citep{Strong04b,Porter08}, it is possible that additional MeV emission will come from unresolved discrete sources \citep[c.f.,][]{Strong00b}.  We do not consider discrete sources further here, except to note them as a great uncertainty.  

\subsection{Brief Summary}
Over much of the range a few hundred keV to a few hundred MeV, IC is expected to be the dominant emission process.  This is because it forms a broad continuum; furthermore, both radiation fields and $e^{\pm}$ of the relevant energies (GeV--TeV) are known to be present in star-forming galaxies.  Nonthermal bremsstrahlung may be important at $\sim 100\ \MeV$, but it appears as a bump in the spectrum that subsides at lower energies.  Line emission from CCSNe, positron annihilation, and $^{26}$Al may be important from a few hundred keV to an MeV.  The tail of the synchrotron spectrum also may be important at a few hundred keV, though its strength is highly uncertain.

\section{Modeling of $\gamma$-Ray Spectra}
\label{sec:Modelling}
To estimate the MeV -- GeV background contribution from normal and starburst galaxies, we need to understand the $\gamma$-ray SEDs of star-forming galaxies -- both the spectral shape and the normalization per unit star-formation.  Since much of this emission is from CRs, this requires an understanding of CR populations in both normal and starburst galaxies.  The CR population is governed by the diffusion-loss equation, which includes terms for CR injection, spatial propagation, cooling, and destruction (see Appendix).  However, for the purposes of understanding other galaxies, we reduce the complicated diffusion-loss equation into a one-zone leaky-box equation, which treats the galaxy (or its starburst core) as one region with an homogeneous environment and CR population.

There are very detailed models of the nonthermal broadband spectrum of the Milky Way, particularly with the GALPROP code \citep[e.g.,][]{Strong10}, and several one-zone and 3D models of the starburst galaxy M82 \citep{Paglione96,Persic08,deCeaDelPozo09,Lacki10a,Paglione12,YoastHull13-M82}.  However, these models do not take into account possible changes in physical conditions at high $z$.  In addition, these models generally do not include positron annihilation or nuclear lines.  We therefore run our own models, using the Milky Way as a spectral template for normal galaxies, and M82 as a spectral template for starbursts.

\subsection{Procedure for One-Zone Models}
\label{sec:OneZoneProcedure}
We solve the steady-state, one zone CR propagation equation (leaky box model) using a Green's function described in \citet{Torres04b}, which we solve with a code described in \citet{Lacki10c}.  We include primary nuclei and $e^{\pm}$, secondary (knock-on from ionization and pionic) $e^{\pm}$, and higher order pair $e^{\pm}$ from $\gamma\gamma$ annihilation.  Our procedure is largely the same as in \citet{LackiXRay}, although with several new features: we include positron annihilation, nuclei heavier than hydrogen (both in the primary CRs and the ISM), and nuclear de-excitation lines (as described in the Appendix).

Our code gives the volumetric emissivities of CR populations in star-forming galaxies.  To convert the volumetric into total luminosities, we model the template galaxies as disks with radius $R$ and and midplane-to-edge scale heights $h$.  We assume neither the radius nor the scale height evolves with redshift.  The main parameters of the models are listed in Table~\ref{table:ModelParameters}.

The CR injection power is proportional to the supernova rate: a fraction $\eta \sim 0.1$ of the $10^{51} E_{51}\ \erg$ in supernova kinetic energy goes into primary protons, while a fraction $\xi \sim 0.01$ goes into primary electrons. For the Milky Way, we set $\eta = 0.1$ and $\xi$ according to a proton/electron ratio given below.  The acceleration efficiency is scaled by some factor to match the pionic $\gamma$-ray luminosity in the \citet{Strong10} models of the Galaxy.  For the M82 template, we let $\eta$ vary and choose $\xi$ to match the 1 GHz radio flux.  In neither template do $\eta$ or $\xi$ evolve with redshift.\footnote{\bfnop When calculating the background, we scale the emission per unit star formation by the cosmic star-formation rate.  Thus, it is not important if the volume and total luminosity of individual galaxies are overestimated, as long as the $\gamma$-ray yield per unit star formation is correct.}

For {\bfnop M82}, we relate the supernova rate directly to the TIR (total infrared: $8 - 1000\ \mu{\rm m}$) luminosity, which should scale almost directly with star-formation rate:
\begin{equation}
\Gamma_{\rm SN}^{\rm SB} = 0.036\ \yr^{-1} \psi_{17} \left(\frac{L_{\rm TIR}}{10^{10.5}\ \Lsun}\right),
\end{equation}
where $\psi_{17} = 1$ is an initial mass function (IMF) dependent constant \citep{Thompson07,LackiXRay}.  Since massive stars dominate both the bolometric luminosity and the supernova rate, the supernova rate inferred from the luminosity should vary by only a few percent for different IMFs \citep{Thompson07}.  For the Milky Way, we relate the supernova rate to an assumed star-formation rate:
\begin{equation}
\Gamma_{\rm SN}^{\rm MW} = 0.0084\ \yr^{-1} \epsilon_{\rm Sal A} \psi_{17} \left(\frac{\rm SFR}{\Msun\ \yr^{-1}}\right).
\end{equation}
where $\epsilon_{\rm Sal A} = \epsilon / (4.9 \times 10^{-4})$ is a conversion between bolometric luminosity and star-formation rate corrected to a ``Salpeter A'' IMF \citep{Kennicutt98,Baldry03}, as used for the cosmic star-formation rate in \citet{Hopkins06}.\footnote{\citet{Hopkins06} find that for a given bolometric luminosity, a ``Salpeter A'' IMF has only 77\% of the star-formation rate of a standard Salpeter IMF from $0.1 - 100\ \Msun$ used in \citet{Kennicutt98}.}  From the supernova rate, we get the volumetric power injection for CR protons:
\begin{equation}
\epsilon_{\rm CR,p} = 10^{51} \erg\ E_{51} \eta \Gamma_{\rm SN} / (2 \pi R^2 h)
\end{equation}
and primary $e^{\pm}$:
\begin{equation}
\epsilon_{\rm CR,e} = 10^{51} \erg\ E_{51} \xi \Gamma_{\rm SN} / (2 \pi R^2 h)
\end{equation}

Primary CRs are injected with a $dQ/dq = C q^{-p}$ power law spectrum, where $q$ is momentum: this is the test-particle approximation, although nonlinear physics in acceleration may alter the injection spectrum at low energies \citep[e.g.,][]{Berezhko99,Ellison00,Malkov01,Blasi05}.  The normalization for protons and $e^{\pm}$ is set by:
\begin{equation}
\epsilon_{\rm CR} = C \int_{q_{\rm min}}^{\rm q_{max}} q^{-p} (\sqrt{q^2 c^2 + m^2 c^4} - m c^2) dq,
\end{equation}
where $m$ is the rest mass of the accelerated particle, $q_{\rm min}$ and $q_{\rm max}$ are the momenta corresponding to the minimum and maximum kinetic energy particles are injected with, and $\epsilon_{\rm CR}$ is the volumetric power injected in CRs of a given species.  We use a minimum injection kinetic energy of 1 MeV; for the protons and nuclei, we use a maximum injection Lorentz factor of $10^6$ ($K_{\rm max} \approx 1\ \PeV$), while for $e^{\pm}$, the maximum injection Lorentz factor ($\gamma_{\rm max}^{\rm prim}$) is either $2 \times 10^6$ ($K_{\rm max} \approx 1\ \TeV$) for the Milky Way, or $10^6$ for M82.  With these injection spectra, the ratio of $dQ/dq$ for primary protons and electrons at energies $> m_p c^2$ is approximately:
\begin{equation}
\tilde\delta \equiv \delta \left(\frac{m_p c^2}{K_{\rm min}}\right)^{p - 2} \approx \frac{C_p}{C_e},
\end{equation}
which is known to be $\sim 50 - 100$ in the Milky Way \citep[e.g.,][]{Ginzburg76,Warren05,Strong10}, and is explicitly set to 75 in our Milky Way model.  The normalization for the nuclei are scaled using the Milky Way CR abundance ratios (see the Appendix).

We run models for hadronic-origin CRs (nuclei, pionic secondaries, and the radiation and $\gamma\gamma$ pair $e^{\pm}$ they generate) and leptonic-origin CRs (primary CR $e^{\pm}$ and the radiation and $\gamma\gamma$ pair $e^{\pm}$ they generate) separately, and scale them by $\eta$ and $\xi$ respectively.  These components are then added together.  Diffusive losses are applied to CRs of all species in the Milky Way.  {\bfnop These diffusive losses have an energy dependence of $t_{\rm diff} \propto E^{-1/2}$.  Since the diffusive losses determine the lifetime of the Milky Way's CRs, the steady-state spectrum is steepened by $E^{-1/2}$ compared to the injection spectrum.} Advective losses are applied in M82. Although the Milky Way does have a wind in its inner regions \citep{Everett08}, {\bfnop GALPROP} models strongly constrain its effects on the bulk of the CRs detected at Earth \citep{Strong98}.  The decreasing grammage with increasing energy of CRs observed at Earth indicates that energy-dependent diffusion governs the CR lifetime in the local Milky Way.  

CR nuclei experience catastrophic inelastic (including pionic) losses and continuous ionization cooling.  CR electrons experience ionization, bremsstrahlung, synchrotron, and IC cooling; CR positrons experience these losses as well as catastrophic annihilation losses.  Secondary nuclei from collisions are not included.  We calculate the emission from pionic collisions, synchrotron, positron annihilation, several of the strongest nuclear lines, IC, and bremsstrahlung.  More technical details are given in the Appendix.

\begin{figure}
\centerline{\includegraphics[width=9cm]{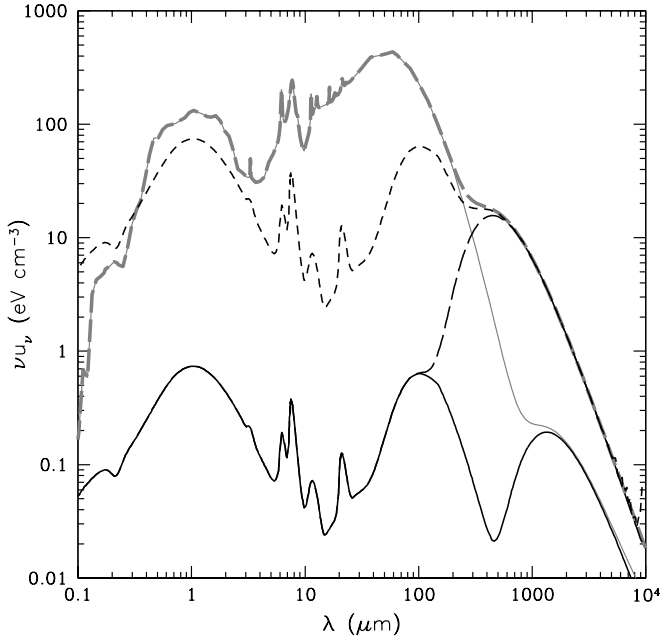}}
\figcaption[figure]{The radiation fields of the Milky Way (black) and M82 (grey) in our models.  The solid lines depict the radiation fields at $z = 0$, while the long-dashed lines depict the radiation fields at $z = 2$ in the non-evolving case.  The short-dashed line shows the $z = 2$ Milky Way radiation field in the evolving models.\label{fig:BackgroundRads}}
\end{figure}

The propagated spectra depend on a number of parameters, many of which are poorly constrained in starburst galaxies.  The GeV-TeV $\gamma$-ray emission, which is pionic in most of our models, constrains the proton acceleration efficiency $\eta$.  The radio data constrains a combination of the electron acceleration efficiency $\xi$ and $B$, but there is a degeneracy between these parameters (see the similar discussion in \citealt{LackiXRay}; also see \citealt{Persic08}, \citealt{deCeaDelPozo09}, and \citealt{Rephaeli09}).  With high $B$, few $e^{\pm}$ are needed to produce the observed GHz synchrotron radio emission, and the $e^{\pm}$ population in starbursts is dominated by hadronic secondaries.  Models with Milky Way values for the proton/electron injection ratio have high $B$.  Models with low $B$ have large $\xi$ and therefore large amounts of leptonic $\gamma$-ray emission in the MeV bands.  In addition, the wind speed $v_{\rm wind}$, gas scale height $h$, primary electron cutoff $\gamma_{\rm max}^{\rm prim}$, and diffusive escape time $t_{\rm diff} (3\ \GeV)$ can all influence the CR spectrum.  Finally, we add a component of free-free emission with spectrum $S_{\nu} = S_{\rm ff} (1\ \GHz) (\nu / 1\ \GHz)^{-0.1}$, which flattens out the radio spectrum.

\begin{deluxetable*}{lcccll}
\tabletypesize{\scriptsize}
\tablecaption{Model Parameters}
\tablehead{\colhead{Symbol} & \colhead{Units} & \multicolumn{2}{c}{Allowed Values} & \colhead{Name} & \colhead{Effect on emission} \\ & & \colhead{MW} & \colhead{M82} & }

$\Sigma_g$                    & $\gcm2$    & 0.0025\tablenotemark{a}      & 0.17     & Gas surface density & Pionic, bremsstrahlung, and $e^+$ annihilation\\
$B$                           & $\muGauss$ & 6\tablenotemark{a}           & 50 -- 400 & Magnetic field strength & Synchrotron\\
$h$                           & $\pc$      & 1000 -- 2000 & 30       & Scale height & Pionic, bremsstrahlung, and $e^+$ annihilation\\
$p$                           & \nodata    & 2.2         & 2.2      & Injection spectral slope & Influences final spectral shape\\
$t_{\rm diff} (3\ \GeV)$      & Myr        & 30          & $\infty$ & Diffusive escape time & Steepens spectrum at high energy\\
$\gamma_{\rm max}^{\rm prim}$ & \nodata    & $2 \times 10^6$ & $10^6$ & Primary $e^-$ spectrum maximum energy & Influences synchrotron X-rays\\
$v_{\rm wind}$                & $\kms$     & 0           & 300      & Wind speed & Limits CR population at all energies, reducing emission\\
$\eta$                        & \nodata    & $0.1 \times {\rm Scale}$ & 0.009 -- 1.1 & CR proton acceleration efficiency & Scales nuclear and pionic $e^{\pm}$ populations and emission\\
$\xi$                         & \nodata    & $0.0052 \times {\rm Scale}$ & \nodata  & Primary CR electron acceleration efficiency & Scales primary electron population and emission\\
$S_{\rm ff} (1\ \GHz)$        & Jy         & \nodata     & 0.06 -- 10  & Free-free radio flux at 1 GHz & Reduces amount of radio emission to be fit as synchrotron
\enddata
\tablenotetext{a}{{\bfnop Indicates} the parameter varies with $z$ in our evolving models of the Milky Way.}
\label{table:ModelParameters}
\end{deluxetable*}

\subsection{The non-cosmic ray emission}

To the total $\gamma$-ray spectra, we then add in the nucleosynthetic $\gamma$-ray lines.  The 1.809 MeV line from $^{26}$Al is scaled from its Milky Way luminosity using the supernova rate: $L_{1.809} = 1.7 \times 10^{37} \ergps (\Gamma_{\rm SN} / \Gamma_{\rm SN, MW})$ \citep{Diehl06a}.  In the Milky Way, the 1.173 and 1.333 MeV $^{60}$Fe lines are observed to have 0.11 photon each for every photon of the 1.809 MeV $^{26}$Al lines \citep{Harris05}.  We therefore set $L_{1.173} = 0.11 (1.173/1.809) L_{1.809}$ and $L_{1.333} = 0.11 (1.333/1.809) L_{1.809}$.  Finally, we scale the 1.157 MeV $^{44}$Ti $\gamma$-ray line directly with the supernova rate: $L_{1.157} = 1.6 \times 10^{38} (M_{Ti-44} / 10^{-4}\ \Msun)\ \ergps (\Gamma_{\rm SN} / \yr^{-1})$, where $M_{Ti-44}$ is the yield of $^{44}$Ti per supernova.  The $^{44}$Ti yield is controversial, because Cas A is bright in the line (\citealt{Iyudin94}; see also \citealt{Vink01} for detections of associated X-ray lines) and SN 1987A's emission indicates large yields \citep[e.g.,][]{Chugai97,Fransson02,Jerkstrand11}, but the Galaxy as a whole is faint in 1.157 MeV \citep{The06}.  We assume $10^{-4}\ \Msun$ per supernova, but estimated values range from $10^{-5} - 2 \times 10^{-4}\ \Msun$.

In addition, we run a one-zone model of nucleosynthetic $e^+$ from radioactive isotopes injected as a delta function at 1 MeV, scaled to the positron generation rate in the Galactic disk from \citet{Weidenspointner08b}: $(dQ/dE)_{\rm SN} = 8 \times 10^{42} \sec^{-1} (\Gamma_{\rm SN} / \Gamma_{\rm SN, MW}) \delta(K - 1\ \MeV) / (2 \pi R^2 h)$.  The $\gamma$-ray spectrum is then the sum of the emission from hadronic-origin CRs, the leptonic-origin CRs, the nucleosynthetic $e^+$, and the nucleosynthetic $\gamma$-ray lines.

The $\gamma$-rays from CCSNe are not added to the models of the individual galaxies (Milky Way {\bfnop and} M82), since supernovae are rare and transient events.  However, we do include them in the $\gamma$-ray backgrounds.  We assume in our calculations that each CCSN produces $0.07\ \Msun$ of $^{56}$Ni \citep[e.g.,][]{Catchpole88,Arnett89,Bouchet93,Hamuy03}.  We include the strongest $\gamma$-ray lines from $^{56}$Co at 0.847 MeV and 1.238 MeV, with time-averaged luminosities $L_{0.847} = 5.9 \times 10^{36} \ergps (\Gamma_{\rm SN} / \Gamma_{\rm SN, MW})$ and $L_{1.238} = 6.7 \times 10^{36} \ergps (\Gamma_{\rm SN} / \Gamma_{\rm SN, MW})$ \citep{RuizLapuente01}.  In addition, we include the continuum emission using the spectrum calculated in \citet{RuizLapuente01}, scaled to the $0.07\ \Msun$ of $^{56}$Ni per supernova.

\subsection{Translating the models to high $z$}

We run selected models at redshifts from 0.0 to 4.9 at intervals $\Delta z = 0.1$.  For the M82 starburst templates, the only difference with redshift is that the radiation field is altered so that the CMB is increased to its value at each redshift.  For the Milky Way, we consider both an unevolved template, where only the CMB is changed between redshifts, and an evolving template.  

In the evolving template, the gas density increases as
\begin{equation}
\Sigma_g (z) = (1 + \tilde{z})^3 \times \Sigma_g (0)
\end{equation}
following \citep{Chakraborty13}, where
\begin{equation}
\tilde{z} = \left\{ \begin{array}{ll} z & (z \le 2.5)\\ 2.5 & (z > 2.5) \end{array} \right.
\end{equation}
We assume that evolution stops at $z = 2.5$, motivated by the \citet{Bethermin12} models of main sequence galaxy evolution.  The ISM composition is assumed to remain constant out to high redshift.  According to the Schmidt law, the star-formation rate surface density increases as $\Sigma_{\rm SFR} \propto \Sigma_g^{1.4}$ \citep{Kennicutt98}, and the starlight radiation field intensity is proportional to the star-formation rate \citep{Lacki10c}.  Thus, in the evolving Milky Way template, we multiply the Milky Way starlight and dust radiation field as 
\begin{equation}
U_{\star} (z) = (1 + \tilde{z})^{4.2} \times U_{\star} (0)
\end{equation}
Finally, the far-infrared radio correlation implies that the magnetic field strength scales as $\Sigma_g^{0.7}$ \citep{Lacki10c}.  Thus, the magnetic field strength in the evolving magnetic field template goes as
\begin{equation}
B (z) = (1 + \tilde{z})^{2.1} \times B (0).
\end{equation}

We also consider galaxy evolution in the form of a decreasing fraction of dense, starburst-like galaxies from high $z$ to low $z$ (see the extensive discussion in section~\ref{sec:StarburstFraction}).

\subsection{Milky Way}
\label{sec:MWModel}

\begin{figure*}
\centerline{\includegraphics[width=9cm]{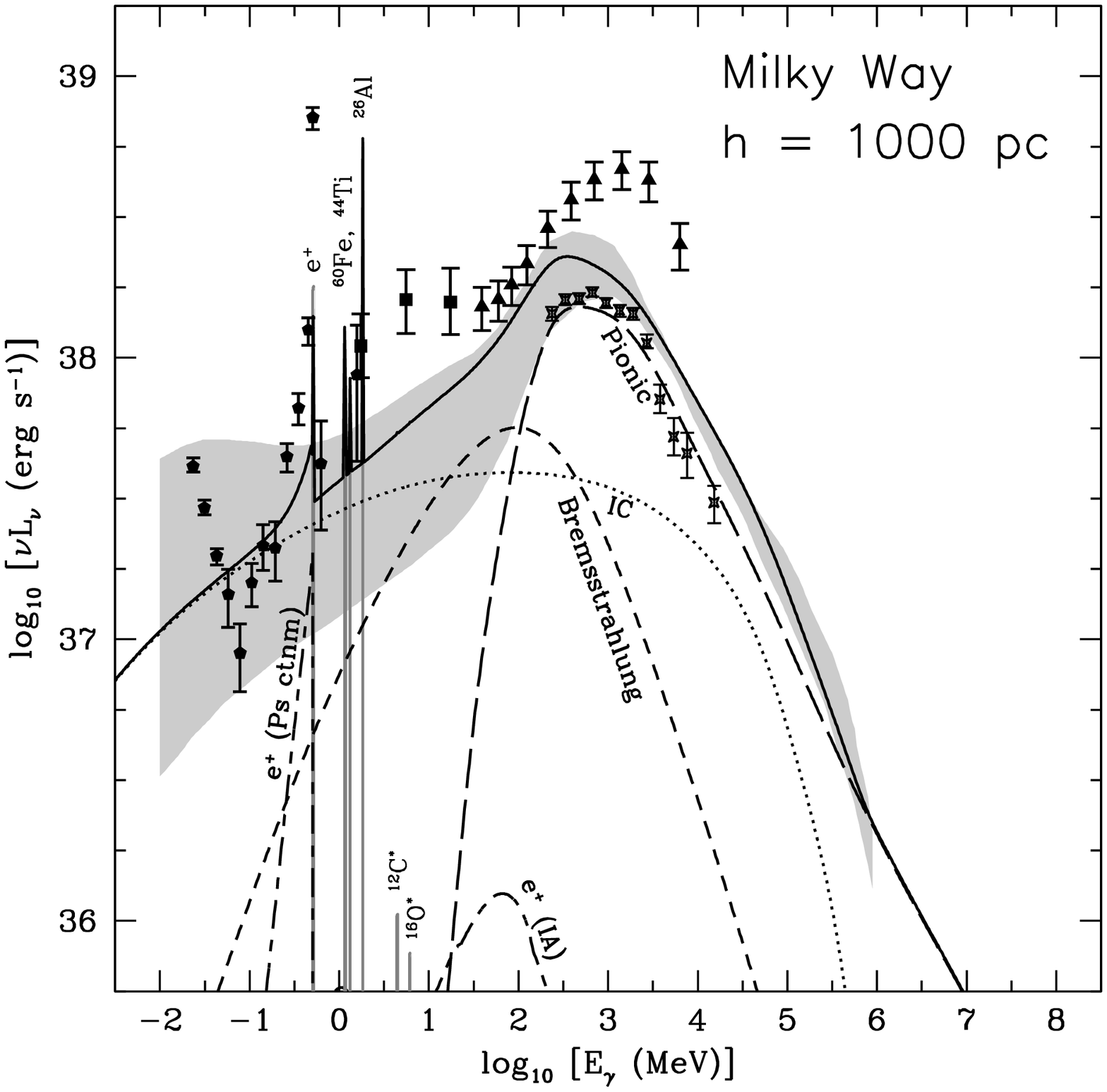}\includegraphics[width=9cm]{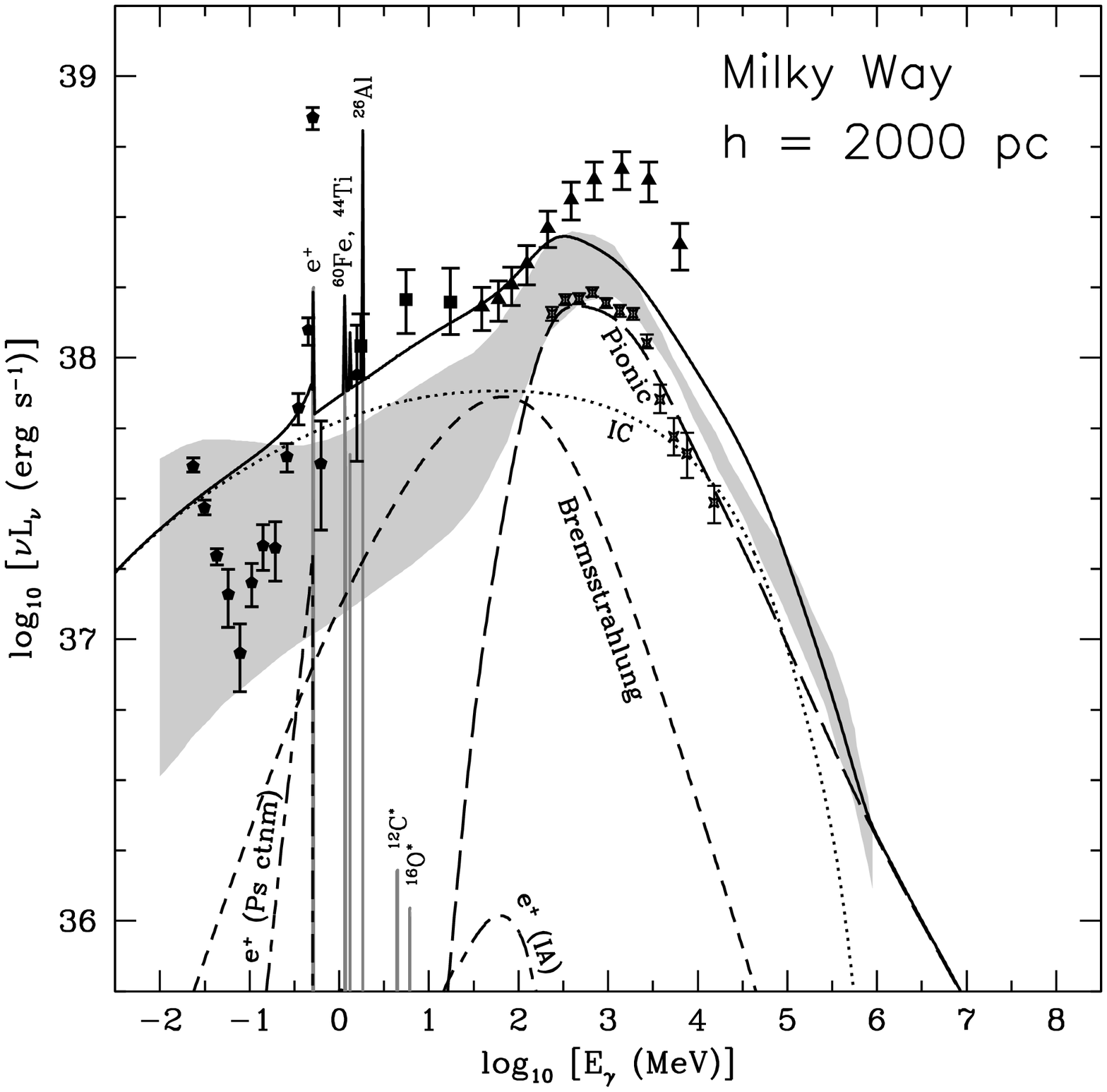}}
\figcaption[simple]{The predicted $\gamma$-ray luminosity spectrum of the Milky Way (solid), compared to extrapolated observations of the inner Galaxy with SPI (pentagons), COMPTEL (squares), EGRET (triangles), and Fermi-LAT (X's) as well as GALPROP models (grey shading; \citealt{Strong10}).  The different components to the emission are positron annihilation (long-dashed/short-dashed), nucleosynthetic $\gamma$-ray lines (grey solid), CR nuclear de-excitation lines (grey solid), bremsstrahlung (short-dashed), IC (dotted), and pionic emission (long-dashed).  Note that the predicted positron annihilation rate only includes the contribution from star-formation, but the observed positron emission includes the Galactic Center source.  Emission from CCSNe is not included.  Also note that the observations, predicted line spectra, and total predicted spectra have different energy binnings; the binning for the total spectrum is $\Delta E / E = 3\%$.  The ``observed'' luminosities depend on unknown conversions between flux or emissivity and luminosity, and so are only suggestive.\label{fig:MWGamma}}
\end{figure*}

\emph{Introduction} -- The Milky Way (MW) is the \emph{only} non-active galaxy observed in MeV $\gamma$-rays, and only one of a handful of normal (non-starburst, non-active) galaxies observed in GeV $\gamma$-rays (with the LMC, SMC, and M31).  At low energies, below $\sim 50\ \keV$, most of the Galactic X-ray ridge emission is resolved into stellar sources, such as cataclysmic variables \citep{Revnivtsev09}.  The energy region up to 511 keV includes a spectral spike from positron annihilation radiation.  This annihilation radiation is concentrated towards the center of the Galaxy, which is not expected if it is correlated with star-formation \citep[see the review by][]{Prantzos10}.  In the energy range from 100 keV to 100 MeV, there appears to be a power law continuum, with $\Gamma \approx 1.5$, thus increasing in luminosity to higher energies.  This power law continuum has been interpreted in the past as bremsstrahlung \citep{Schlickeiser82,Sacher83}, though it is now believed to be Inverse Compton emission \citep{Porter08,Bouchet11}.  However, there also appears to be an excess of $\gamma$-rays above the power law detected by COMPTEL in the $\sim 5 - 20\ \MeV$ range towards the inner Galaxy \citep{Strong04b,Porter08}.  The Galactic $\gamma$-ray luminosity peaks around a GeV, a pionic feature expected from theoretical expectations of CR propagation \citep[e.g.,][]{Stecker70,Hunter97,Strong04b}.  Finally, the emission drops off rapidly as $E^{-2.7}$ to $E^{-2.6}$, continuing to TeV energies in at least some regions of the Galactic Plane \citep{Prodanovic07,Abdo08}.

In modeling the Milky Way, we use a radius of 10 kpc.  We consider scale heights (midplane to edge) of 1 kpc and 2 kpc.  The gas surface density $\Sigma_g$ is measured to be $\sim 0.0025\ \gcm2$ in the Solar neighborhood and through much of the Galactic disk \citep{Boulares90,Yin09}.  For the radiation field, we average the maximum metallicity gradient radiation fields of \citet{Porter08} for $R = 4\ \kpc$ and $R = 8\ \kpc$, reflecting the fact that much of the star-formation (and therefore CR generation) occurs at $R \approx 5 - 7\ \kpc$ \citep{Yin09}.  {\bfnop This radiation field is plotted in Figure~\ref{fig:BackgroundRads}.  The radiation field is dominated by three peaks, one in the optical/near-infrared and one in the far-infrared from the Galaxy's luminosity, and the CMB.}  In addition, we use a magnetic field strength of $6\ \muGauss$ \citep[e.g.,][]{Strong00,Beck11}, a diffusive escape time of $t_{\rm diff} (E) = 30\ \Myr\ (E / 3\ \GeV)^{-1/2}$ \citep{Connell98,Webber03}, and a maximum primary electron Lorentz factor of $\gamma_{\rm max}^{\rm prim} = 2 \times 10^6$ as inferred from the apparent cutoff in the primary electron spectrum at TeV energies \citep[e.g.,][]{Aharonian08,Aharonian09}.

We assume that the total Galactic star-formation rate is $2\ \Msun\ \yr^{-1}$, giving us a supernova rate of $1.6\ {\rm century}^{-1}$ for the SalA IMF.  Estimates of the star-formation rate and supernova rate of the Galaxy span nearly an order of magnitude, though most of the larger ones are from older works \citep[see Table 1 of ][]{Diehl06a}.  Recent estimates of the star-formation rate from free-free emission or direct counting of young stellar objects find star-formation rates of $0.9 - 2.2\ \Msun\ \yr^{-1}$ \citep{Murray10} and $0.7 - 1.5\ \Msun\ \yr^{-1}$ \citep{Robitaille10}, respectively; therefore, our star-formation rate estimate may be somewhat high.  \citet{Chomiuk11} compile several star-formation rate estimators for the Galaxy and conclude that it is $\sim 1.9\ \Msun\ \yr^{-1}$ for a Kroupa IMF.  For a given star-formation rate, and for stellar masses greater than $1\ \Msun$ where most of the starlight used to infer the star-formation rate history of the Universe comes from, the IMFs differ by less than 30\%.  At $10\ \Msun$, the approximate mass of most supernova progenitors, the IMFs differ by only 13\%.  The result of $2\ \Msun\ \yr^{-1}$ should thus be applicable for the SalA IMF that we use, to within a factor of $<2$.  On the other hand, we use the star-formation rate solely to calculate the supernova rate, which is about a factor $\sim 1.2$ lower than in \citet{Diehl06a} (estimated from the $^{26}$Al $\gamma$-ray line).  

We set the 1 MeV positron injection rate from radioactive elements for the entire Galaxy to be $8.1 \times 10^{42}\ \sec^{-1}$, as determined for the Galactic disk by \citet{Weidenspointner08} with their ``bulge + thick disk'' model.  This does not include the positron rate from the bulge, which is substantially higher ($\sim 2 \times 10^{43}\ \sec^{-1}$).  In using this value for the nucleosynthetic positrons, we assume that CR positrons contribute a minority of this rate, which is verified by our modeling and GALPROP models \citep{Porter08}.  

\emph{Comparison with $\gamma$-ray expectations} -- We scale our predicted $\gamma$-ray spectra by comparing our models' 100 MeV -- 100 GeV pionic $\gamma$-ray luminosities to the more detailed predictions in \citet{Strong10}.  We find using standard acceleration efficiencies ($\eta = 0.1$) and our fiducial supernova rate that our models' $\gamma$-ray luminosity is too small.  We find we must scale up $\eta$ (and $\xi$ using $\tilde\delta = 75$) in the $h = 1000\ \pc$ $\gamma$-ray luminosity by a factor 2.2 and the $h = 2000\ \pc$ luminosity by a factor 4.3 to match the average pionic luminosity in \citet{Strong10}.  These factors vary from 1.8 to 2.6 and 3.6 to 5.1, respectively, depending on which GALPROP model we compare to.  These we take to bound the uncertainty in the acceleration efficiency scaling of our models.

The average density that CRs experience in our models is $0.24\ \cm^{-3}$ in the $h = 1000\ \pc$ model and $0.12\ \cm^{-3}$ in the $h = 2000\ \pc$ model; CRs observed at Earth are inferred to traverse mean densities of $0.25\ \cm^{-3}$ in the Milky Way \citep{Connell98,Schlickeiser02}.  The $h = 2000\ \pc$ model therefore is too $\gamma$-ray dim partly because it is too low density; this is not the case for $h = 1000\ \pc$.  The CR proton luminosity in our models is $5.1 \times 10^{40}\ \erg\ \sec^{-1}$; \citet{Strong10} find CR proton luminosities of $\sim 7 \times 10^{40}\ \erg\ \sec^{-1}$ for their $h = 2\ \kpc$ models ($6.0 - 7.4 \times 10^{40}\ \erg\ \sec^{-1}$ for all of their listed models).  Since we are comparing to their $\gamma$-ray luminosity, our CR luminosity is $\sim 40\%$ too low; after accounting for this, the $\gamma$-ray luminosity of the $h = 1000\ \pc$ model is within a factor 2 of that of the \citet{Strong10} model.  We conclude that the need for the scale factor arises because the CR luminosity is too low in our models, and also because the mean density is too low in the $h = 2000\ \pc$ model.  

The rescaled $\gamma$-ray spectra are plotted in Figure~\ref{fig:MWGamma}.  Pionic emission dominates the $\gamma$-ray emission at most energies above 100 MeV, leading to the observed pionic bump.  At MeV energies, the relative strengths of the leptonic emission processes depend on the assumed scale height.  In the smaller scale height ($1000\ \pc$; left panel in Figure~\ref{fig:MWGamma}) models, CR $e^{\pm}$ experience higher gas densities than the large scale height models ($2000\ \pc$; right panel in Figure~\ref{fig:MWGamma}), but the radiation field is the same.  Therefore, bremsstrahlung receives more power in the $h = 1000\ \pc$ model than the $h = 2000\ \pc$ model.  We find that in the $h = 1000\ \pc$ models, bremsstrahlung is the most important emission from $\sim 10 - 100\ \MeV$ and IC is the most important continuum emission at energies below that.  In the $h = 2000\ \pc$ models, IC is the dominant continuum emission process at energies below 100 MeV, although bremsstrahlung nearly equals it at 100 MeV.  This is in accordance with expectations that the truly diffuse Galactic nonthermal continuum in the $100\ \keV - \MeV$ range is IC emission \citep{Porter08,Bouchet11}.  The $h = 1000\ \pc$ model spectrum is essentially in agreement with GALPROP luminosity spectra within uncertainties, although the $h = 2000\ \pc$ model has more leptonic emission (Figure~\ref{fig:MWGamma}; \citealt{Strong10}).

\begin{deluxetable*}{lccccc}
\tabletypesize{\scriptsize}
\tablecaption{Models used as spectral templates\tablenotemark{a}}
\tablehead{\colhead{Name} & \colhead{$\eta$} & \colhead{$\xi$} & \colhead{$S_{\rm ff} (1\ \GHz)$} & $\tilde\delta$ & $\chi^2$ \\ & & & \colhead{$\Jy$} & & }
\cutinhead{Milky Way}
${\bf h = 1000\ \pc}$ & {\bf 0.22} & {\bf 0.011} & {\bf \nodata} & {\bf 75} & {\bf \nodata\tablenotemark{b}}\\
$h = 2000\ \pc$ & 0.43 & 0.022 & \nodata & 75 & \nodata\\
\cutinhead{M82}
$B = 150\ \muGauss$ & 0.0089 & 0.050  & 0.45 & 0.70 & 86.2\\
    & 0.013  & 0.049  & 0.45 & 1.0  & 83.8\\
    & 0.018  & 0.048  & 0.45 & 1.4  & 81.0\\
    & 0.025  & 0.048  & 0.44 & 2.1  & 78.6\\
    & {\bf 0.035}  & {\bf 0.046}  & {\bf 0.44} & {\bf 3.0}  & {\bf 77.8\tablenotemark{c}}\\
    & 0.050  & 0.044  & 0.46 & 4.4  & 82.0\\
$B = 200\ \muGauss$ & 0.013  & 0.032  & 0.46 & 1.5  & 94.4\\
    & 0.018  & 0.031  & 0.47 & 2.2  & 89.4\\
    & 0.025  & 0.030  & 0.48 & 3.3  & 84.0\\ 
    & 0.035  & 0.029  & 0.48 & 4.8  & 78.8\\
    & 0.050  & 0.027  & 0.47 & 7.3  & 76.7\\
    & 0.071  & 0.024  & 0.47 & 11.5 & 84.5\\
$B = 250\ \muGauss$ & 0.025  & 0.021  & 0.51 & 4.7  & 91.3\\
    & 0.035  & 0.020  & 0.51 & 7.1  & 83.4\\
    & 0.050  & 0.018  & 0.52 & 11   & 78.7\\
    & 0.071  & 0.015  & 0.50 & 18   & 82.0\\
$B = 300\ \muGauss$ & 0.035  & 0.014  & 0.59 & 9.8  & 88.6\\
    & 0.050  & 0.012  & 0.58 & 16   & 81.6\\
    & 0.071  & 0.0097 & 0.58 & 29   & 82.3\\
$B = 400\ \muGauss$ & 0.050  & 0.0064 & 0.72 & 31   & 87.9\\
    & {\bf 0.071}  & {\bf 0.0038} & {\bf 0.68} & {\bf 73}   & {\bf 85.1\tablenotemark{d}}
\enddata
\label{table:BestFitModels}
\tablenotetext{a}{\bfnop Fiducial models are indicated in bold.}
\tablenotetext{b}{\bfnop Fiducial model for Milky Way.}
\tablenotetext{c}{Fiducial low-$B$ model for M82.}
\tablenotetext{d}{Fiducial high-$B$ model for M82.}
\end{deluxetable*}

To compare the spectral \emph{shapes} of the model to data, we show flux data from INTEGRAL SPI \citep{Bouchet08}, COMPTEL, and EGRET \citep{Porter08} as plotted in \citet{Turler10}.  We roughly convert the fluxes from the inner Galaxy to Galactic luminosities by assuming that CR $\gamma$-ray emission is distributed spatially the same as $^{26}$Al nuclear line emission (which is associated with massive stars).  We also estimate the luminosity from \emph{Fermi} measurements of the local emissivity per hydrogen atom \citep{Ackermann12-MW}, normalized to a gas mass of $3.8 \times 10^9\ \Msun = \pi (10\ \kpc)^2 \times 0.0025\ \gcm2$.  The observations display a prominent pion bump as in our models, falling above and below a GeV.  However, the COMPTEL data indicate a spectral plateau at $\sim 10$ MeV, which is present in neither of our models.  Note that the SPI and COMPTEL data may have systematic errors that are larger than reported, which are difficult to quantify.  The normalization of the data does not exactly match either our models or those of \citet{Strong10}.  {\bfnop But note that the data points shown in the Figure can only be suggestive, since they are actually flux data points and not luminosity data points.  Our one-zone modeling approach only gives the total luminosity of the Galaxy.  Without information on the proportion of emission from each distance on the sightline, it is not possible to invert flux into luminosity with certainty.  Since the GALPROP models are fit to the flux seen from Earth, and they do model the structure of the Galaxy, we essentially use the \citet{Strong10} models to convert between luminosity and flux, and only fit our models' normalization to \citet{Strong10}.}

We use the $h = 1000\ \pc$ model with redshift evolution as our fiducial model of the Milky Way.  

\emph{Results for MeV $\gamma$-rays} -- Most of the star-formation associated positron annihilation rate comes from nucleosynthetic positrons.  Of the $8.1 \times 10^{42}\ \sec^{-1}$ nucleosynthetic positrons injected in the Galactic disk at 1 MeV, we find that $6.8 \times 10^{42}\ \sec^{-1}$ (83\%) survive to low energies.  In addition, we find that $7.5 - 7.6 \times 10^{41}\ \sec^{-1}$ positrons are injected through pion production, about $40 - 80\%$ of the rate calculated by GALPROP models \citep{Porter08}.  Of these, $4.7 \times 10^{41}\ \sec^{-1}$ (63\%) annihilate near rest in the $h = 1000\ \pc$ model and $4.1 \times 10^{41}\ \sec^{-1}$ (54\%) annihilate near rest in the $h = 2000\ \pc$ model. The remaining positrons either escape or annihilate in flight.  Our models indicate that $\sim 7\%$ of the disk positrons are hadronic in origin.  The luminosity in positronium continuum and the 511 keV line is $\sim 9 \times 10^{36}\ \ergps$, whereas the $\nu L_{\nu}$ luminosity of the nonthermal Galaxy at that energy is $4 \times 10^{37}\ \ergps$ in the $h = 1000\ \pc$ model and $7 \times 10^{37} \ergps$ in the $h = 2000\ \pc$.  Therefore positron annihilation near rest associated with star-formation is only $\sim 10\%$ of the luminosity of the Galaxy near 511 keV.

The total positron in-flight annihilation luminosity is one half of the luminosity of positrons annihilating at rest, $\sim 3 - 4 \times 10^{36}\ \ergps$.  While by number, most positrons survive to low energy, pionic positrons have much more energy when they are injected than when they are near rest.  Therefore, most of the power from pionic positron annihilation comes out near $\sim 65\ \MeV$.  On the other hand, the other nonthermal Galactic emission is also greater at these energies, with $\nu L_{\nu} \approx 1.2 - 1.8 \times 10^{38}\ \ergps$ so that in-flight annihilation is just a few percent of the Galactic luminosity at these energies.

The nucleosynthetic $\gamma$-ray lines are a significant but minority contributor to the MeV $\gamma$-ray luminosity of the Galaxy.  The strongest line, 1.809 MeV emission from $^{26}$Al, has a luminosity of $1.7 \times 10^{37}\ \ergps$.    The other nucleosynthetic lines considered ($^{60}$Fe 1.173 and 1.333 MeV, $^{44}$Ti 1.157 MeV) have luminosities of $1.5 - 2.5 \times 10^{36}\ \ergps$, for a total nucleosynthetic line luminosity of $2.6 \times 10^{37}\ \ergps$. The nonthermal luminosity at 1.5 MeV is $\nu L_{\nu} \approx 4 \times 10^{37}\ \ergps$ for $h = 1\ \kpc$ and $8 \times 10^{37}\ \ergps$ for $h = 2\ \kpc$.  Nucleosynthetic lines therefore make up about $\sim 30\%$ of the luminosity at energies around 1 MeV.

More exotic processes, such as synchrotron and nuclear de-excitation lines, are insignificant in our models of the Milky Way.

\begin{figure}
\centerline{\includegraphics[width=9cm]{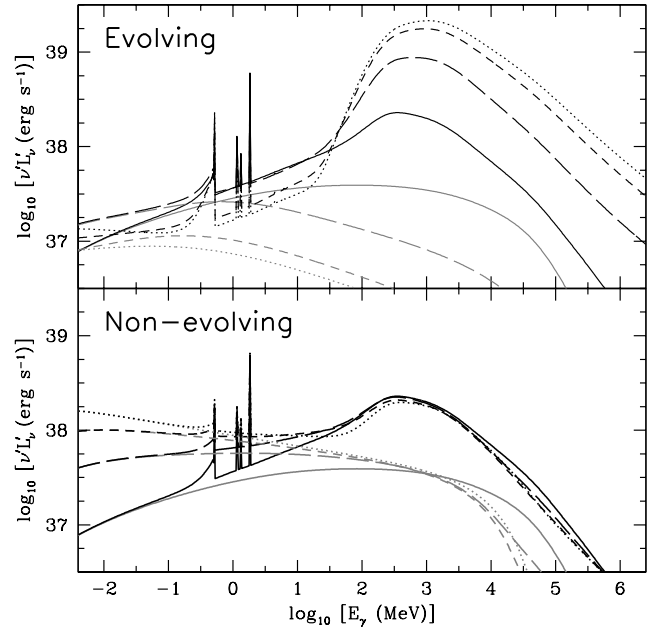}}
\figcaption[figure]{Rest-frame evolution of the total (black) and IC (grey) $\gamma$-ray spectrum of the Milky Way (normalized to its current SFR) placed at different redshifts (0: solid; 1: long-dashed; 2: short-dashed; 4: dotted). In the evolving model, the high-$z$ Galaxy is brighter at GeV energies as normal galaxies are denser.  However, the MeV luminosity slightly decreases, as the stronger bremsstrahlung, ionization, and synchrotron losses cool electrons before they radiate via IC.  In the non-evolving model, the stronger CMB enhances Inverse Compton emission in the MeV band until it is at least $\sim 1/2$ of the GeV luminosity, partly flattening the pion bump.\label{fig:MWatHighz}}
\end{figure}

\emph{High z} -- Figure~\ref{fig:MWatHighz} shows how the $\gamma$-ray SED of the fiducial Milky Way template changes with redshift.  In the evolving model, the increased gas density leads to more effective pionic losses for CR protons.  This leads to a brightening per unit star-formation at GeV energies: the pion bump rises.  But the MeV emission actually falls, despite the increasing CMB strength.  This is because non-Inverse Compton losses for electrons also get stronger in the evolving model.  Thus, there is less energy left to go into Inverse Compton.  

In the non-evolving model, though, the increasing CMB strength at high $z$ leads to a stronger Inverse Compton component.  This enhances the MeV emission considerably in Milky Way-like galaxies: in fact, at $z = 2$, the 1 MeV $\nu L_{\nu}$ luminosity is 76\% of the 1 GeV $\nu L_{\nu}$ luminosity. Thus the enhancement of IC at high $z$ must be accounted for in calculating the MeV background from MW like galaxies.  

An interesting effect that occurs in Milky Way-like galaxies at high $z$ is that the fraction of hadronic positrons surviving to rest increases from 50 -- 60\% to $\sim$ 75 -- 85\%, enhancing the hadronic positron annihilation rate.  This is because of the stronger electron losses in high-$z$ galaxies, so that CR positrons are trapped more effectively in the host galaxy.  However, in the non-evolving models, given that hadronic positrons form only a small fraction of the star-formation associated positrons, the total Galactic disk positron annihilation rate is only enhanced by 2\%.

\subsection{M82}
\label{sec:M82Model}

\emph{Introduction} -- To find a $\gamma$-ray SED template for starburst galaxies, we choose M82, a nearby starburst galaxy ($D \approx 3.6\ \Mpc$; \citealt{Freedman94}), the brightest in the GeV and TeV $\gamma$-ray sky \citep{Abdo10f,Acciari09}, and one of the brightest galaxies observed in the IR \citep{Sanders03}.  The starburst lies in the central regions of the galaxy ($R \le 250\ \pc$), where a large amount of gas, infrared emission, and radio emission is observed \citep{Goetz90,Williams10}.  About $2.3\ \times 10^8\ \Msun$ of gas is observed in the central regions, giving a gas surface density of $0.17~\gcm2$ (\citealt{Weiss01}; although they find $2.7 \times 10^8\ \Msun$ of gas with a second radiative transfer model).  M82 is also observed to host a starburst wind.  Models of NGC 253, the other TeV-detected starburst, generally find similar propagation environments as M82 \citep{Domingo05,Rephaeli09,Paglione12,YoastHull13}; furthermore, a substantial fraction of its radio and infrared emission comes from outside the starburst core \citep{Melo02,Williams10}.  To avoid redundancy, we therefore focus our efforts on M82, which is indeed dominated by the starburst region \citep{Adebahr13}.

Using the observed TIR luminosity of $5.9 \times 10^{10}\ \Lsun$ from \citep{Sanders03}, we take the supernova rate as $0.067\ \yr^{-1}$, compared to typically quoted values in the literature of $\sim 0.1\ \yr^{-1}$ with factor $\sim 2 - 3$ uncertainty (see the summary in \citealt{Lacki10a}).  This corresponds to a star-formation rate of $8.0\ \Msun\ \yr^{-1}$.

Essentially nothing is known observationally about M82's MeV emission.  It has been detected in 10 -- 50 keV X-ray emission by Suzaku and Swift-BAT \citep{Cusumano10}.  Very weak upper limits on the continuum MeV emission exist from non-detections by OSSE \citep{Bhattacharya94}.   Searches for nuclear lines from M82 and NGC 253 with OSSE have also found nothing yet \citep{Bhattacharya94}.  M82 is detected with $\ga 5\sigma$ significance at energies down to 100 MeV, although the \emph{Fermi}-LAT data does not yet go down below that \citep{Ackermann12}.  There are, however, several theoretical models for M82's multiwavelength nonthermal emission, including MeV energies \citep[e.g.,][]{Persic08,deCeaDelPozo09,Paglione12,YoastHull13-M82}.  

There are potentially a large number of parameters to fit.  In order to prevent combinatorial explosion and keep the number of templates manageable, we use $p = 2.2$, $v_{\rm wind} = 300\ \kms$, $h = 30\ \pc$, and $\gamma_{\rm max}^{\rm prim} = 10^6$.  We also ignore diffusive escape, motivated by the lack of observed spectral breaks in starbursts \citep{Abramowski12}.  {\bfnop We calculate the radiation field from a GRASIL model of M82's spectral energy distribution, to which we add the CMB.\footnote{\bfnop Available at: http://adlibitum.oat.ts.astro.it/silva/grasil/modlib/fits/fits.html.}  The radiation field is dominated by a far-infrared peak as plotted in Figure~\ref{fig:BackgroundRads}.} We consider magnetic field strengths of 50, 100, 150, 200, 250, 300, and 400 $\muGauss$.

\emph{Constraints from radio and GeV-TeV $\gamma$-rays} -- The most relevant uncertainty is due to a degeneracy between the magnetic field strength and the electron acceleration efficiency.  The observed amount of radio emission could either be produced by a large electron population in a relatively weak magnetic field, or a small electron population in a strong magnetic field.  Since the MeV leptonic emission is proportional to the number of electrons, the two scenarios predict very different amounts of MeV emission.  We could constrain the magnetic field if we assume that CR acceleration works the same as it does in the Milky Way, with a similar proton/electron injection ratio, but we do not know this at present with the data we have.

\begin{figure}
\centerline{\includegraphics[width=9cm]{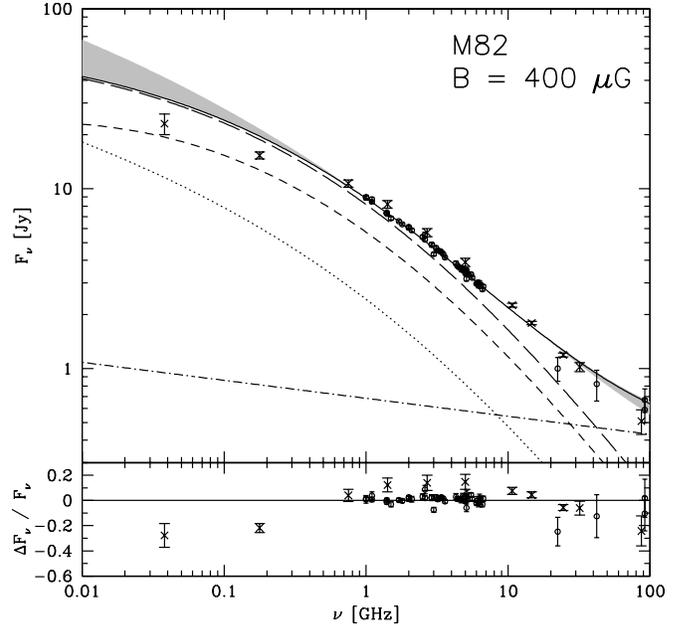}}
\figcaption[figure]{Predicted radio spectrum of M82 in the fiducial high-$B$ model.  Plotted lines are total radio emission (solid), total synchrotron emission (long-dashed), synchrotron from secondaries (short-dashed), synchrotron from primaries (dotted), and free-free emission (dash-dotted).  The grey band is the range spanned by all selected models of M82.  Plotted data are the interferometric data compiled in \citet{Williams10} (open circles), and data from \citet{Klein88} (X's).  The bottom panel shows the residuals to the radio data.  The discrepancy at low frequencies (data at $<1\ \GHz$ is not included in our fitting) is probably due to free-free absorption. \label{fig:M82Radio}}
\end{figure}

\begin{figure*}
\centerline{\includegraphics[width=9cm]{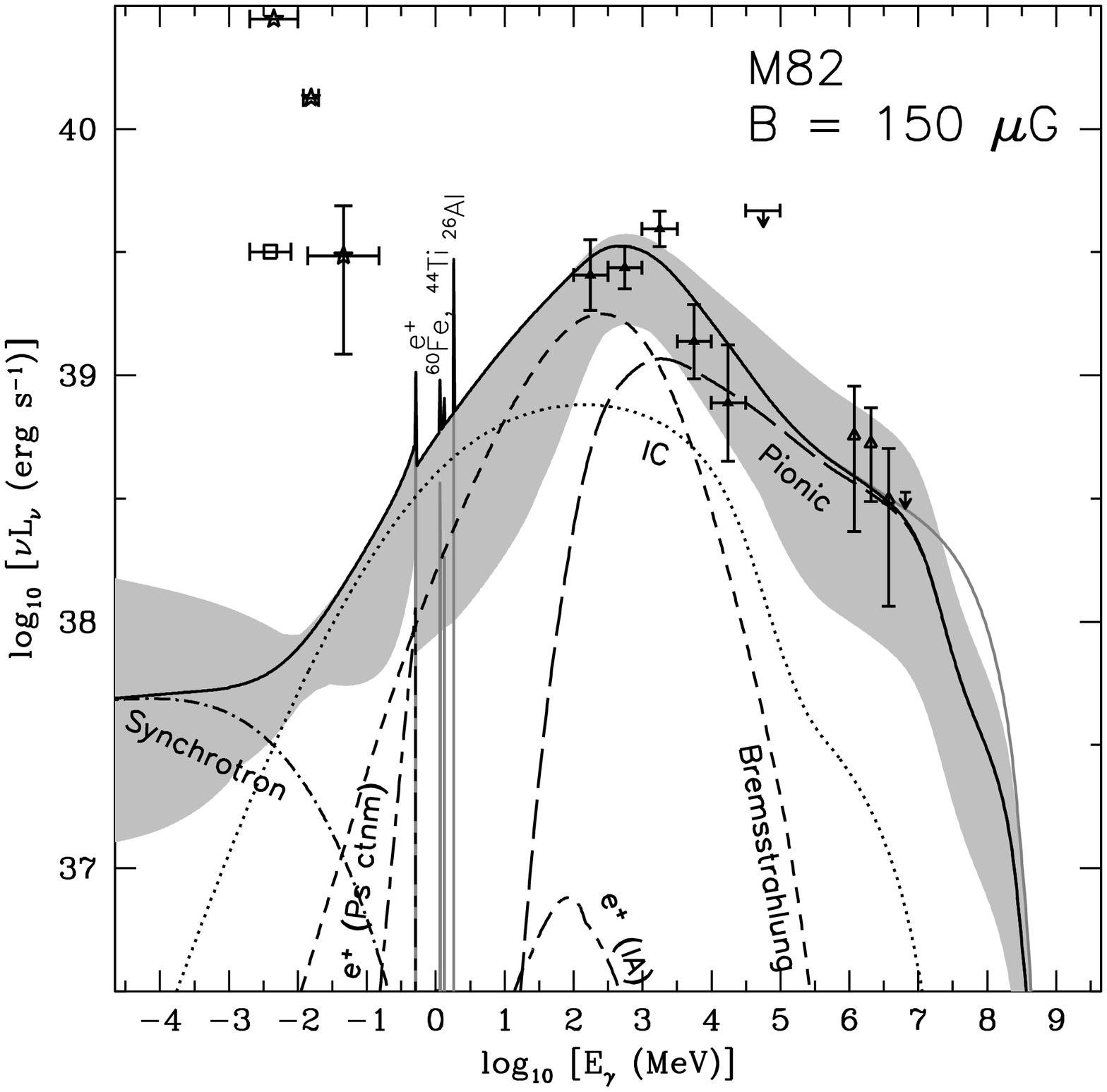}\includegraphics[width=9cm]{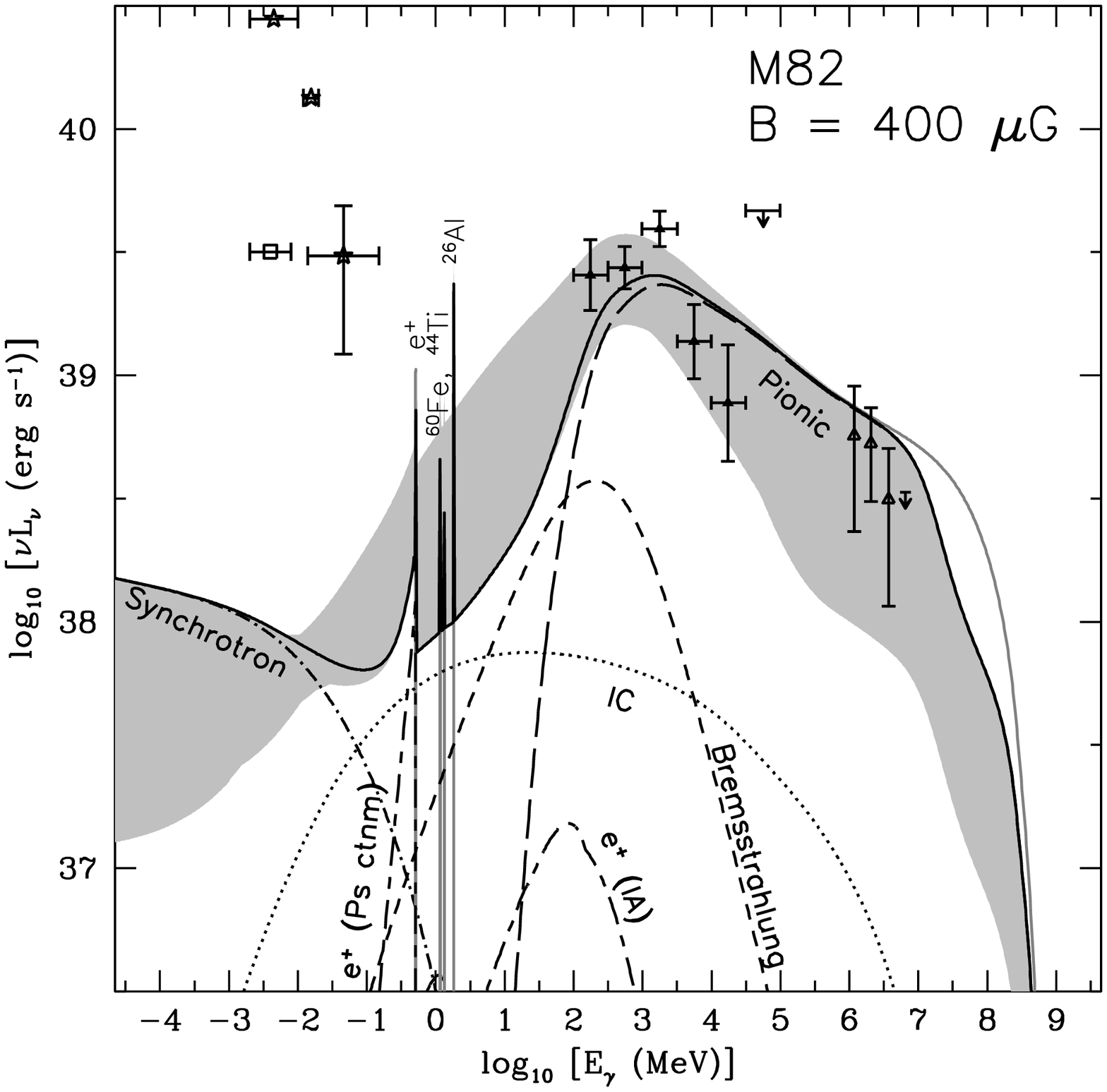}}
\figcaption[simple]{The predicted $\gamma$-ray luminosity spectrum of M82, compared to X-ray (total, stars: \citealt{Cappi99,Miyawaki09,Cusumano10}; diffuse, square: \citealt{Strickland07}) and $\gamma$-ray \citep{Abdo10f,Acciari09} observations.  The different components to the emission are synchrotron (dash/dotted), positron annihilation (long-dashed/short-dashed), nucleosynthetic $\gamma$-ray lines (grey lines), bremsstrahlung (short-dashed), IC (dotted), and pionic emission (long-dashed).  The total emission are the solid lines, black including $\gamma\gamma$ absorption on the sightline to Earth and grey without it.  The grey bands are the range spanned by all selected models of M82.   Emission from CCSNe is not included.  The X-ray emission mostly comes from X-ray binaries and thermal emission from gas and is not included in our fitting.  As in Figure~\ref{fig:MWGamma}, note the different energy binnings for the lines.\label{fig:M82Gamma}}
\end{figure*}

We use the interferometric radio data compiled in \citet{Williams10}, combined with the $\gamma$-ray data from \citet{Ackermann12} and \citet{Acciari09}.  There are 49 radio data points and 8 $\gamma$-ray data points versus 4 fit parameters ($B$, $\eta$, $\xi$, and $S_{\rm ff} (1\ \GHz)$) for 53 degrees of freedom.  The combined radio (Fig.~\ref{fig:M82Radio}) and $\gamma$-ray (Fig.~\ref{fig:M82Gamma})  data constrain the combinations of allowed $\eta$ and $\xi$ at each $B$.  We also plot X-ray data {\bfnop from BeppoSAX \citep{Cappi99}, \emph{Chandra} \citep{Strickland07}, \emph{Suzaku} \citep{Miyawaki09}, and SWIFT-BAT \citep{Cusumano10}} on Figure~\ref{fig:M82Gamma}, although we do not include {\bfnop them} in the fitting.  This emission largely comes from discrete X-ray binaries, and even the diffuse emission likely comes from hot thermal gas which would not contribute at MeV energies \citep[e.g.,][]{Strickland07}.

Our best-fit model is for $B = 200\ \muGauss$ with $\eta = 0.05$ and $\xi = 0.026$, which has $\chi^2 = 76.7$ (a reduced $\chi^2$ of 1.4, or 2.3$\sigma$ higher than expected for a good fit).  There are 21 considered models within $2\sigma$ of this best-fit model, with $\Delta\chi^2 \le 20.6$.  These models are listed in Table~\ref{table:BestFitModels}.  The allowed {\bfnop range in $B$ is between} $150\ \muGauss$ {\bfnop and} $400\ \muGauss$.  As expected, {\bfnop there are fewer primary electrons (small $\xi$) in high-$B$ models.}  {\bfnop The emission in} low{\bfnop-}$B$ models {\bfnop is} largely leptonic, {\bfnop and few protons are present in these models} (small $\eta$).

If the primary CR proton/electron injection ratio is the same in M82 as in the Milky Way, then the $B = 400\ \muGauss$ model with $\eta = 0.071$ is preferred, as it has $\tilde\delta = 73$, which is in the range 50 -- 100.  This is our ``fiducial'' or ``fiducial high-$B$'' model.  To demonstrate the effects of the degeneracy between $B$ and $\xi$, we also choose the $150\ \muGauss$ model with the smallest $\chi^2$ ($\eta = 0.035$) as our ``fiducial low-$B$'' model.

\emph{Results for MeV $\gamma$-rays} -- The main source of MeV $\gamma$-ray emission is that from Inverse Compton and bremsstrahlung.  In the fiducial model, the MeV emission is $\sim 10\%$ of the GeV luminosity (Figure~\ref{fig:M82Gamma}, right panel).  We find that while the 0.1 -- 10 GeV luminosity of M82 is $9.6 \times 10^{39}\ \ergps$ in our fiducial (high-$B$) model, the 1 -- 100 MeV luminosity is only $1.3 \times 10^{39}\ \ergps$.  

The low-$B$ models have stronger leptonic emission, in some models erasing the pion bump entirely.  Even in these models, the bremsstrahlung bump still causes the peak of the emission to be at 300 MeV (Figure~\ref{fig:M82Gamma}).  However, the relative dominance of IC emission means that $\nu L_{\nu}$ is only $\sim 6$ times smaller at 1 MeV than at 300 MeV in the fiducial low-$B$ model.  In this model, the 0.1 -- 10 GeV luminosity is $1.3 \times 10^{40}\ \ergps$ while the 1 -- 100 MeV luminosity is $6.4 \times 10^{39}\ \ergps$.  

\begin{figure}
\centerline{\includegraphics[width=9cm]{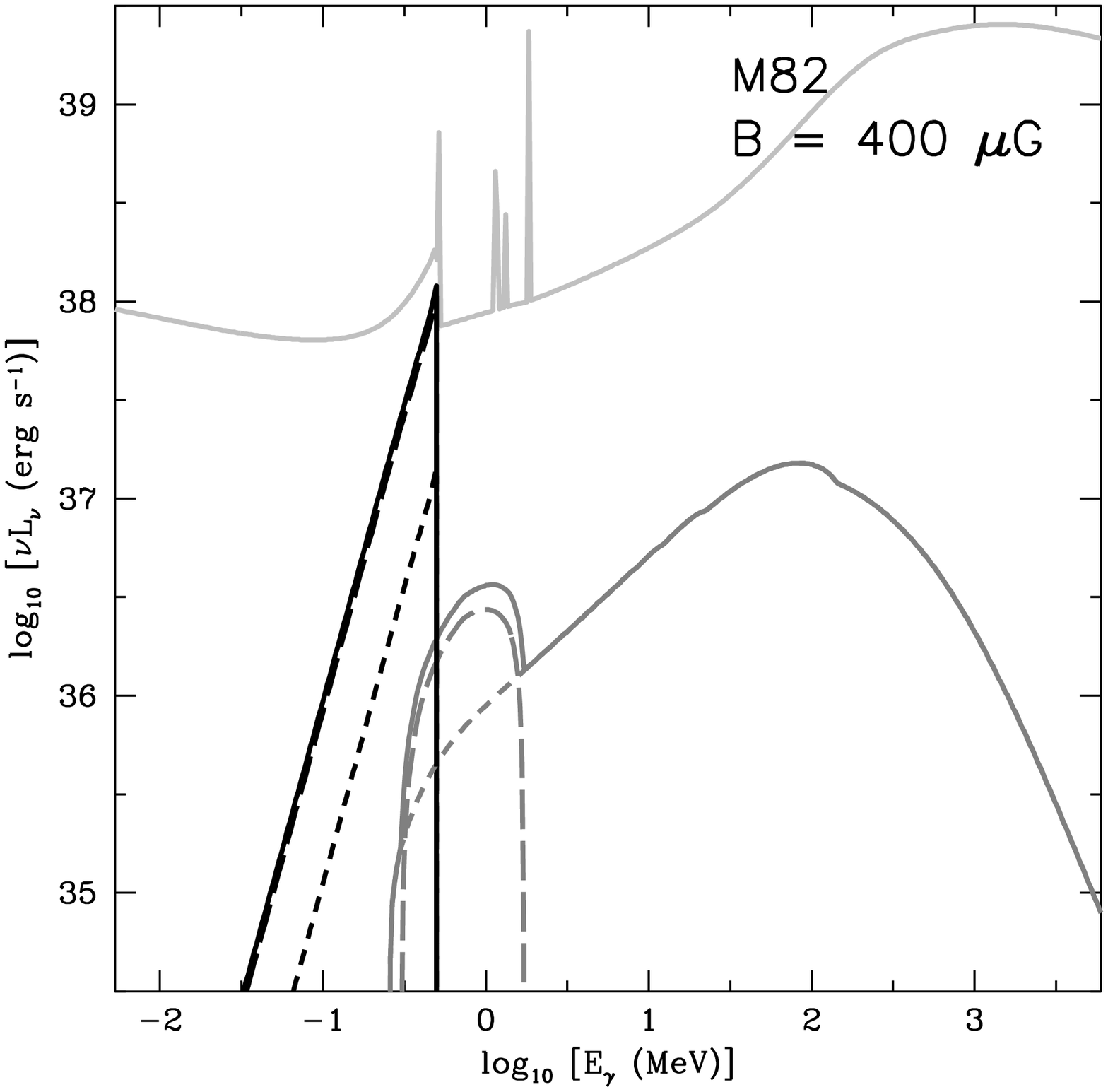}}
\figcaption[figure]{The continuum positron annihilation radiation (solid), both positronium continuum (black) and in-flight annihilation (grey), for the fiducial high-$B$ model.  The contributions from hadronic $e^+$ are short-dashed and nucleosynthetic $e^+$ from supernovae are long-dashed.  Nucleosynthetic $e^+$ dominate over the hadronic $e^+$ below 1 MeV.  For comparison, the total emission from M82 in this model is shown as the thick solid light grey line. \label{fig:M82ePlus}}
\end{figure}

The positron annihilation radiation is a possible contributor to the $\gamma$-ray flux in the 100 -- 500 keV range (Figure~\ref{fig:M82ePlus}).  The nucleosynthetic positrons are the dominant source of positrons.  However, the relatively calorimetric environment means that there is a large rate of hadronic positrons cooling all the way down to thermal energies: roughly $1/7$ of the positron annihilation radiation in our fiducial model is from pionic $e^+$.  Thus starbursts have slightly enhanced positron annihilation radiation associated with star-formation compared to normal galaxies (although non-star-formation processes are relatively weaker).  In the fiducial model, the total positron annihilation energy luminosity is $6.0 \times 10^{37}\ \ergps$ from positronium continuum and 511 keV line emission, plus $5.3 \times 10^{37}\ \ergps$ from in-flight annihilation radiation.  Hence the positron annihilation emission is only $\sim 5\%$ of the total MeV band emission; only near 511 keV does it actually dominate.

Finally, the $^{26}$Al and other nuclear lines are significant contributors to the $\sim 1\ \MeV$ luminosity of M82.  The 1.809 MeV line is the strongest $\gamma$-ray line except possibly the 511 keV annihilation line.  However, the photon flux is actually small, only $\sim 2 \times 10^{-8}\ \cm^{-2}\ \sec^{-1}$.  As for the nuclear de-excitation lines, they are completely negligible in M82 and far beyond any proposed detection capability (Table~\ref{table:GammaLineFluxes}).

\begin{deluxetable*}{lllcccc}
\tabletypesize{\scriptsize}
\tablecaption{Gamma-ray Line Strengths {\bfnop from M82}}
\tablehead{\colhead{Line\tablenotemark{a}} & \colhead{$E_{\gamma}$} & \colhead{Source\tablenotemark{b}} & \multicolumn{2}{c}{{\bfnop Fiducial} Low{\bfnop-}$B$} & \multicolumn{2}{c}{{\bfnop Fiducial} High{\bfnop-}$B$}\\ & & & \colhead{Luminosity} & \colhead{Photon Flux} & \colhead{Luminosity} & \colhead{Photon Flux} \\ & \colhead{MeV} & & \colhead{$\ergps$} & \colhead{$\phFluxUnits$} & \colhead{$\ergps$} & \colhead{$\phFluxUnits$}}
$e^+$ annih.           & 0.511 & CR (hadr.) & \SciNot{1}{36} & \SciNot{1}{-9} & \SciNot{2}{36} & \SciNot{2}{-9}\\
                       &       & Nucleo.    & \SciNot{2}{37} & \SciNot{1}{-8} & \SciNot{2}{37} & \SciNot{1}{-8}\\
$^{26}$Al decay\tablenotemark{c}        & 1.809 & Nucleo. & \SciNot{7}{37} & \SciNot{2}{-8} & \SciNot{7}{37} & \SciNot{2}{-8}\\
$^{60}$Fe decay\tablenotemark{c}        & 1.173 & Nucleo. & \SciNot{5}{36} & \SciNot{2}{-9} & \SciNot{5}{36} & \SciNot{2}{-9}\\
                       & 1.333 & Nucleo. & \SciNot{5}{36} & \SciNot{2}{-9} & \SciNot{5}{36} & \SciNot{2}{-9}\\
$^{44}$Ti decay\tablenotemark{c}        & 1.157 & Nucleo. & \SciNot{1}{37} & \SciNot{4}{-9} & \SciNot{1}{37} & \SciNot{4}{-9}\\
$^4$He de-ex.   & 0.429 & CR (broad)    & \SciNot{2}{32} & \SciNot{2}{-13} & \SciNot{4}{32} & \SciNot{4}{-13}\\
$^4$He de-ex.   & 0.478 & CR (broad)    & \SciNot{2}{32} & \SciNot{2}{-13} & \SciNot{5}{32} & \SciNot{4}{-13}\\
$^{12}$C de-ex. & 4.438 & CR (narrow)   & \SciNot{3}{34} & \SciNot{2}{-12} & \SciNot{5}{34} & \SciNot{5}{-12}\\
                &       & CR (broad)    & \SciNot{4}{34} & \SciNot{4}{-12} & \SciNot{8}{34} & \SciNot{7}{-12}\\
$^{16}$O de-ex. & 6.129 & CR (narrow)   & \SciNot{1}{34} & \SciNot{9}{-13} & \SciNot{3}{34} & \SciNot{2}{-12}\\
                &       & CR (broad)    & \SciNot{2}{34} & \SciNot{1}{-12} & \SciNot{3}{34} & \SciNot{2}{-12}\\
$^{16}$O de-ex. & 6.916 & CR (narrow)   & \SciNot{3}{33} & \SciNot{2}{-13} & \SciNot{6}{33} & \SciNot{4}{-13}\\
                &       & CR (broad)    & \SciNot{4}{33} & \SciNot{3}{-13} & \SciNot{9}{33} & \SciNot{5}{-13}\\
$^{16}$O de-ex. & 7.115 & CR (narrow)   & \SciNot{3}{33} & \SciNot{2}{-13} & \SciNot{7}{33} & \SciNot{4}{-13}\\
                &       & CR (broad)    & \SciNot{4}{33} & \SciNot{3}{-13} & \SciNot{9}{33} & \SciNot{5}{-13}\\
$^{12}$C de-ex. & 15.10 & CR (narrow)   & \SciNot{5}{32} & \SciNot{1}{-14} & \SciNot{1}{33} & \SciNot{3}{-14}\\
                &       & CR (broad)    & \SciNot{1}{33} & \SciNot{3}{-14} & \SciNot{2}{33} & \SciNot{7}{-14}
\enddata
\label{table:GammaLineFluxes}
\tablenotetext{a}{``Annih.'' abbreviates annihilation and ``de-ex.'' abbreviates de-excitation.}
\tablenotetext{b}{``Nucleo.'' stands for $\gamma$-ray lines produced by short-lived radioisotopes produced directly by stellar isotopes, while ``CR'' stands for $\gamma$-ray lines powered by CRs interacting with the ISM.  For de-excitation lines, ``narrow'' stands for the line component produced by the ISM atoms after being hit by a CR, while ``broad'' stands for the Doppler-shifted line component emitted by the CRs after hitting an ISM atom.  For the positron annihilation line, ``hadr.'' stands for hadronic positrons.}
\tablenotetext{c}{The nucleosynthetic $\gamma$-ray lines have exactly the same luminosities and fluxes in the high{\bfnop-}$B$ and low{\bfnop-}$B$ models.}
\end{deluxetable*}

\emph{High $z$} -- The CMB has much less of an effect on a starburst like M82 than in the Milky Way because the CMB is subdominant with respect to the high magnetic fields and starlight energy density (synchrotron and IC are the dominant loss processes at high electron energy).  We find that even at $z = 4.9$, the 1 -- 100 MeV emission is enhanced only by 6\% in our fiducial high-$B$ model.  In the fiducial low-$B$ model, the MeV enhancement at $z = 4.9$ is $10\%$ between 1 and 100 MeV.

\subsection{Brief Summary}
The starburst M82 is more luminous per unit star-formation than the Milky Way at GeV energies, but this is due to strong pion losses for protons, leading to a prominent pion bump in the high-$B$ models.  At present, we have essentially no MeV data for any starburst, so we cannot empirically rule out leptonic low-$B$ models with strong MeV emission.  However, low-$B$ models require that the ratio of accelerated electrons to protons be much higher than the Milky Way.

Compared to the high-$B$ starburst models, the Milky Way has a higher MeV/GeV ratio, due to relatively efficient conversion of primary $e^-$ energy into IC and bremsstrahlung radiation.  In the evolving model, the bigger gas density at high $z$ leads to much stronger pionic losses for CR protons, increasing the GeV emission, but the MeV emission remains comparable to its present levels.  In the non-evolving model, the Milky Way's MeV--GeV spectrum becomes flat from increased Inverse Compton losses off the CMB.  We therefore \emph{a priori} do not expect starbursts or evolving normal galaxies (with their prominent pion bumps) or unevolving normal galaxies (which are relatively dim at GeV to begin with) to be particularly strong MeV sources.

\section{M\lowercase{e}V emission from intergalactic cascades}
\label{sec:Cascades}
Even if the $\gamma$-ray spectra of star-forming galaxies were purely pionic, MeV emission is produced during the propagation of $\gamma$-rays from distant star-forming galaxies to Earth through the cascade process.  About half of the star-formation in the Universe occurs before $z \approx 1$.  At these distances, the Universe is opaque above 100 GeV from pair production processes ($\gamma + \gamma \to e^+ + e^-$) on the infrared {\bfnop portion of the} Extragalactic Background Light \citep[EBL;][]{Stecker92,Gilmore09,Finke10}.  The $e^{\pm}$ have a typical energy comparable to the original $\gamma$-ray, and cool mainly by Inverse Compton scattering of the CMB.

The typical rest-frame energy of upscattered CMB light is then
\begin{equation}
E_{\gamma}^{\prime\prime} \approx 32\ \MeV (1 + z) \left(\frac{E_e}{100\ \GeV}\right)^2,
\end{equation}
where $z$ is the redshift where the scattering takes place.  Thus intergalactic cascades shift energy from TeV bands, where pionic emission can be important, to MeV bands.  Normal star-forming galaxies have a pionic spectrum that goes as $E^{-2.7}$, with little $\gamma$-ray emission above 100 GeV, so they will contribute little cascade emission.  Starburst galaxies, on the other hand, are observed to have hard $E^{-2.2}$ spectra between GeV and TeV energies.  Nearly $1/3$ of their $\ge \GeV$ luminosity is above 100 GeV and can be affected by the cascade process.

Recently, \citet{Broderick12} have proposed that plasma instabilities {\bfnop stop cascade $e^{\pm}$ from bright TeV sources like blazars before they radiate IC.  There have been conflicting conclusions about the viability of this mechanism \citep[see][]{Miniati13,Venters13,Schlickeiser12}, although recent particle-in-cell simulations suggest that the instabilities saturate before quenching a significant fraction of the cascade emission \citep{Sironi13}.  In any case,} the TeV luminosities required for this process are much higher than from M82 and NGC 253.  \citet{Chang12} conclude that only HyperLIRGs would be affected by the plasma instability by assuming that the IR/$\gamma$-ray ratio is the same as in M82 and NGC 253.  We note that purely proton calorimetric galaxies would be expected to have IR/$\gamma$-ray ratios $\sim 2 - 3$ times higher than M82 and NGC 253's starburst, and so TeV emission from brighter ULIRGs (which are numerous at $z \approx 2$) may not cascade.  However, since most star-forming galaxies are relatively faint, we ignore the effects of plasma instabilities.  

\subsection{Calculation of Cascade Radiation}
Suppose the star-forming galaxy lies at redshift $z_s$ and emits a spectrum $dN_{\gamma}^{\rm source}/dE^{\prime}$ of photons over some time interval.  Then the cascade spectrum observed at redshift $z_{\rm obs}$ can be calculated as 
\begin{eqnarray}
\nonumber \frac{dN_{\rm casc}}{dE_{\gamma}^{\prime\prime}} (E_{\gamma}^{\prime\prime}, z_{\rm obs}) & = & \int_{z_{\rm obs}}^{z_s} dz \int_0^{E_{\gamma}(1+z)/(1 + z_{\rm obs})} \frac{dN_e}{dz dE_e^{\prime\prime}} \frac{dQ_{\rm IC}}{dE_{\gamma}^{\prime\prime}} (E_e^{\prime\prime}, E_{\gamma}^{\prime\prime})\\
& & \times t_{\rm life} \exp(-\tau_{\gamma\gamma} (E_{\gamma}, z_{\rm obs}, z)) dE_e^{\prime\prime}
\end{eqnarray}
where $E_{\gamma}^{\prime\prime} = (1 + z) E_{\gamma}$ is the cascade-frame energy, $E_{\gamma}^{\prime} = (1 + z_{\rm obs}) E_{\gamma}$ is the source-frame energy, and $E_{\gamma}$ is the Earth-frame energy.  The key ingredients in this calculation are $dN_e/(dE_e^{\prime\prime} dz)$, the spectrum of pair $e^{\pm}$ generated in the redshift step $z$ to $z - dz$ from both the primary $\gamma$-ray spectrum and the cascade $\gamma$-rays at previous redshift steps; $dQ_{\rm IC}/dE_{\gamma} (E_e, E_{\gamma})$, the IC emission at $E_{\gamma}$ from one CR $e^{\pm}$ of energy $E_e$; $t_{\bfnop \rm life}$, the cooling time for $e^{\pm}$; and $\tau_{\gamma\gamma} (E_{\gamma}, z_{\rm obs}, z)$, the $\gamma\gamma$ optical depth to a photon observed at $E_{\gamma}$ from redshift $z_{\rm obs}$ to $z$.

Accurate source functions for pair production are given in \citet{Aharonian83} and \citet{Boettcher97}.  We consider the emission from star-forming galaxies at redshift $z_s$ separately.  We then calculate the cascade emission in redshift steps, each from $z_i$ to $z_i - dz$.  At each step, the input photon spectrum (in the cascade rest frame) is
\begin{equation}
\frac{dN_{\gamma}}{dE_{\gamma}^{\prime\prime}} = \frac{dN_{\gamma}^{\rm source}}{dE_{\gamma}^{\prime}} \frac{1 + z_s}{1 + z} \exp(-\tau_{\gamma\gamma} (z_{\rm obs}, z)) + \frac{dN_{\rm casc}}{dE_{\gamma}^{\prime\prime}} (E_{\gamma}^{\prime\prime}, z)
\end{equation}
which includes both the attenuated primary spectrum and the attenuated cascade spectrum from previous redshift steps.  Within the redshift step, the UV-IR background effectively forms a screen to the $\gamma$-rays with optical depth:
\begin{equation}
\Delta \tau_{\gamma\gamma} = \tau_{\gamma\gamma} (E_{\gamma}^{\prime\prime}, z_i - \Delta z_i, z_i)
\end{equation}
So the generated pairs are calculated as
\begin{equation}
\frac{dN_e}{dz dE_e^{\prime\prime}} = \int \frac{dN_{\gamma}}{dE_{\gamma}^{\prime\prime}} \frac{dQ_e(E_{\gamma}^{\prime\prime})}{dE_e^{\prime\prime}} \frac{dt}{dz} \frac{1 - \exp[-\Delta\tau_{\gamma\gamma}]}{\Delta \tau_{\gamma\gamma}} dE_{\gamma}^{\prime\prime},
\end{equation}
where $dQ(E_{\gamma}^{\prime\prime})/dE_e^{\prime\prime}$ is the rate at which pair $e^{\pm}$ at energy $E_e$ are produced by a photon of energy $E_{\gamma}$, $dt/dz$ puts the redshift step in terms of time.  The exponential term guarantees energy conservation even if the Universe is opaque through the redshift step; in the limit of large $\tau$ the $dz (dt / dz) (1 - \exp(-\tau))/(\tau)$ expression reduces simply to the timescale for photon-photon annihilation.\footnote{The term comes from the time-averaged mean of the photon flux within the redshift step, which goes as $\int_{0}^{\Delta\tau} e^{-\tau} d\tau$ / $\int_0^{\Delta\tau} d\tau$.}  \citet{Boettcher97} give the \citet{Aharonian83} expression for the rate of $e^{\pm}$ pair production per high energy photon of energy $\varepsilon_1 m_e c^2$ as:
\begin{align}
\nonumber \frac{dQ_e(E_{\gamma}^{\prime\prime})}{dE_e^{\prime\prime}} & = \frac{3}{16} \frac{c \sigma_T}{m_e c^2} \int_{\frac{\varepsilon_1}{4 \gamma (\varepsilon_1 - \gamma)}}^{\infty} d\varepsilon_2 \frac{1}{(\varepsilon_2)^2} \frac{dn_{\rm back}}{d{\varepsilon_2}} \times \\
\nonumber & \left[\frac{4 \varepsilon_1^2}{\gamma (\varepsilon_1 - \gamma)} \ln \left(\frac{4\varepsilon_2 \gamma (\varepsilon_1 - \gamma)}{\varepsilon_1}\right) - 8 \varepsilon_1 \varepsilon_2 \right. \\
& \left. + \frac{2(2\varepsilon_1 \varepsilon_2 - 1) \varepsilon_1^2}{\gamma(\varepsilon_1 - \gamma)} - \left(1 - \frac{1}{\varepsilon_1 \varepsilon_2} \right) \frac{\varepsilon_1^4}{\gamma^2 (\varepsilon_1 - \gamma)^2} \right],
\end{align}
where $\varepsilon_2$ is the cascade-frame energy of a background photon in units of $m_e c^2$ and $dn_{\rm back} / d{\varepsilon_2}$ is the cascade-frame number spectrum of background photons. \footnote{Note that formulas 26 and 32 of \citet{Boettcher97} are for electrons only, not for both electrons and positrons (B\"ottcher, personal communication; Aharonian \& Khangulyan, personal communication).}  

The rate of Inverse Compton cascade emission from each pair $e^{\pm}$ is calculated as
\begin{equation}
\frac{dQ_{\rm IC}}{dE_{\gamma}^{\prime\prime}} = c E_{\gamma}^{\prime\prime} \int_0^{\infty} d\epsilon^{\prime} \frac{dn(\epsilon^{\prime}, z)}{d\epsilon^{\prime}} \sigma_{\rm IC} (E_{\gamma}^{\prime\prime}, \epsilon^{\prime}, E_e)
\end{equation}
where $dn (\epsilon^{\prime}, z)/d\epsilon^{\prime}$ is the rest-frame background radiation spectrum at redshift $z$ and rest-frame energy $\epsilon^{\prime}$ \citep{Schlickeiser02}.  The IC cross section is equal to
\begin{equation}
\sigma_{\rm IC} (E_{\gamma}, \epsilon^{\prime}, E_e) = \frac{3 \sigma_T}{4 \epsilon^{\prime} \gamma^2} \left[2q \ln q + (1 + 2q)(1 - q) + \frac{(\Gamma_e q)^2 (1 - q)}{2 (1 + \Gamma_e q)}\right]
\end{equation}
with $q = E_{\gamma} / [\Gamma_e (E_e - E_{\gamma})]$ ($1/(4\gamma^2) \le q \le 1$) and $\Gamma_e = 4 \epsilon^{\prime} \gamma / (m c^2)$ \citep{Schlickeiser02}.

In the intergalactic medium, electrons and positrons are mainly cooled by Inverse Compton losses and the adiabatic losses from the expansion of the Universe (again ignoring any plasma instability losses).  The electron lifetime to Inverse Compton emission is
\begin{equation}
t_{\rm IC} (E_e^{\prime\prime}) = E_e^{\prime\prime} \left[ \int_0^{\infty} E_{\gamma}^{\prime\prime} \frac{dQ_{\rm IC}}{dE_{\gamma}^{\prime\prime}} dE_{\gamma}^{\prime\prime} \right]^{-1},
\end{equation}
while the loss time due to the expansion of the Universe are simply
\begin{equation}
t_H = [H_0 \sqrt{\Omega_M (1 + z)^3 + \Omega_{\Lambda}}]^{-1}.
\end{equation}
The final $e^{\pm}$ lifetime is then just found as $t_{\rm life}^{-1} = t_{\rm IC}^{-1} + t_H^{-1}$.  

We use the extragalactic background light presented in \citet{Finke10} in these calculations.

\subsection{Results of Cascade Calculation}

\begin{figure}
\centerline{\includegraphics[width=9cm]{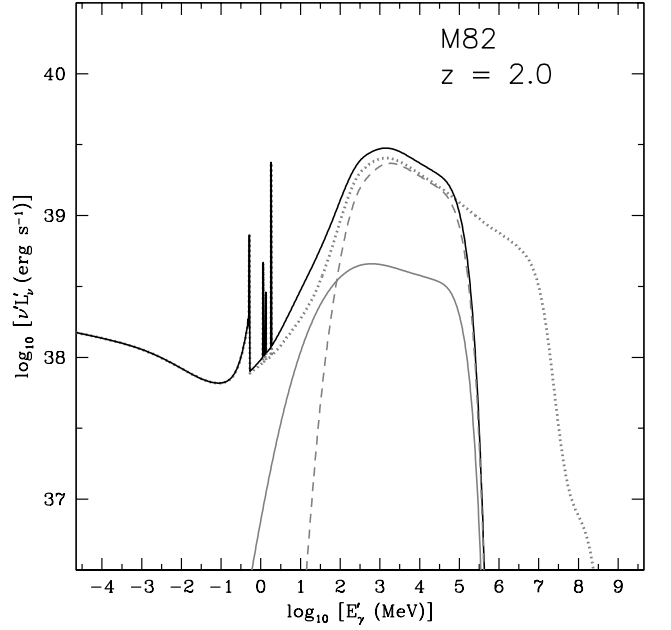}}
\figcaption[figure]{The effects of intergalactic cascades on M82's apparent rest-frame spectrum shape, for the fiducial model placed at $z = 2$.  The grey dotted line is the unpropagated input spectrum (after absorption within M82 itself), the grey solid line is the cascade emission, and the solid black line is the attenuated plus cascade emission. For comparison, the unpropagated pionic spectrum is shown as a dashed grey line.  The cascade spectrum sticks out of the pion bump, but is less strong than the leptonic emission.\label{fig:M82Cascade}}
\end{figure}

The effects of cascades on the spectra of Milky Way-like galaxies is relatively weak.  In the evolving galaxy model, cascade emission does extend below the pion bump, reaching a maximum of 40\% of the total observed emission at $\sim $5 MeV.  In the non-evolving model, though, the MeV Inverse Compton emission from within the galaxy swamps the tiny cascade contribution.  Cascades do have a small but noticeable effect on the spectrum of our high-B M82 spectra, as shown in Figure~\ref{fig:M82Cascade}.  Cascade emission makes up $\sim 20\%$ of the 100 MeV -- 10 GeV emission of starburst galaxies at $z = 2$.  There is a spectral downturn below 100 MeV, however the intrinsic emission of these models falls even faster because of the pion bump.  Thus, the fraction of emission due to cascades peaks near $\sim 10\ \MeV$, where it makes up about a third of the observed emission from starbursts.

We conclude that cascade emission alters the MeV spectrum of starburst galaxies at high $z$, but nowhere near enough to fill in the pion bump by itself.  

\section{Constraints on cosmic M\lowercase{e}V emission from star-forming galaxies}
\label{sec:MeVConstraints}
\subsection{Using the templates to predict the background}

The number density of $\gamma$-rays observed at energy $E_{\gamma}$ is 
\begin{equation}
\frac{dn}{dE_{\gamma}} (E_{\gamma}) = \int dt \frac{d^2 q_{\rm com}}{dV_{\rm com} dE_{\gamma}}
\end{equation}
where $dt = dz / [H_0 (1 + z) \sqrt{\Omega_{\Lambda} + \Omega_M (1 + z)^3}]$ describes the lookback time, $V_{\rm com}$ is the comoving volume, and $d^2 q_{\rm com} / dV_{\rm com} dE_{\gamma}$ is the injection rate of $\gamma$-rays per unit volume per unit observed energy.  The number density is converted into a photon flux by multiplying by $c / (4\pi)$.  This gives us
\begin{equation}
\label{eqn:BackgroundCalculation}
\frac{d\Phi}{dE_{\gamma}} (E_{\gamma}) = \int_0^{z_{\rm max}} dz \frac{c}{4 \pi H_0 (1 + z) [\Omega_M (1 + z)^3 + \Omega_\Lambda]^{1/2}} \frac{d^2 q_{\rm com}}{dV_{\rm com} dE_{\gamma}}.
\end{equation}
This expression can also be derived using the photon number luminosity distance.  The energy flux per (natural) log bin of energy is $\nu I_{\nu} = E_{\gamma}^2 d\Phi/dE_{\gamma}$.  

We calculate the injection rate of $\gamma$-rays as
\begin{equation}
\frac{d^2 q_{\rm com}}{dV_{\rm com} dE_{\gamma}} = \frac{\rho_{\rm com}^{\rm SFR} (z) f(z)}{{\rm SFR}_{\rm model}} \frac{dQ}{dE_{\gamma}^{\prime}}[E_{\gamma} (1 + z)] (1 + z),
\end{equation}
using $\rho_{\rm com}^{\rm SFR}$ is the star-formation rate per unit volume, $f(z)$ is the fraction of star-formation from galaxies like the model galaxy, $dQ/dE_{\gamma}^{\prime}$ is the total source-frame photon generation rate in the model galaxy (after cascades), and ${\rm SFR}_{\rm model}$ is the star-formation rate of the model galaxy.  The comoving star-formation rate comes from \citet{Hopkins06}.

\begin{figure}
\centerline{\includegraphics[width=9cm]{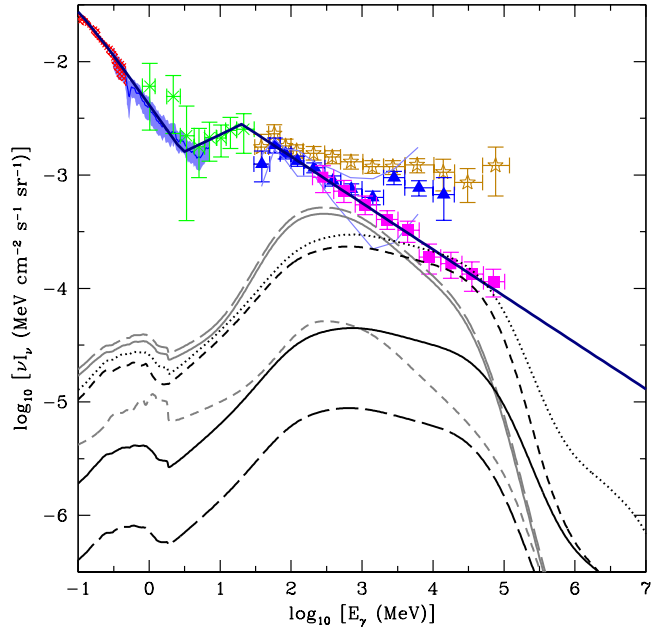}}
\figcaption[figure]{The effects of starburst fraction on the predicted $\gamma$-ray background.  Starbursts (fiducial {\bfnop high-$B$} M82) are in black and normal galaxies  (fiducial Milky Way) are in grey.  The long-dashed line is the low starburst fraction history, the solid line is the fiducial medium starburst fraction, the short-dashed line is the high starburst fraction, and the dotted line is if all of the star formation is in starbursts.  The normal galaxy contribution is only different for the high starburst fraction, whereas the starburst contribution can vary by a factor $\sim 30$ between the different starburst fractions.\label{fig:fSBBack}}
\end{figure}

To compare the predicted background to the observed background, we link together fits to the X-ray background and the GeV background.  At X-ray energies, \citet{Ajello08} fit the X-ray background with a double power law, while at GeV energies, \citet{Abdo10a} fit the GeV background with a single power law.  We connect the two with a power law (linear fit in $\log\ E$ and $\log\ \nu I_{\nu}$) at intermediate energies, which we take to be 3 to 20 MeV by visual inspection.  We use the resulting function to represent the strength of the X-ray--$\gamma$-ray background:
\begin{equation}
\nu I_{\nu} = \left\{ \begin{array}{ll}
                     \displaystyle \frac{0.102 E_{\rm keV}^2}{\left(\displaystyle \frac{E_{\rm keV}}{29.99}\right)^{1.32} + \left(\displaystyle  \frac{E_{\rm keV}}{29.99}\right)^{2.88}} & (E \le 3\ \MeV)\\
		     1.58 \left(\displaystyle \frac{E_{\rm keV}}{3000}\right)^{0.303} & (3\ \MeV \le E \le 20\ \MeV)\\
		     1.45 \left(\displaystyle \frac{E_{\rm keV}}{100000}\right)^{-0.41} & (E \ge 20\ \MeV) \end{array} \right.		     
\end{equation}
in units of $\keV\ \cm^{-2} \sec^{-1} \sr^{-1}$ for a photon of energy $E$, with $E_{\rm keV} = E / \keV$ \citep{Ajello08,Abdo10a}.  The function is plotted in Figure~\ref{fig:GammaBackground} (and Figures~\ref{fig:fSBBack} and \ref{fig:MWScaledBackground}) as the solid dark blue line.  

\subsection{The question of the starburst fraction of the cosmic SFR}
\label{sec:StarburstFraction}
A large uncertainty is the fraction of cosmic star-formation history that occurs in starburst galaxies.  Indeed, discussions of the starburst contribution to the $\gamma$-ray background must be clear about what ``starburst'' even means for these calculations.  We can simply define starburst as having a similar $\gamma$-ray SED (and thus, CR propagation environment) as M82.  In particular, it means (1) similar $\gamma$-ray to SFR ratios as M82 and (2) hard GeV -- TeV $\gamma$-ray spectra, from winds and pionic losses.  At the high end of starburst fraction estimates, \citet{Loeb06} and \citet{Thompson07} argued that most of the star-formation of the Universe was in proton calorimetric galaxies, and thus, the $\gamma$-ray background from starbursts is high.  Further down in starburst fraction, \citet{Stecker07} argued that most of the star-formation at $z < 1$ was in normal galaxies, implying a small $\gamma$-ray background from starbursts.  At the low end, \citet{Stecker10} have recently used semi-analytic models which argue that the fraction of cosmic star-formation in pure merger-driven starbursts is very low, $\sim 1 - 2\%$ at $z = 0$, $\sim 3\%$ at $z = 1$, and $\sim 5\%$ at $z = 4$ \citep{Hopkins10}.  

There are potential issues with each of these approaches.  \citet{Loeb06} and \citet{Thompson07} argue that normal galaxies have $\sim 10$ times more gas at high $z$ than at low $z$; therefore they should be $\sim 10$ times more efficient at pionic losses and nearly proton calorimetric.  However, at sufficiently high gas surface densities, advective losses may be strong enough to negate this effect.  In fact, the Galactic Center has similar gas surface densities ($\Sigma_g \approx 0.03\ \gcm2$), but is in fact \emph{fainter} in $\gamma$-rays per unit star-formation than the surrounding Galaxy \citep{Crocker11-Wind,Crocker11}.  This is apparently because of a strong wind in the region and also because CRs do not appear to be fully sampling the gas \citep{Crocker11}.  Therefore, normal galaxies at high $z$ may form stars quickly and have large amounts of gas without having M82-like $\gamma$-ray spectra (although the pionic $\gamma$-ray spectrum will likely be hard because both pionic and advective losses harden the spectrum).  However, galaxies at high $z$ appear to lie on the far-infrared correlation \citep[e.g.,][]{Appleton04,Garn09,Sargent10,Mao11}, whereas the Galactic Center does not \citep{Crocker11-Wind}, which is consistent with CRs in high $z$ galaxies interacting with average density gas and radiating before escaping \citep{Lacki10c}.  

The predictions of \citet{Stecker10} are based on the semi-analytic models of \citet{Hopkins10} (H10).  In these models, galaxies have a normal star-formation mode most of the time, but have a finite probability of merging; when they merge, they enter into a brief ($\sim 100\ \Myr$) starburst mode of star-formation whose evolution is guided by gas hydrodynamics simulations.  The simulations of \citet{Hopkins10} do not include processes like stellar bars which may feed the lower luminosity starbursts.  For example, M82 (the starburst with the largest $\gamma$-ray flux) is interacting with its neighbor M81 and has a compact starburst driven by a stellar bar \citep{Telesco91}, but is not yet merging, so it might not be counted by H10.   Also, a normal galaxy with a high enough gas density and weak enough advective losses will become proton calorimetric and will have a starburst-like $\gamma$-ray spectrum, even if it is not formally starbursting.  Thus defining starbursts purely as $\sim 100$ Myr long mergers may underestimate the contribution of starburst-like $\gamma$-ray emitters.  It is nonetheless likely that the strongest compact starbursts like Arp 220, which arise from true mergers, are a small fraction of the cosmic star-formation rate.

Observationally, a number of lines of evidence point to a starburst fraction of $\sim 10\%$, in terms of star-formation mode.  The total core-collapse supernova rate within 10 Mpc is $\sim 0.3 - 2\ \yr^{-1}$ \citep[e.g.,][]{Kochanek08,Kistler11,Li11,Horiuchi11}.  The compact starbursts in M82, NGC 253, and NGC 4945 ($D \approx 3.9\ \Mpc$) have a total supernova rate of $\sim 0.12\ \yr^{-1}$ from their FIR luminosity \citep{Sanders03}.  Further out, there is as nuclear starburst in NGC 4631 ($D = 7.7\ \Mpc$) with $\Gamma_{\rm SN} = 0.01 - 0.06\ \yr^{-1}$ \citep{Golla99}, and smaller nuclear starbursts in M51, M83, Maffei 2, NGC 2903, NGC 4736, and IC 342 \citep{Kennicutt98}.  Conservatively assuming a total starburst supernova rate of $0.1\ \yr^{-1}$ within 10 Mpc, we find a starburst fraction of $\sim 5 - 30\%$.  Radio and IR luminosity functions can be fit with two populations of star-forming galaxies, identified as normal galaxies and starbursts \citep{Yun01}.  These luminosity functions indicate that starbursts make up $\sim 10 - 20\%$ of the star-formation rate at $z \approx 0$ \citep[e.g.,][]{Yun01,Bothwell11}.  Whether these starbursts are all compact and $\gamma$-ray bright like M82 is unclear.  \citet{Rodighiero11} find that only 10\% of the cosmic star-formation is in starbursts that lie off the ``main sequence'' of galaxies, a relation between the star-formation rate and stellar mass (see also \citealt{Sargent12}).  \citet{Hopkins10b} derive a starburst fraction of $\sim 5 - 10\%$ by using the  surface brightness profiles of spheroidal galaxies.  In addition to the compact starbursts at $z \approx 0$, there are the submillimeter galaxies (SMGs) observed at $z \ga 2$.  There is still some debate whether SMGs are simply disk galaxies with very high star-formation rates or mergers, but they appear to make up 2 -- 10\% of the cosmic star formation  \citep[e.g.,][]{Chapman05,Dye07,Michalowski10}.  With vast amounts of gas, SMGs are likely to have efficient CR proton losses, although their large sizes means that the physical conditions (and therefore leptonic emission) could be different than M82.  

In this work, we consider three cases: (1) the low starburst fraction case, calculated by integrating over the Schechter functions given in H10 (similar to the \citealt{Stecker10} calculation), rising from $1.6\%$ at $z = 0$ to $3.8\%$ at $z = 2$ to $5.1\%$ at $z = 4$ and beyond\footnote{Since the H10 luminosity functions are not designed for use at $z > 4$, we use the $z = 4$ starburst fraction at these redshifts.}; (2) a high starburst fraction case as given in \citet{Thompson07}; and (3) our fiducial medium starburst fraction case, which is 15\% at all redshifts, since that is roughly the geometric mean of the low and high cases and it is fairly close to the $\sim 10\%$ that seems to hold at low $z$. 

\subsection{The predicted MeV -- GeV $\gamma$-ray background}
Using the cosmic star-formation rate, assumptions about the contribution of starbursts to the cosmic star-formation history, and our Milky Way and M82 templates, we can tentatively predict the $\gamma$-ray background with eqn.~\ref{eqn:BackgroundCalculation}.  We list our results for different galaxy spectral templates and starburst fractions in Table~\ref{table:GammaBackSummary}.  

\begin{deluxetable*}{llcccccccc}
\tabletypesize{\scriptsize}
\tablecaption{Predicted $\gamma$-Ray Background}
\tablehead{\colhead{Model} & \colhead{$f_{\rm SB}$\tablenotemark{a}} & \multicolumn{8}{c}{Predicted fraction of $\gamma$-ray background} \\ & & \colhead{100 keV} & \colhead{300 keV} & \colhead{1 MeV} & \colhead{10 MeV} & \colhead{100 MeV} & \colhead{1 GeV} & \colhead{10 GeV} & \colhead{100 GeV} \\ & & \colhead{$\%$} & \colhead{$\%$} & \colhead{$\%$} & \colhead{$\%$} & \colhead{$\%$} & \colhead{$\%$} & \colhead{$\%$} & \colhead{$\%$}}
\cutinhead{Normal Galaxies: Milky Way template\tablenotemark{b}}
$h = 1000\ \pc$ (Non-evolving)  & H10  & 0.13  & 0.40 & 1.2  & 2.1  & 7.1 & 14 & 9.7 & 2.8\\
                                & 15\% & 0.11  & 0.35 & 1.1  & 1.8  & 6.2 & 13 & 8.5 & 2.5\\
                                & T07  & 0.017 & 0.06 & 0.26 & 0.41 & 1.5 & 3.9 & 3.1 & 1.4\\
$h = 1000\ \pc$ (Evolving)      & H10  & 0.071 & 0.27 & 0.90 & 2.3  & 28  & 73  & 58  & 8.9\\
                                & 15\% & 0.062 & 0.24 & 0.79 & 2.1  & 25  & 65  & 51  & 7.7\\
                                & T07  & 0.015 & 0.065 & 0.26 & 0.44 & 2.4 & 7.7 & 6.0 & 2.7\\
$h = 2000\ \pc$ (Non-evolving)  & H10  & 0.23  & 0.65  & 2.0  & 3.6  & 8.8 & 17  & 12  & 3.3\\
                                & 15\% & 0.20  & 0.57  & 1.7  & 3.1  & 7.7 & 15  & 10  & 2.9\\
                                & T07  & 0.031 & 0.10  & 0.40 & 0.73 & 1.9 & 4.6 & 3.9 & 1.7\\
$h = 2000\ \pc$ (Evolving)      & H10  & 0.11  & 0.37  & 1.2  & 3.2  & 37  & 86  & 63  & 40\\
                                & 15\% & 0.098 & 0.33  & 1.1  & 2.8  & 32  & 76  & 55  & 8.0\\
                                & T07  & 0.028 & 0.10  & 0.38 & 0.71 & 2.9 & 8.4 & 6.6 & 2.9\\
\cutinhead{Starbursts: M82 template}
$B = 150\ \muGauss$ & H10  & 0.0041 & 0.16          & 0.066 & 0.23 -- 0.24 & 0.56 -- 0.76 & 1.0 -- 1.9 & 0.86 -- 2.5 & 0.26 -- 1.3\\
                    & 15\% & 0.019  & 0.080 -- 0.081 & 0.33  & 1.1 -- 1.2   & 2.8 -- 3.8   & 5.4 -- 10  & 4.6 -- 13   & 1.6 -- 8.3\\
                    & T07  & 0.11   & 0.43 -- 0.44   & 1.7   & 6.1 -- 6.4   & 15 -- 20     & 27 -- 49   & 22 -- 66    & 6.1 -- 32\\
$B = 200\ \muGauss$ & H10  & 0.0028 & 0.012         & 0.042 -- 0.043 & 0.14 -- 0.16 & 0.37 -- 0.65 & 0.72 -- 1.9 & 0.75 -- 3.1 & 0.33 -- 1.8\\
                    & 15\% & 0.013 -- 0.014 & 0.057  & 0.22          & 0.70 -- 0.76 & 1.9 -- 3.2   & 3.8 -- 10   & 4.0 -- 16   & 2.1 -- 11\\
                    & T07  & 0.074 -- 0.075 & 0.30 -- 0.31 & 1.1 -- 1.2 & 3.8 -- 4.2  & 10 -- 17     & 19 -- 51    & 20 -- 81    & 7.9 -- 43\\
$B = 250\ \muGauss$ & H10  & 0.0022        & 0.0090 -- 0.0091 & 0.031 -- 0.032 &  0.10 -- 0.11 & 0.33 -- 0.54 & 0.80 -- 1.7 & 1.1  -- 2.9 & 0.63 -- 1.8\\
                    & 15\% & 0.010 -- 0.011 & 0.045       & 0.16     & 0.49 -- 0.54 & 1.6 -- 2.6   & 4.1 -- 8.9  & 5.9 -- 15   & 4.0 -- 11\\
                    & T07  & 0.057 -- 0.058 & 0.24        & 0.82 -- 0.84 & 2.7 -- 3.0 & 8.8 -- 15 & 21 -- 46 & 30 -- 77 & 15 -- 43\\
$B = 300\ \muGauss$ & H10  & 0.0018 -- 0.0019 & 0.0076    & 0.024 -- 0.025 & 0.078 --  0.086 & 0.32 -- 0.48 & 0.91 -- 1.6 & 1.5 -- 2.9 & 0.88 -- 1.7\\
                    & 15\% & 0.0086 -- 0.0088 & 0.038     & 0.13     & 0.38 -- 0.41 & 1.6 -- 2.3   & 4.7 -- 8.4  & 7.6 -- 15 & 5.5 -- 11\\
                    & T07  & 0.047 -- 0.048   & 0.20      & 0.65 -- 0.66 & 2.1 -- 2.3 & 8.5 -- 13 & 24 -- 43 & 39 -- 75 & 21 -- 42\\
$B = 400\ \muGauss$ & H10  & 0.0014 -- 0.0015 & 0.0060    & 0.017 -- 0.018 & 0.054 -- 0.060 & 0.32 -- 0.42 & 1.1 -- 1.5 & 2.0 -- 2.8 & 1.2 -- 1.7\\
                    & 15\% & 0.0068 -- 0.0070 & 0.030     & 0.093 -- 0.095 & 0.26 -- 0.28 & 1.6 -- 2.0 & 5.7 -- 7.8 & 10 -- 14 & 7.7 -- 11\\
                    & T07  & 0.037 -- 0.038   & 0.16      & 0.46 -- 0.47   & 1.4 -- 1.6   & 8.7 -- 11  & 29 -- 41   & 52 -- 74 & 29 -- 42
\enddata
\label{table:GammaBackSummary}
\tablenotetext{a}{Starburst fraction used in the model.  H10: our low starburst fraction case, with the starburst fraction in the \citet{Hopkins10} semianalytic models; 15\%: our fiducial medium starburst fraction case, where 15\% of the cosmic star formation rate at all $z$ is in starbursts; T07: our high starburst fraction case, using the values in \citet{Thompson07}.}
\tablenotetext{b}{Due to the uncertainty in the \citet{Strong10} pion lumiosity, all Milky Way template values are uncertain at the 20\% level.}
\end{deluxetable*}

In general, we find that starburst galaxies have a lot of \emph{absolute} power per unit star-formation at GeV energies.  Using the evolving normal galaxy template, high redshift normal galaxies also are bright at GeV energies (Figure~\ref{fig:GammaBackground}).  With the non-evolving template, normal galaxies are faint overall but have much \emph{relative} power per unit star-formation at MeV energies.  Using the fiducial evolving models and starburst fraction history, we find that normal galaxies contribute 25-65\% of the 0.1 -- 10 GeV background.  Their contribution falls off rapidly at lower energies due to the pion bump to $\sim 2\%$ of the 10 MeV background, $\la 1\%$ of the 1 MeV background, and $\sim 0.1\%$ of the the 100 keV background (neglecting discrete hard X-ray sources in these galaxies; Table~\ref{table:GammaBackSummary}).  In the non-evolving models, by contrast, the normal galaxies are a minority of the GeV background ($\sim 10\%$), but the expected contribution to the MeV background is of the same order as for the evolving templates.  In the non-evolving case, the small contribution to the MeV background is not so much due to the pion bump, but instead arises because the $\gamma$-ray background is steeply falling with energy.

The starburst contribution to the $\gamma$-ray background depends on the uncertain contribution of starbursts to the cosmic star-formation rate (\S~\ref{sec:StarburstFraction} and Figure~\ref{fig:fSBBack}) and, for MeV energies in particular, the magnetic field strengths of the starbursts.  We find the starburst contribution at 1 -- 10 GeV is only $\sim 1 - 3\%$ with the low starburst fraction, $\sim 4 - 16\%$ with the medium starburst fraction, and from $\sim 19 - 81\%$ with the high starburst fraction (Table~\ref{table:GammaBackSummary}).  The high starburst fraction case gives similar results similar to the predictions for the evolving normal galaxy template: either way assumes that most star-formation at high $z$ occurs in very dense galaxies.  The starburst contribution rapidly falls off at lower energies because of the prominent pion bump: with the medium starburst fraction and the fiducial M82 spectral template, the contribution falls to $2\%$ at 100 MeV, $0.3\%$ at 10 MeV, $0.09\%$ at 1 MeV, and $0.007\%$ at 100 keV (again, ignoring discrete sources within the galaxies).  Note that the pion bump in high-$B$ models does not mean that starbursts are MeV dim per unit star-formation compared to normal galaxies; it arises because starbursts are GeV-TeV \emph{bright} through their higher proton calorimetry fraction.  Of course, in the fiducial low-$B$ models, where there is a greater population of CR electrons, the starburst contribution at MeV energies is higher: 3\% at 100 MeV, 1\% at 10 MeV, 0.3\% at 1 MeV, and 0.02\% at 100 keV or the medium starburst history.

The $\gamma$-ray lines, whether from CCSNe, $e^+$ annihilation from nucleosynthetic positrons, or unstable isotopes in the ISM, do not constitute the majority of the MeV contribution from star-forming galaxies; continuum processes are the dominant source of MeV luminosity of star-forming galaxies.  The lines do make up $\sim 1/2$ of the star-forming galaxy contribution at 0.5 MeV and $\ga 10\%$ from $\sim 0.2 - 1$ MeV.  The fraction of the MeV emission from star-forming galaxies from the transient CCSNe emission peaks at $\sim 17\%$ at 0.4 MeV. 

In the fiducial model, the 1--30 GeV background is slightly steeper than the background observed by \emph{Fermi}, with a spectral slope $\Gamma_{1-30} = 2.52$, compared to $2.41$ for the observed background \citep{Abdo10a}.  The predicted and observed spectrum strongly diverge at energies below those observed with \emph{Fermi}-LAT.

Summarizing our fiducial results, we find that the evolving normal galaxies dominate the unresolved 1 GeV background.  Starbursts contribute at the 15\% level.  Neither makes up an appreciable amount of the MeV background.

However, there are a number of caveats with this prediction.  Most importantly, we have assumed that there are only two types of galaxies: those like the Milky Way (possibly with some evolution), and those like M82.  In reality, galaxies are probably more diverse in their $\gamma$-ray properties.  Very weak starbursts like the Galactic Center region may have hard TeV $\gamma$-ray spectra like M82, but very low $\gamma$-ray luminosities because of winds.  Very extreme starbursts like Arp 220 would probably be true proton calorimeters and be even more efficient at producing $\gamma$-rays than M82, although these are a small fraction of the cosmic star-formation rate. Finally, we have ignored IC emission from $e^{\pm}$ that have escaped normal galaxies, which will increase their MeV emission at low $z$.

\subsection{Comparison with previous predictions}
As a check, we can compare our predictions to other recent predictions of the cosmic GeV background.  

Using the high starburst fraction history, \citet{Thompson07} predicted that $\sim 20\%$ of the Fermi 1 GeV $\gamma$-ray background ($\sim 10\%$ of the EGRET GeV background) is from starbursts.  We in fact predict $\sim 1 - 3$ times more GeV $\gamma$-rays with the T07 starburst history; with the fiducial starburst template, starbursts make up $41\%$ of the \emph{Fermi} 1 GeV background.  The difference arises because \citet{Thompson07} assumed smaller values of $\eta$ than we find, thus underpredicting the $\gamma$-ray luminosity of M82. \citet{Lacki10a} noted this effect and argued that a high starburst fraction would lead to a $\ge \GeV$ starburst contribution of order $\sim 50\%$, more in line with our predictions here.

The pre-\emph{Fermi} calculation of \citet{Bhattacharya09} essentially scales the $\gamma$-ray luminosity of a galaxy to its TIR luminosity (roughly proportional to the star-formation rate), with different constants of proportionality for normal galaxies and starbursts, an approach comparable to ours.  They find the total $\gamma$-ray background by integrating over infrared luminosity functions.  They find that only $\sim 1\%$ of the $\ge 0.1\ \GeV$ background is from normal galaxies.  This is probably because they used the luminosity functions of \citet{Lagache03}, in which normal galaxies make up only a small fraction of the cosmic star-formation rate at high $z$.  (For comparison, the T07 starburst fraction gives us a $\sim 2 - 3\%$ normal galaxy contribution to the 100 MeV $\gamma$-ray background.)  They also used a hard $E^{-2.2}$ spectrum for normal galaxies, although that only increases their predictions.  They predict the starburst contribution to be $\la 6\%$; the small estimate likely arises because they derive (ultimately from equipartition arguments about M82's radio luminosity from \citealt{Akyuz91}, with a volume scaling applied) that $L_{\rm GeV} / L_{\rm IR}$ is only $30\%$ higher for starbursts than for normal galaxies, whereas the \emph{Fermi} observations indicate $L_{\rm GeV} / L_{\rm IR}$ is about 5 -- 10 times higher.  

\citet{Fields10} calculated the pionic $\gamma$-ray background from normal star-forming galaxies, relating the $\gamma$-ray emissivity of a galaxy to its gas mass, which in turn is related to the star-formation rate of a galaxy through the Schmidt Law and empirical trends in the average radius of galaxies.  Their varying assumptions about what drives the evolution in the cosmic star-formation rate lead to predictions that normal galaxies are at least $\sim 20\%$ and possibly all of the unresolved $\gamma$-ray background.  Because they used ${\rm SFR}_{\rm MW} = 1\ \Msun\ \yr^{-1}$ {\bfnop (compared to our value of $2\ \Msun\ \yr^{-1}$)}, their normal galaxy contribution to the $\gamma$-ray background is twice as large as we would predict.  After taking this into account, the ``density evolution'' case of \citet{Fields10} gives similar prediction of the normal galaxy contribution as our non-evolving case, while the ``luminosity evolution'' case with increased gas masses at high $z$ predicts a contribution similar to evolving case. 

\citet{Makiya11} predicted that $\sim 5\%$ of the $\ge 100\ \MeV$ $\gamma$-ray background is from (mostly normal) star-forming galaxies, if the $\gamma$-ray luminosity per unit star-formation is the same as in the Milky Way.  This is somewhat lower than our estimate of $6-17\%$ of the $0.1 - 10\ \GeV$ background coming from normal galaxies in the non-evolving case.  They also find that if gas evolution is taken into account, the contribution increases to $\sim 10\%$.  By using the Mitaka models of galaxy evolution \citep{Nagashima04,Nagashima05} and by approximating the starburst galaxy spectrum with the Milky Way spectrum plus a power law, \citet{Makiya11} also calculated the starburst contribution to GeV background.  They found the starburst contribution was small, about 1\% of 100 MeV background and a few percent of the 100 GeV background.  According to the Mitaka models, starbursts are only a few percent of the $z \approx 1$ star-formation rate \citep{Nagashima05}, which is of the same order as our low starburst fraction (H10).  By comparison, we find similar results for the starburst contribution to the GeV background using the H10 {\bfnop starburst} fraction.  

\citet{Stecker10} performed a similar calculation of the GeV background from star-forming galaxies to \citet{Makiya11}, although they used the H10 galaxy evolution models, and they assumed that starbursts had a soft, Milky Way-like spectrum.  They found that normal star-forming galaxies were anywhere between $\sim 10\%$ and all of the GeV background depending on how the $\gamma$-ray luminosity scales with star-formation rate and/or gas mass.  {\bfnop The spread in $\gamma$-ray background flux is comparable to the spread in flux in our models between evolving and non-evolving models.}  They find a starburst contribution peaking at $\sim 1\%$ to the GeV background, which is somewhat lower than our results for an H10 star-formation rate, and dropping off more quickly at high energies than we predict because of the soft $\gamma$-ray spectrum they used.

After fitting a relationship between the observed $\gamma$-ray and infrared luminosities of galaxies, \citet{Ackermann12} present an estimate of the pionic star-forming galaxy contribution to the $\gamma$-ray background.  They conclude that star-forming galaxies make up $\sim 4 - 23\%$ of the $\gamma$-ray background.  This is smaller than most of our estimates, namely those with large starburst fraction or with evolving normal galaxies.  If our estimates are correct, then galaxies at high redshift are brighter per unit star-formation than would be expected from the \citet{Ackermann12} relation.

\citet{Chakraborty13} presented a calculation of the pionic and IC backgrounds from star-forming galaxies.  Overall, their results are similar to our results with the evolving normal galaxy template.  They found that the pionic background is a substantial fraction of the unresolved GeV background, with uncertainties at the order of magnitude level.  The Inverse Compton contribution is a distinct minority of the resulting $\gamma$-ray background, with an intensity only $\sim 10\%$ that of the pionic background, for an intensity $\nu I_{\nu} \approx 2 \times 10^{-5}\ \MeV\ \cm^{-2}\ \sec^{-1}$ at energies of about 100 MeV, slightly lower than our predictions for the total MeV background including bremsstrahlung and $\gamma$-ray lines.

Overall, given the differing methods of other calculations and some large uncertainties in the input parameters, our results are broadly consistent with previous predictions.  

\subsection{Constraints from spectral shape}
The star-forming galaxy contribution to the GeV $\gamma$-ray background remains unclear, despite the detection of several star-forming galaxies (both normal and starburst) by \emph{Fermi}.  In addition, our predictions rely on assumptions about the star-formation rate of the model galaxies, and the Milky Way needs an additional scaling factor to match its known $\gamma$-ray luminosity.  However, star-forming galaxies obviously contribute $\le 100\%$ of the GeV background.  Assuming we know the SED of star-forming galaxies from MeV to GeV energy ranges, we can set an upper limit on the star-forming galaxy contribution to the MeV background by normalizing the SED so it touches but never exceeds the GeV background.

Table~\ref{table:SEDConstraints} lists the maximum fractions of the observed MeV background using different models normalized to the observed GeV background at $E_{\rm norm}$.  

\begin{figure*}
\centerline{\includegraphics[width=9cm]{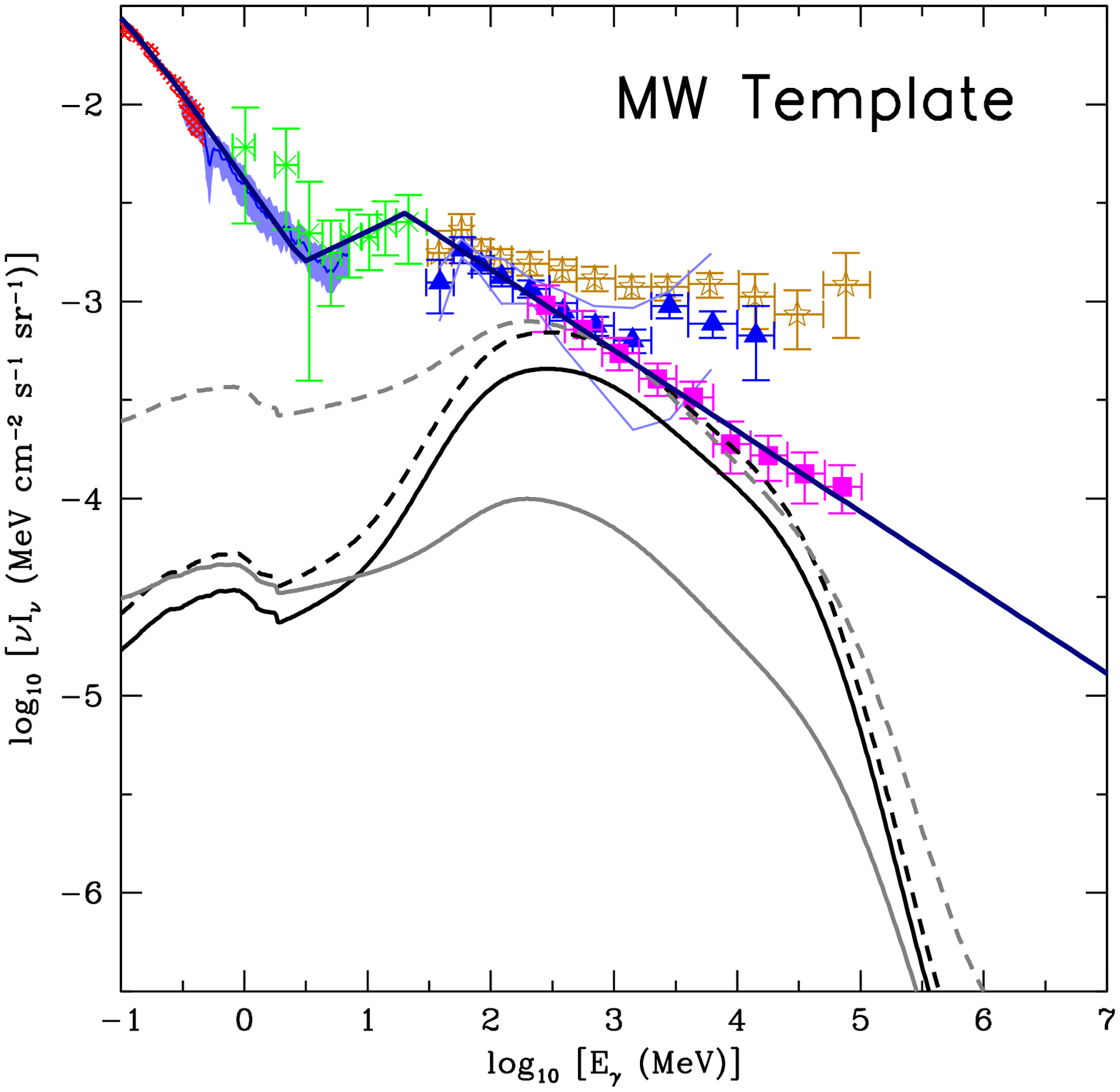}\includegraphics[width=9cm]{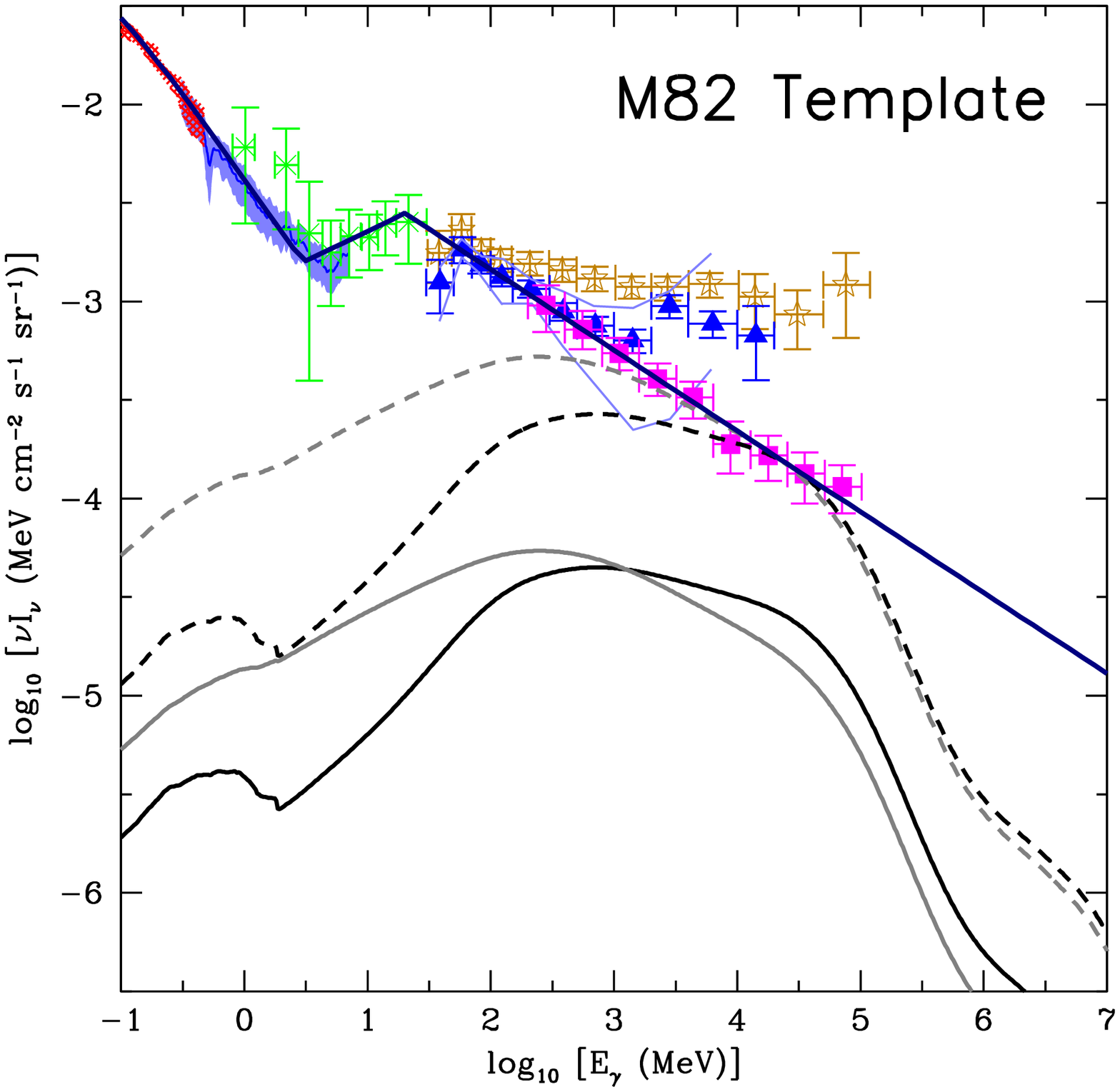}}
\figcaption[figure]{Predicted $\gamma$-ray background (solid) from normal galaxies (left) and starbursts (right) using our medium starburst fraction.  On left, we show the $h = 1000\ \pc$ model of the Milky Way with (black) and without (grey) evolution.  On the right, we show {\bfnop the} fiducial low-$B$ (grey) and high-$B$ (black) models.  Also shown are the spectra normalized to fit the observed background (dashed).  In each case, another large component at energies below 100 MeV is needed to get the shape of the background correct.\label{fig:MWScaledBackground}\label{fig:M82ScaledBackground}}
\end{figure*}

The shape of the normal galaxy $\gamma$-ray SED is incompatible with normal galaxies being the majority of the MeV background.  The shape of the spectrum constrains normal galaxies to be a half or less of the 100 MeV background, $\la 25\%$ of the 30 MeV background, $\la 4 - 20\%$ of the 10 MeV background, and $\la 1 - 10\%$ of the 1 MeV background, with the evolving templates on the low end of these ranges.  The discrepancy in shape is shown in Figure~\ref{fig:MWScaledBackground}: there is a large hump of additional energy near 10 MeV that normal galaxies cannot explain.

For starburst galaxies, the MeV contribution is typically even smaller, although the exact fraction of the MeV background depends on the magnetic field strength.  The contribution from starbursts does not need to be multiplied by much (factor $\la 2$) to account for the GeV background, as long as the starburst fraction is high; with the medium starburst fraction, the emission must be scaled up by a factor $\sim 6 - 25$ to reach the GeV background.  In high{\bfnop-}$B$ (hadronic) models, starbursts are only $\la 2\%$ of the 10 MeV background (black lines in Figure~\ref{fig:M82ScaledBackground}).  Only a fraction of one percent of the 1 MeV background could be nonthermal emission from high{\bfnop-}$B$ starburst galaxies.  Low{\bfnop-}$B$ (leptonic) models have larger allowed MeV fractions, but still make up a small minority of the MeV background: $\la 10 - 20\%$ of the 10 MeV background, and $\la 2 - 7\%$ of the 1 MeV background (grey lines in Figure~\ref{fig:M82ScaledBackground}).  As Figure~\ref{fig:M82ScaledBackground} shows, starbursts cannot be the main source of the MeV background either.

\begin{deluxetable*}{llllccccc}
\tabletypesize{\scriptsize}
\tablecaption{Constraints From Spectral Shape}
\tablehead{\colhead{Model} & \colhead{$f_{\rm SB}$\tablenotemark{a}} & \colhead{$E_{\rm norm}$\tablenotemark{b}} & \colhead{Scale\tablenotemark{c}} & \multicolumn{5}{c}{Maximum fraction of $\gamma$-ray background\tablenotemark{d}} \\ & & & & \colhead{100 keV} & \colhead{300 keV} & \colhead{1 MeV} & \colhead{10 MeV} & \colhead{100 MeV} \\ & & \colhead{GeV} & & \colhead{$\%$} & \colhead{$\%$} & \colhead{$\%$} & \colhead{$\%$} & \colhead{$\%$}}
\cutinhead{Normal Galaxies: Milky Way templates}
$h = 1000\ \pc$ (Non-evolving)  & H10  & 1.1 & 7.0 & 0.90 & 2.8 & 8.5 & 15 & 49 \\
                                & 15\% & 1.1 & 8.0 & 0.90 & 2.8 & 8.5 & 15 & 49 \\ 
                                & T07  & 1.5 & 25  & 0.41 & 1.6 & 6.6 & 10 & 36 \\
$h = 1000\ \pc$ (Evolving)      & H10  & 1.2 & 1.3 & 0.096 & 0.36 & 1.2 & 3.1 & 38 \\
                                & 15\% & 1.2 & 1.5 & 0.095 & 0.36 & 1.2 & 3.2 & 38 \\
                                & T07  & 1.5 & 13  & 0.19  & 0.82 & 3.3 & 5.6 & 31 \\
$h = 2000\ \pc$ (Non-evolving)  & H10  & 1.1 & 6.0 & 1.4   & 3.9  & 12  & 21  & 52 \\
                                & 15\% & 1.1 & 6.8 & 1.4   & 3.9  & 12  & 21  & 52 \\
                                & T07  & 1.7 & 21  & 0.66  & 2.2  & 8.4 & 15  & 40 \\
$h = 2000\ \pc$ (Evolving)      & H10  & 1.0 & 1.2 & 0.13  & 0.43 & 1.4 & 3.6 & 42 \\
                                & 15\% & 1.0 & 1.3 & 0.13  & 0.43 & 1.4 & 3.6 & 42 \\
                                & T07  & 1.4 & 12  & 0.33  & 1.2  & 4.4 & 8.3 & 34 \\
\cutinhead{Starbursts: M82 templates}
$B = 150\ \muGauss$ & H10  & 1.5 -- 18 & 38 -- 97  & 0.15 -- 0.40 & 0.63 -- 1.6 & 2.5 -- 6.4 & 9.2 -- 23 & 29 -- 55\\
                    & 15\% & 1.6 -- 21 & 7.2 -- 18 & 0.14 -- 0.36 & 0.58 -- 1.5 & 2.4 -- 6.1 & 8.5 -- 21 & 27 -- 52\\
                    & T07  & 1.4 -- 19 & 1.5 -- 3.7 & 0.16 -- 0.40 & 0.63 -- 1.6 & 2.5 -- 6.5 & 9.2 -- 23 & 29 -- 55\\
$B = 200\ \muGauss$ & H10  & 7.6 -- 23 & 30 -- 130 & 0.085 -- 0.38 & 0.34 -- 1.6 & 1.3 -- 5.8 & 4.7 -- 19 & 19 -- 50\\
                    & 15\% & 7.4 -- 25 & 5.6 -- 25 & 0.077 -- 0.34 & 0.31 -- 1.4 & 1.2 -- 5.6 & 4.3 -- 18 & 18 -- 47\\
                    & T07  & 7.8 -- 23 & 1.1 -- 5.1 & 0.084 -- 0.38 & 0.34 -- 1.6 & 1.3 -- 5.8 & 4.7 -- 19 & 19 -- 50\\
$B = 250\ \muGauss$ & H10  & 21 -- 24  & 31 -- 83  & 0.069 -- 0.18 & 0.28 -- 0.76 & 0.95 -- 2.6 & 3.4 -- 8.4 & 17 -- 27\\
                    & 15\% & 24 -- 27  & 5.8 -- 16 & 0.061 -- 0.16 & 0.26 -- 0.71 & 0.93 -- 2.6 & 3.1 -- 7.8 & 15 -- 26\\
                    & T07  & 21 -- 23  & 1.2 -- 3.1 & 0.068 -- 0.18 & 0.28 -- 0.76 & 0.96 -- 2.6 & 3.5 -- 8.4 & 17 -- 28\\
$B = 300\ \muGauss$ & H10  & 23       & 31 -- 62  & 0.058 -- 0.11 & 0.24 -- 0.47 & 0.76 -- 1.5 & 2.7 -- 4.8 & 15 -- 20\\
                    & 15\% & 26 -- 27  & 5.9 -- 12 & 0.052 -- 0.10 & 0.22 -- 0.44 & 0.75 -- 1.5 & 2.4 -- 4.4 & 14 -- 18\\
                    & T07  & 23 -- 24  & 1.2 -- 2.3 & 0.057 -- 0.11 & 0.24 -- 0.47 & 0.76 -- 1.5 & 2.7 -- 4.8 & 15 -- 20\\
$B = 400\ \muGauss$ & H10  & 24       & 32 -- 45  & 0.047 -- 0.064 & 0.19 -- 0.27 & 0.55 -- 0.79 & 1.9 -- 2.4 & 13 -- 15\\
                    & 15\% & 28       & 6.0 -- 8.5 & 0.042 -- 0.057 & 0.18 -- 0.25 & 0.56 -- 0.80 & 1.7 -- 2.2 & 12 -- 13\\
                    & T07  & 24       & 1.2 -- 1.7 & 0.046 -- 0.063 & 0.19 -- 0.27 & 0.56 -- 0.80 & 1.9 -- 2.5 & 14 -- 15
\enddata
\label{table:SEDConstraints}
\tablenotetext{a}{Starburst fraction used in the model.  See Table~\ref{table:GammaBackSummary}.}
\tablenotetext{b}{The energy where the ratio of the predicted $\gamma$-ray background to the observed $\gamma$-ray background reaches its maximum.}
\tablenotetext{c}{Maximum allowed scaling of the predicted $\gamma$-ray background for that model that does not overproduce the observed $\gamma$-ray background; it is simply the reciprocal of the maximum ratio of the predicted $\gamma$-ray background to the observed $\gamma$-ray background.}
\tablenotetext{d}{Maximum fraction of the $\gamma$-ray background at each energy allowed by the spectral shape of each model.  It reaches 100\% at $E_{\rm norm}$.}
\end{deluxetable*}

We therefore conclude that star-forming galaxies cannot make up a significant portion of the MeV background, even if the overall $\gamma$-ray production efficiency is much higher than we suppose.  Either (1) some other source of MeV emitters is needed, (2) some unknown emission process make star-forming galaxies much brighter in MeV emission than we expect (and much brighter per unit star-formation than the Milky Way appears to be), or (3) the claimed cosmic MeV background is incorrect.

\section{Conclusions}
\label{sec:Conclusion}

The origin of the MeV background remains mysterious, assuming the claimed intensity of the background is correct.  Star-forming galaxies are an attractive source for the GeV background because they are ubiquitous, the physics is at least partly understood (especially in the Milky Way), and several have now been detected by \emph{Fermi} and TeV telescopes.  However, most of their GeV emission is likely pionic, and therefore rapidly falls off below $\sim 100\ \MeV$; hence this process cannot be responsible for the MeV background.  We have considered possible processes that can provide MeV emission, including (1) leptonic Inverse Compton and bremsstrahlung emission produced by CR $e^{\pm}$ in star-forming galaxies, (2) $\gamma$-ray line and positron annihilation radiation, and (3) intergalactic cascades that shift energy from the TeV band to the MeV-GeV band.  Leptonic radiation overwhelmingly dominates the MeV emission of star-forming galaxies, although intergalactic cascades contribute significantly for starburst galaxies which are observed to have hard spectra extending to the TeV.

We have modeled the Milky Way and the starburst M82 placed at various redshifts.  From these models, we have assembled the expected spectrum of the $\gamma$-ray background, taking into account redshift and intergalactic cascades.  Our primary results are:

\begin{itemize}

\item If we use our evolving model Milky Way templates, then the increased density of gas in high-$z$ galaxies leads to a prominent pion bump.  The MeV emission per unit star-formation slightly decreases with redshift in this model.  In contrast, in the non-evolving models of normal galaxies, the MeV emission of normal galaxies is greatly enhanced by the increased Inverse Compton losses from the CMB at high $z$, ``filling in'' the pion bump (Figure~\ref{fig:MWatHighz}).

\item Starburst galaxies have relatively low MeV emission compared to their strong GeV emission.  The relatively strong proton losses imply a strong pion bump, since there is more power in CR protons than in CR $e^{\pm}$.  Cascade emission increases the MeV emission by a factor $\la 2$, nowhere near enough to fill in the pion bump.  This conclusion can be altered in ``low-$B$'' models of starbursts, in which the GeV emission comes mostly from Inverse Compton emission (Figure~\ref{fig:M82Gamma}).

\item We find with our ``medium'' starburst fraction (15\% of star-formation at all $z$) and using the evolving normal galaxy templates, that star-forming galaxies are the majority of the GeV $\gamma$-ray background, with most coming from the normal galaxies (Figure~\ref{fig:GammaBackground}).  However, the effective starburst fraction is highly uncertain and depends on gas evolution in normal galaxies.  As the fraction of cosmic star formation in starbursts approaches unity, the observed GeV $\gamma$-ray background is produced even if the normal galaxies do not evolve (Figure~\ref{fig:fSBBack}).  The uncertainty in the GeV background prediction is roughly an order of magnitude.

\item In our fiducial model, star-forming galaxies make up only $\sim 1\%$ of the 1 MeV background and $\sim 3\%$ of the 10 MeV background (Figure~\ref{fig:GammaBackground}).  Much of this emission is IC from normal galaxies, but there are also important contributions from $\gamma$-ray lines from CCSNe, positron annihilation, and $^{26}$Al decay.  As with the GeV background, we find an order of magnitude spread in the MeV background predictions depending on our assumptions.

\item To account for the uncertainties in $\gamma$-ray luminosity normalization, we derive more robust limits on the MeV contribution from the spectral shape of star-forming galaxies.  According to these constraints, normal galaxies could make at most half of the 100 MeV background and 1 (evolving) to 9\% (non-evolving) of the 10 MeV background (Figure~\ref{fig:MWScaledBackground}). Similarly, high{\bfnop-}$B$ starbursts could make up at most $\sim 15\%$ of the 100 MeV background and 2\% of the 10 MeV background.  If starbursts have weak $B$ and substantial leptonic emission, these fractions could be higher, with $\sim 10 - 20\%$ of the 10 MeV background allowed to be from starbursts (Figure~\ref{fig:M82ScaledBackground}).
\end{itemize}

In short, while star-forming galaxies can (and according to our models, do) provide a substantial fraction of the GeV $\gamma$-ray background, they are a minor contributor of the MeV background.  The reason why the MeV background cannot be explained by star-forming galaxies is simply that the $\gamma$-ray background is \emph{very steep}: a lot more power is required at MeV energies than at GeV energies.  Thus, even if star-forming galaxies do have enough luminosity at GeV energies, and even if the spectrum is flat at MeV energies (as is the case for the non-evolving normal galaxy templates), this is insufficient for the observed background.

\begin{figure}
\centerline{\includegraphics[width=9cm]{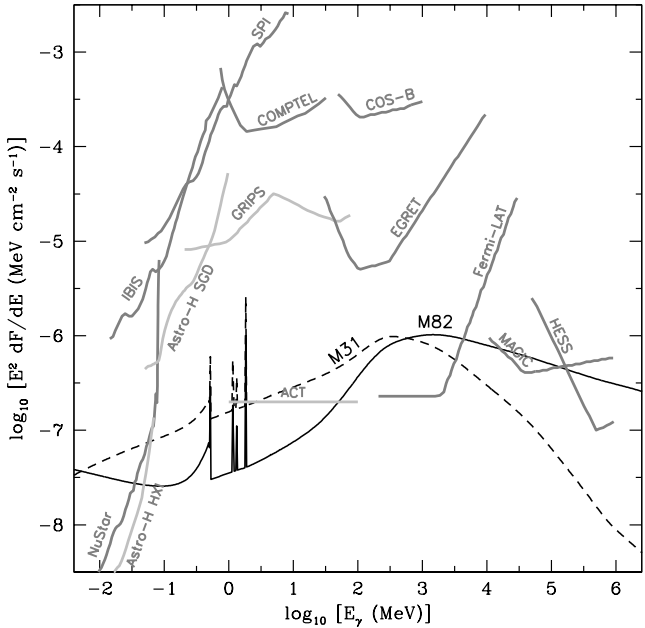}}
\figcaption[figure]{Fiducial spectra of M82 ({\bfnop high-$B$}; black solid) and M31 (black dashed; rescaled from MW) compared to existing (grey) and proposed (light grey) $\gamma$-ray instruments.  Even with a $10^3$ increase in MeV sensitivity over COMPTEL, detection of these galaxies will be difficult.  The sensitivity curves assume exposure times of $10^6$ seconds and that $\Delta E = E$, and are taken from \citet{Koglin05}, \citet{Kanbach04}, \citet{Tajima05}, \citet{Bloser06}, and \citet{Greiner11}. The proposed MEGA instrument would have a similar sensitivity as GRIPS. \label{fig:FutureMeV}}
\end{figure}

Thus, there must be some other explanation for the MeV background.  The simplest explanation would be some other source that peaks at MeV energies.  But this just again raises the mystery of why the MeV-GeV $\gamma$-ray background appears to be a featureless power law if there are different sources responsible at different energies \citep{Stecker10}.  It is possible this is sheer coincidence.  For example, current blazar models of the $\gamma$-ray background invoke a coincidence with soft Flat Spectrum Radio Quasars at low energies and hard BL Lacs at high energies \citep[e.g.,][]{Venters11}.  We should also keep in mind that the featureless power law in Figure~\ref{fig:GammaBackground} is actually for the currently unresolved background.  The distinction between unresolved and resolved sources is an experimental one (see the huge sensitivity gap in Figure~\ref{fig:FutureMeV} between COMPTEL and \emph{Fermi}-LAT), not necessarily one with fundamental astrophysical significance.  The resolution of sources is much better at GeV energies than MeV energies, and the \emph{resolved} $\gamma$-ray blazars would add a large component on top of the unresolved background.  Finally, the reported MeV background could simply be incorrect (due to detector backgrounds) or Galactic in origin.

Our conclusions can be evaded if there are extremely high amounts of low energy CRs below the pion production threshold, or extremely large contributions from discrete MeV sources.  IC emission from escaping $e^{\pm}$ in the halos around normal galaxies can increase the MeV emission, but since the IC emission is a broad continuum extending to GeV energies, the spectral shape constraints imply that this emission does not make up all of the MeV background.  Low energy spikes of CR electrons can enhance the bremsstrahlung and IC emission, but can be constrained with synchrotron radio.  An extreme enhancement of MeV nuclei might even increase the nuclear line emission enough to alter the SED, although the ionization rate from these nuclei should also be extreme.  Discrete sources are not expected to be dominant, based on observations of the Milky Way, so a new significant MeV source population would have to scale non-linearly with star-formation rate.  

There are several indirect tests of our predictions.  A key distinction between our evolving and non-evolving templates for normal galaxies is that Inverse Compton on the CMB losses overwhelmingly dominate the electron lifetime in the latter case.  This comes at the expense of other losses -- particularly synchrotron.  Thus at high $z$, it has long been expected that galaxies should be radio-faint compared to expectations from their infrared luminosities \citep{Carilli99,Murphy09}.  In contrast, observations indicate that galaxies lie on the empirical $z = 0$ infrared--radio correlation to relatively high $z$ \citep[e.g.,][]{Appleton04,Garn09,Sargent10,Mao11}, or even are radio-bright \citep[e.g.,][]{Kovacs06,Vlahakis07,Murphy09,Ivison10,Michalowski10}.  At high $z$, these observations probe galaxies with high star-formation rates and possibly more starburst-like, where the effects of the CMB are expected to be weak.  However, new instruments like {\bfnop the Karl G. Jansky Very Large Array} and eventually {\bfnop Square Kilometer Array} may probe Milky Way--like galaxies, and determine the effects of the stronger CMB.

Further measurements extending the hard X-rays and the high energy $\gamma$-rays towards the MeV range may prove constraining for the starbursts.  \emph{Fermi}-LAT should ultimately reach down to 30 MeV.  At these energies the pion bump should fall off in the high-$B$ models, but the leptonic emission peaks in the low-$B$ models.  Thus \emph{Fermi}-LAT can help constrain $B$ in the nearest starbursts.  Further \emph{Fermi}-LAT data will resolve out additional blazars \citep{Abazajian10}, which would constrain the true contribution of star-forming galaxies to the GeV background and limit their contribution to the MeV background.  In the hard X-ray regime, measurements of the 10 -- 100 keV luminosity of starbursts would limit their IC emission, helping constrain $B$, and synchrotron emission, which can be significant to $\sim 1\ \MeV$ in some of our models.   

Better MeV instruments may also help us understand the emission mechanisms of $\gamma$-rays in this energy range within the Milky Way, particularly the contribution of discrete sources.  As noted in section~\ref{sec:MWModel}, there is a spectral plateau around $\sim 10\ \MeV$ observed by COMPTEL that is not present in either GALPROP or our models (Figure~\ref{fig:MWGamma}).  The discrepancy is poorly understood, and implies that either there is something we do not know about galactic MeV emission, or there are unaccounted for systematics in the COMPTEL spectrum of the Milky Way.  The discrepancy by itself cannot explain the MeV background -- the Milky Way is less luminous at 10 MeV than at 1 GeV, unlike the background -- but it may increase the predicted background by a factor of a few.

However, direct measurements of the 1 -- 30 MeV emission of star-forming galaxies beyond the Milky Way and possibly the Magellanic Clouds seem far off.  In Figure~\ref{fig:FutureMeV}, we plot the fiducial {\bfnop high-$B$} model of M82 compared to the sensitivities of current (grey) and possible future (light grey) $\gamma$-ray instruments.  In X-rays and GeV-TeV $\gamma$-rays, the highly sensitive instruments we have now allow detection of these starbursts.  However, COMPTEL, the best instrument over much of the MeV band, is about three orders of magnitude behind \emph{Fermi}-LAT.  Nor would proposed next-generation instruments like MEGA \citep[e.g.,][]{Bloser02,Bloser06} or GRIPS \citep{Greiner11} be able to detect the nonthermal emission from M82.  But even a MeV instrument as sensitive as \emph{Fermi}-LAT, such as ACT \citep{Milne02,Boggs06}, might not detect M82 from 1 -- 10 MeV: the pion bump in the fiducial model means that these galaxies are much fainter in the MeV band than in the GeV band.  We also consider M31 by scaling the Milky Way fiducial model to the star-formation rate of M31 ($1\ \Msun\ \yr^{-1}$; \citealt{Williams03}) and using a distance of 780 kpc \citep{Stanek98}.  It may be possible to detect M31 with ACT, because the MeV-GeV spectrum is softer than M82.  Thus, vastly more powerful MeV telescopes are required to directly observe the MeV emission of individual star-forming galaxies; such improvements may be feasible with new $\gamma$-ray optics technologies, as in \citet{Habs12}.

The origin of the MeV (and GeV) background can be tested by the anisotropies in the background \citep{Zhang04}.  Normal galaxies are expected to have a very weak anisotropy signal \citep{Ando09}, while blazars have strong anisotropy signals \citep{Ando07}.  The signal for starbursts has not yet been calculated, but is probably stronger than normal galaxies; anisotropies in the submillimeter associated with submillimeter galaxies have been detected \citep{Amblard11}.  In our fiducial model, normal galaxies contribute more to the background than starbursts at both MeV and GeV energies.  A future MeV experiment could therefore search for the star-forming galaxy contribution by looking for a uniformly distributed component to the MeV background.  Further studies of the starburst contribution to background anisotropies, and a better treatment of galaxy evolution, is necessary to evaluate this effect.

In short, the sheer power of the MeV background as reported requires many luminous sources, and star-forming galaxies simply are not bright enough to be its origin.  How the MeV and GeV backgrounds join up so seamlessly is unclear, since star-forming galaxies seem to make up at least $\sim 10\%$ and plausibly a majority of the GeV background.  Future MeV instruments must be improved, possibly by many orders of magnitude, to solve the riddle of its source.

\acknowledgments
We thank Felix Aharonian, Markus B\"ottcher, Brian Fields, Dmitry Khangulyan, Kohta Murase, Vasiliki Pavlidou, and especially Todd Thompson for discussions.  We thank the referee for detailed and helpful comments on the manuscript.  For the second half of this project, BCL was supported by a Jansky fellowship from the National Radio Astronomy Observatory.  The National Radio Astronomy Observatory is operated by Associated Universities, Inc., under cooperative agreement with the National Science Foundation.  For the first half of this project, BCL was supported in part by an Elizabeth Clay Howald Presidential Fellowship from the OSU.  SH is supported by the Center for Cosmology and Astro-Particle Physics (CCAPP) at the OSU.  JFB was supported by NSF CAREER Grant PHY-0547102 and NSF Grant PHY-1101216.

\appendix
\emph{General approach} -- Our approach to one-zone models is ultimately based on the extensive discussion in \citet{Torres04b}, and is described in detail in \citet{Lacki10c} and \citet{LackiXRay}.

The population of CRs in galaxies is governed by the diffusion-loss equation:
\begin{equation}
-D \nabla^2 N(E) + \frac{N(E)}{\tau(E)} - \frac{d}{dE} [b(E) N(E)] - Q(E) = -\frac{\partial N(E)}{\partial t}
\end{equation}
where $D$ is the spatial diffusion constant, $b(E)$ is the sum of all of the energy loss rates, $Q(E)$ is the CR injection rate for CRs at the energy $E$, $\tau(E)$ includes all catastrophic and escape losses, and $N(E)$ is the CR energy spectrum \citep[see also][]{Strong07}.  Our steady-state one-zone models assume the CR population has no time or spatial dependence within the modeled region, so that the diffusion-loss equation reduces to the leaky box equation:
\begin{equation}
\frac{N(E)}{\tau(E)} - \frac{d}{dE} [b(E) N(E)] - Q(E) = 0.
\end{equation}
This may be solved by the Green's function method described in \citet{Torres04b}.

\emph{CR nuclei heavier than hydrogen} -- In order to estimate the strengths of the CR de-excitation nuclear lines, we treat primary helium-4, carbon-12, and oxygen-16 ions in addition to protons, electrons, and positrons.  However, secondary nuclei are not included.

We normalize these nuclei by the ratio of their rigidity spectra with CR protons' rigidity spectrum, comparing at equal rigidities.  For our momentum power law spectra $dQ/dq = C q^{-p}$, it can be shown that
\begin{equation}
\frac{C_i}{C_p} = \frac{(dQ/dR)_i}{(dQ/dR)_H} \left(\frac{Z_i}{Z_p}\right)^{p - 1},
\end{equation}
where $R = q c / (Ze)$ is the rigidity, and $Z$ is the electric charge of the nucleus ($Z_p = 1$).  The ratio of the helium-4 and hydrogen rigidity spectra is assumed to be 8, as observed in the Milky Way \citep{Webber74,Webber87}.  The abundances of carbon and oxygen are then normalized to helium-4 using the values in \citet{Meyer98}. 

Heavier nuclei can interact with gas in the ISM and not only produce pions, but undergo nuclear reactions as well.  To calculate these lifetimes, we use a modified version of the inelastic cross sections of \citet{Letaw83}.  The cross section for interactions of CR nuclei and ISM protons is
\begin{equation}
\sigma_{\rm inel}^{pZ} (E) = 45 A_{\rm CR}^{0.7} [1 + 0.016\ \sin (5.3 - 2.63 \ln A_{\rm CR})] [1 - 0.62 {\rm e}^{-K / 200\ {\rm MeV}} \sin (10.9 (K / \MeV)^{-0.28})]\ \mbarn
\end{equation} 
for CRs with an atomic mass $A_{\rm CR}$ and kinetic energies $K$ above 10 MeV.  For interactions between CR nuclei and ISM helium nuclei, we replace the $A_{\rm CR}^{0.7}$ term with $[4^{0.35} + A_{\rm CR}^{0.35} - 1]^2$ to get a cross section $\sigma_{\rm inel}^{\alpha Z} (E)$, by analogy with the Glauber rule.  The lifetime of a CR nucleus to nuclear reactions is then 
\begin{equation}
t_{\rm nuclear} = [(\sigma_{\rm inel}^{pZ} (E) n_H + \sigma_{\rm inel}^{\alpha Z} (E) n_{\rm He}) \beta c]^{-1},
\end{equation} 
where $\beta c$ is the speed of the CR.  

\emph{Pionic emission from nuclei} -- While pionic cross sections for proton-proton collisions are given in \citet{Kamae06}, no cross sections are given for proton-helium or helium-helium collisions in that reference.  Since the helium is expected to be a minor contributor to the pionic emission ($\sim 10\%$), we scale the proton-proton cross sections using the \citet{Orth76} approximation:
\begin{equation}
\sigma^{ZZ}_{\pi} (E_Z) = \sigma^{pp}_{\pi} (E_Z / A_{\rm CR}) [A_{\rm CR}^{3/8} + A_{\rm ISM}^{3/8} - 1]^2
\end{equation}
which is applied to the differential cross sections for production of pionic $\gamma$-rays, $e^{\pm}$, and neutrinos.\footnote{The \citet{Orth76} approximation is also used by the GALPROP code for simulations of the Milky Way \citep{Moskalenko98}.}  Roughly speaking, this rule takes into account that nuclei are not transparent to hadronic interactions, with some of the nucleons being shadowed by the others.

\emph{Nuclear lines} -- Nuclei colliding with each other can excite nuclei; the de-excitations then generate $\gamma$-ray lines in the rest-frame of the emitting nucleus.  There are three different cases: a CR nucleus can hit an ISM proton and become excited, a CR proton can hit an ISM nucleus and excite it, and a CR nucleus can penetrate a dust grain.  Excited CR nuclei will generate a broad continuum as their Doppler shift smears out the $\gamma$-ray line, whereas excited ISM nuclei will generate a narrow line.  We neglect the dust grain case, but it would generate an even narrower line as energy losses in the dust grain dampened the nuclear recoil.  

Following \citet{Meneguzzi75}, we calculate the line emissivity of CR nuclei on ISM p/$\alpha$ collisions as 
\begin{equation}
\frac{dQ}{dE} = \frac{1}{2} \sum_{k,i} n_k \int dE^{\prime} \int_{E_{\rm min}^{\prime\prime}}^{E_{\rm max}^{\prime\prime}} dE^{\prime\prime} \frac{dN_i}{dE^{\prime}} (E^{\prime}) c \sigma_{ki} (E^{\prime}) \frac{E^{\prime\prime}}{\gamma_i^{\prime} E^2} \delta (E^{\prime\prime} - E_L),
\end{equation}
where the emissivity is in units of ph $\cm^{-3}\ \sec^{-1}\ \ \erg^{-1}$ (contrast with \citet{Meneguzzi75}, where it is photons per ISM hydrogen atom per steradian), $E_{\rm min}^{\prime\prime} = E / (\gamma (1 + \beta))$ and $E_{\rm max}^{\prime\prime} = E / (\gamma (1 - \beta))$. 

For the line emissivity of CR p/$\alpha$ on ISM nuclei, we use
\begin{equation}
\frac{dQ}{dE} = \sum_{k,i} \frac{n_{k}}{2 \Delta E_L} \int dE^{\prime} \sigma_{ik} (E^{\prime}) c \beta_i^{\prime} \frac{dN_i}{dE^{\prime}}(E^{\prime}) \times (|E - E_L| \le \Delta E_L).
\end{equation}
For simplicity, we assume the nuclear line has a line half-width of $\Delta E_L$, which we take to be 40 keV.  Since we are interested in the cosmic background, redshift will smooth out narrow features in any case, so the exact line width is not that important.

We use the line cross sections from \citep{Kozlovsky02}.  Included lines are 429 and 478 keV lines from $^4$He (broad components only since the projectile and target are both the same), 4.438 MeV and 15.1 MeV lines from $^{12}$C (where the 4.438 MeV line can be created in collisions with $^{14}$N and $^{16}$O as well because $^{12}$C is created in the collision), and 6.129 MeV, 6.916 MeV, and 7.115 MeV lines from $^{16}$O.  The 4.438 MeV $^{12}$C line is generally expected to be the strongest \citep{Ramaty79}.  

\emph{Bremsstrahlung} -- In this paper, our models use the bremsstrahlung cross sections from \citet{Strong00}.  An electron with kinetic energy $K_e$ interacts with an atom of number $Z$ and atomic mass $N$, produces a photon with energy $k m_e c^2$, and leaves with a kinetic energy $K_e - k m_e c^2 > 0$.  We include the effects of hydrogen and helium in the ISM; metals have a negligible effect on the total cross section.  \citet{Strong00} consider three regimes for the electron kinetic energy: 10 keV -- 70 keV, 70 keV -- 2 MeV, and more than 2 MeV.  They also consider the Fano-Sauter limit when $E_{\gamma} \approx K_e$, because the other approximations to the bremsstrahlung cross section drop off too quickly.  We use the Fano-Sauter limit when it gives a bigger cross section than the other approximations, although we find in practice this only happens when $k m_e c^2$ is very nearly $K_e$.  

The bremsstrahlung loss time is then calculated by directly integrating up the cross sections:
\begin{equation}
-\frac{dE}{dt} = c \sum_j n_j \int_0^E dE_{\gamma} \frac{E_{\gamma}}{m_e c^2} \frac{d\sigma_j (E_{\gamma}, E_e)}{dk}
\end{equation}
\citep{Schlickeiser02}.

\emph{Positron annihilation} -- Positrons have a small but significant probability of annihilating in-flight, before thermalizing.  The positron annihilation cross section is given as 
\begin{equation}
\sigma_{\rm ann} = \frac{2 \pi r_0^2}{\tau^2 (\tau - 4)} \left((\tau^2 + 4\tau - 8) \log \frac{\sqrt{\tau} + \sqrt{\tau - 4}}{\sqrt{\tau} - \sqrt{\tau - 4}} - (\tau + 4) \sqrt{\tau (\tau - 4)}\right)
\end{equation}
where $\tau = 2 (1 + \gamma)$ \citep[e.g.,][]{Beacom06,Vietri08}.  Then the in-flight annihilation time is $(n_e \sigma_{\rm ann} \beta_e c)^{-1}$, where $n_e$ is the number density of both free and bound electrons.  

The inflight annihilation spectrum is then
\begin{equation}
\frac{dQ_{\rm ann}(\epsilon_{\gamma})}{d\epsilon_{\gamma}} = \frac{3 \sigma_T^2 c n_e}{8 \gamma_+ p_+} \left[\left(\frac{\epsilon_{\gamma}}{\gamma_+ + 1 - \epsilon_{\gamma}} + \frac{\gamma_+ + 1 - \epsilon_{\gamma}}{\epsilon_{\gamma}}\right) + 2\left(\frac{1}{\epsilon_{\gamma}} + \frac{1}{\gamma_+ + 1 - \epsilon_{\gamma}}\right) - \left(\frac{1}{\epsilon_{\gamma}} + \frac{1}{\gamma_+ + 1 - \epsilon_{\gamma}}\right)^2\right].
\end{equation}
where $\epsilon_{\gamma} = E_{\gamma} / (m_e c^2)$, $p_+ = \sqrt{\gamma_e^2 - 1}$, and $\gamma_+ \ge (1 - 2 \epsilon_{\gamma} + 2 \epsilon_{\gamma}^2) / (2 \epsilon_{\gamma} - 1)$ \citep[e.g.,][]{Aharonian04-Book}.

Positrons are more likely to cool by ionization and annihilate after thermalizing than at relativistic speeds.  Positrons are conveyed towards rest in energy space at a ``speed'' $|dE/dt|$.  The rate at which positrons annihilate at low energies in equilibrium is therefore $Q_+^{\rm surv} = dN/dE \times |dE/dt| \times f_{\rm surv} (E)$, where $f_{\rm surv} (E)$ is the probability that positrons survive cooling to rest without being catastrophically lost (through inflight annihilation or escape).  We take our lowest energy bin (with kinetic energy $K_0 = 100\ \keV$) and assume $f_{\rm surv} (K_0) = 1$ to calculate the positron annihilation rate as
\begin{equation}
Q_+^{\rm surv} = \frac{dN}{dE}(K_0) b(K_0).
\end{equation}
Note that this is not the rate at which positrons are injected at all energies: positrons can escape the system through diffusive and advective losses at high energy and avoid annihilation, or annihilate in-flight.  

Annihilation at rest can either occur through direct annihilation or through positronium formation.  Direct annihilation leads to two 511 keV line photons.  Positronium formation, which occurs $f_{\rm Ps} \approx 90 - 95\%$ of the time in the Milky Way, also leads to two 511 keV line photons about 1/4 of the time.  We arbitrarily assume a line half-width of $\Delta E = 5\ \keV$ (which will be smeared out for galaxies at different $z$ anyway) to calculate the 511 keV annihilation line from CR positrons as:
\begin{equation}
\frac{dQ}{dE} = \frac{(1 - 3 f_{\rm Ps} / 4) Q_+^{\rm surv}}{2 \Delta E}
\end{equation}
We assume $f_{\rm Ps} = 0.9$ for this work.  Positronium annihilates into three photons with a continuum distribution of energies below 511 keV 3/4 of the time.  The positronium continuum spectrum is given in \citet{Ore49} and \citet{Guessoum91}:
\begin{equation}
\frac{dQ}{dE} = \frac{6}{(\pi^2 - 9) m_e c^2} \left[\frac{\epsilon (1 - \epsilon)}{(2 - \epsilon)^2} + \frac{2 (1 - \epsilon)}{\epsilon^2} \log(1 - \epsilon) - 2 \frac{(1 - \epsilon)^2}{(2 - \epsilon)^3} \log(1 - \epsilon) + \frac{2 - \epsilon}{\epsilon}\right] \times \frac{3 Q_+^{\rm surv} f_{\rm Ps}}{4},
\end{equation}
where $\epsilon = E / (m_e c^2)$ and we have already taken into account that each annihilation produces three photons.

\emph{$\gamma\gamma$ Absorption} --  At TeV energies, $\gamma\gamma$ absorption process may become important.  This process converts $\gamma$-rays into $e^{\pm}$, which can radiate in Inverse Compton \citep{Inoue11-Pairs} or synchrotron \citep{LackiXRay}.  The interior $\gamma$-ray photon density is calculated using the uniform slab model, where the absorbing and emitting regions are cospatial: 
\begin{equation}
N_{\gamma} = \frac{Q_{\gamma} h}{c \tau_{\gamma\gamma} (h)} [1 - \exp(-\tau_{\gamma\gamma}(h))]
\end{equation}
where $Q_{\gamma}$ is the rate at which $\gamma$-rays are injected per unit volume, $h$ is the midplane-to-edge scale height, and $\tau_{\gamma\gamma} (h)$ is the midplane-to-edge optical depth to $\gamma\gamma$ absorption.  To calculate $\tau_{\gamma\gamma} (h) = h \int n(\epsilon) \sigma_{\gamma\gamma} (\epsilon, E_{\gamma}) d\epsilon$ for a radiation field $n(\epsilon)$, we use the approximation for the $\gamma\gamma$ cross sections $\sigma_{\gamma\gamma} (\epsilon, E_{\gamma})$ given in \citet{Aharonian04-Book}.  The spectrum of these pairs is then calculated using the \citet{Aharonian83} source functions.  

For the Earth-observed spectrum of M82, we again use a uniform slab model, but substitute the galaxy radius $R$ for $h$, since it is viewed nearly edge-on:
\begin{equation}
F_{\gamma}^{\oplus} = F_{\gamma}^{\rm unabs} [1 - \exp(-\tau_{\gamma\gamma}(R))] / \tau_{\gamma\gamma} (R).
\end{equation}
where $F_{\gamma}^{\rm unabs}$ is the unabsorbed $\gamma$-ray flux.

Finally when constructing the background, we assume that a galaxy is typically observed face-on.  This means that the sightline within the disk will typically have length of order $\sim h$.  However, the radiation field of a galaxy extends out to a distance $\sim R > h$.  Thus $\gamma$-rays in the halo of the galaxy/starburst may be absorbed by the $\gamma\gamma$ process.  The pair $e^{\pm}$ in the halo will not radiate in bremsstrahlung; though they will radiate IC and possibly synchrotron, depending on how far out the magnetic field extends.  To be conservative we do not include any emission from these halo $e^{\pm}$.  Instead, we apply the uniform slab model with sightline $h$ to the galaxy itself, and then treat the halo as a foreground screen of length $R$:
\begin{equation}
F_{\gamma}^{\rm back} (E_{\gamma}) = F_{\gamma}^{\rm unabs} \frac{1 - \exp(-\tau_{\gamma\gamma}(h))}{\tau_{\gamma\gamma} (h)} \exp(-\tau_{\gamma\gamma}(R)),
\end{equation}
where we assume the radiation field is constant out to a radius $R$.

\end{document}